\documentclass[10pt]{article}
\usepackage{bm,amsmath,epsf,epsfig,float,amssymb,latexsym}
\usepackage{graphics,psfrag}
\usepackage{cite}
\usepackage{citesort}
\newcommand{\meas}{\frac{d^3k}{(2\pi)^3}}
\newcommand{\mk}{|\vk|}

\newcommand{\va}{\vec{\lambda}}
\newcommand{\val}{\mathbf{a}}

\newcommand{\cT}{{\cal T}}

\newcommand{\vp}{{\mathbf{p}}}
\newcommand{\vq}{{\mathbf{q}}}
\newcommand{\vx}{{\mathbf{x}}}
\newcommand{\vy}{{\mathbf{y}}}
\newcommand{\vv}{{\mathbf{v}}}
\newcommand{\vQ}{{\mathbf{Q}}}

\newcommand{\vk}{{\mathbf{k}}}

\newcommand{\vr}{\mathbf{r}}
\newcommand{\vsi}{\vec{\sigma}}

\newcommand{\Opd}{{\mathcal{O}}(p^2)}

\newcommand{\bg}{\begin{align}}
\newcommand{\eeg}{\end{align}}
\newcommand{\be}{\begin{equation}}
\newcommand{\ee}{\end{equation}}
\newcommand{\ba}{\begin{eqnarray}}
\newcommand{\ea}{\end{eqnarray}}

\newcommand{\nn}{\nonumber}

\newcommand{\barr}[1]{\not\mathrel #1}

\newcommand{\ve}{\varepsilon}

\newcommand{\vs}{\vspace{-0.15cm}}
\newcommand{\la}{\langle}
\newcommand{\ra}{\rangle}
\newcommand{\si}{\sigma}
\newcommand{\vz}{\mathbf{z}}

\newcommand{\ep}{\epsilon}

\begin{document}

\thispagestyle{empty}

\vspace{0.5cm}

\hfill{\tiny HISKP-TH-09/20, FZJ-IKP(TH)-2009-18}

\vspace{2cm}

\begin{center}
{\Large{\bf Non-perturbative methods for a chiral effective field
    theory\\[0.3em] 
   of finite density nuclear systems}}
\end{center}
\vspace{.5cm}

\begin{center}
{\Large  A.~Lacour$^{a}$,  J.~A.~Oller$^{b}$,  and U.-G.~Mei{\ss}ner$^{a,c}$}
\vskip 10pt
{\it  $^a$Helmholtz-Institut f\"ur Strahlen- und Kernphysik (Theorie) and
  Bethe Center for Theoretical Physics\\ Universit\"at Bonn,
D-53115 Bonn, Germany}\\
{\it  $^b$Departamento de F\'{\i}sica, Universidad de Murcia, E-30071 Murcia, 
Spain}\\
{\it $^c$Institut f\"ur Kernphysik, Institute for Advanced Simulation and
J\"ulich Center for Hadron Physics\\Forschungszentrum J\"ulich, D-52425
J\"ulich,
Germany}
\end{center}

\vspace{1cm}
\noindent
\begin{abstract}

Recently we have developed a novel chiral power counting scheme for an 
effective
field theory of nuclear matter with nucleons and pions as degrees of 
freedom
\cite{nlou}. It allows for a systematic expansion taking into account both 
local as well as pion-mediated multi-nucleon interactions. We apply this 
power
counting in the present study to the evaluation of the pion self-energy 
and
the  energy density in nuclear and neutron matter at next-to-leading order. 
To implement this power counting in actual calculations we develop here a 
non-perturbative
method based on Unitary Chiral Perturbation Theory for performing the 
required resummations. We show explicitly that the contributions to the 
pion
self-energy with in-medium nucleon-nucleon interactions to this 
order cancel. The main trends for the energy density of symmetric nuclear 
and 
neutron matter are already reproduced at next-to-leading order. 
In addition, an accurate description of the neutron matter equation of 
state,  
as compared with  sophisticated many-body calculations, is obtained by 
varying 
only slightly a subtraction constant around its expected value. The case 
of symmetric nuclear matter requires the introduction of an additional
fine-tuned subtraction constant, parameterizing the effects from higher 
order
contributions. With that, the empirical saturation point and 
the nuclear matter incompressiblity are well reproduced while the energy 
per 
nucleon as a function of density closely  agrees with sophisticated  
calculations in the literature. 

\end{abstract} 

\newpage

\section{Introduction}
\def\theequation{\arabic{section}.\arabic{equation}}
\setcounter{equation}{0}
\label{sec:int}

In the last decades Effective Field Theory (EFT) has been applied to an
increasingly wider range of phenomena, e.g. in condensed matter, nuclear and
particle physics. An EFT is based on a power counting that establishes a 
hierarchy between the infinite amount of contributions. At a given
order in the expansion only a finite amount of them has to be considered. The
others are
suppressed and constitute higher order contributions.  In this work we employ  
Chiral Perturbation Theory (CHPT) \cite{wein,wein1,wein2},  which is the
low-energy EFT of  QCD and takes pions (and nucleons as well for our present
interests) as the degrees of
freedom. 
CHPT  is related to the underlying theory of strong interactions, QCD,
because it shares the same symmetries, their breaking and low-energy spectrum. 
It has been successfully applied to the lightest nuclear systems of two,
three and four nucleons
\cite{ordo,kolck,entem,epe,epeprl,eperp,Epelbaum:2008ga}. Nonetheless,
still some issues are raised concerning the full consistency of the approach
and variations of the power counting have been suggested
\cite{kaplan,kswnnlo,bean,timm,epereply,pavon,pera,soto}. A common technique for
heavier nuclei
is to employ the chiral nucleon-nucleon potential delivered by CHPT in
standard many-body algorithms \cite{krew,majo}, sometimes supplied with
renormalization group 
techniques \cite{schaefer,furni}. One issue of foremost present interest  is
the role of multi-nucleon interactions involving three or more nucleons  
in nuclear matter and nuclei \cite{nogga,kaiser,krew,Epelbaum:2008ga,furni}.

Ref.~\cite{prcoller} derived many-body field theory from quantum field
theory by considering nuclear matter as a continuous set of free nucleons at 
asymptotic times. The generating functional of CHPT in the presence of
external  sources was deduced, similarly as in the pion and pion-nucleon 
sectors \cite{gl1,sainio}. These results were applied in ref.~\cite{annp} to
study CHPT in nuclear matter, but including only nucleon interactions due to 
pion exchanges. Thus, the local nucleon-nucleon (and multi-nucleon)
interactions were neglected. This approach was later extended in
ref.~\cite{girnalda} to finite nuclei and the
pion-nucleus optical potential is calculated up to ${\cal O}(p^5)$.  In
ref.~\cite{nlou} an extended power counting
is derived that takes into account simultaneously short- and long-range 
multi-nucleon interactions. Notice that  many present applications  of CHPT to 
nuclei and nuclear matter
\cite{kaiser,kai1,annp,kai2,kai3,wirzba,lutz,hardrock,osetdo,osetnie,kalas} 
only consider meson-baryon Lagrangians.  Short-range interactions  are
included without being fixed from the free nucleon-nucleon
scattering. E.g. \cite{lutz} fits the in-medium local nucleon-nucleon
interaction, in terms of just one free parameter, to reproduce the saturation
properties of symmetric nuclear matter. In addition, the nucleon propagators do
not always count as
$1/k$, with $k$ a typical nucleon three-momentum,  but often they do 
as the inverse of a nucleon kinetic energy, $m/k^2$
(with $m$ the nucleon mass), so that they are unnaturally large. This is well
known since the seminal papers of 
Weinberg \cite{wein1,wein2}.  This fact invalidates the straightforward
application of the pion-nucleon 
power counting valid in vacuum as applied e.g. in
refs.~\cite{annp,korean,kai1,kai2,kai3}.

We implement here non-perturbative methods to perform actual calculations
employing the power counting of ref.~\cite{nlou}, which requires the
resummation of  some series of in-medium two-nucleon reducible diagrams. We
employ the  techniques of Unitary CHPT (UCHPT) \cite{npa,nd,higgs,meis} that
are extended to the nuclear medium systematically in a way consistent  with
the chiral power counting of ref.~\cite{nlou}. Our theory is applied to the
problem of calculating up to next-to-leading order (NLO) the pion self-energy
and energy density  in asymmetric nuclear matter. The former problem is  related to that of pionic
atoms 
since the pion self-energy and the pion-nucleus optical potential are tightly 
connected \cite{ericeric,galrep}. The issues of the pion-nucleus S-wave
missing repulsion, the renormalization of the isovector scattering length $a^-$ in the 
medium \cite{chanfray,osetdo} and the energy dependence of the isovector 
amplitude \cite{galrep} are not settled yet, despite the recent progresses
\cite{weiprl,galrep,annp}. Ref.~\cite{nlou} found that the leading corrections
to the linear density approach for calculating the pion self-energy in nuclear 
matter are zero. We show here these cancellations explicitly within the
developed non-perturbative techniques. We also show the related 
cancellation between some next-to-next-to-leading order (N$^2$LO) pieces. The
calculation of the energy density of nuclear matter starting from nuclear forces
is a venerable problem in nuclear physics \cite{panda,day,wal,polls,majo2,urbana,kai1,kai3,lutz}. Our calculation of  the energy density ${\cal E}$ to NLO already leads to  saturation for symmetric nuclear
matter  and repulsion for neutron matter.  Indeed, an accurate reproduction of the equation of state for neutron matter can be achieved by varying slightly one subtraction constant around its expected value. For the case of symmetric nuclear matter an additional  fine-tuning of a subtraction constant is necessary to obtain a remarkable good agreement between our results and previous existing sophisticated many-body calculations \cite{panda,urbana,pollmach}. The saturation point and nuclear matter incompressibility  are reproduced in good agreement with experiment.  

After this introduction, we briefly review in section~\ref{sec:pw} the novel 
chiral power counting in the medium developed in ref.~\cite{nlou}. The
contributions to the pion self-energy in 
the nuclear medium that arise from tree-level pion-nucleon scattering
diagrams and from the one-pion loop nucleon self-energy are the subject of
section~\ref{sec.meson-baryon}.  We
dedicate section \ref{sec:sigma8} to the evaluation of the part of the pion
self-energy due to the dressing of the nucleon propagators in the medium
because of the nucleon-nucleon interactions. For their calculation one
requires the nucleon-nucleon scattering amplitudes in the nuclear medium which 
are calculated in the preceding section~\ref{sec:nn-int}. The terms  of the
pion self-energy due to the nucleon-nucleon interactions that are not part of
the nucleon self-energy are calculated in section~\ref{sec:sig.10} at NLO,
where we also give some N$^2$LO contributions. The derivation of the necessary
loops involved in this calculation is performed in Appendix~\ref{app:explicit}. 
 Section~\ref{sec:energy} is dedicated to the evaluation up to NLO of the
energy density. Section \ref{sec:conc} contains a short summary
and the conclusions. In the Appendices we derive various results that are used
in the main body of the paper. Appendix~A offers a derivation  of the partial
wave expansion of nucleon-nucleon scattering in the  nuclear medium and 
vacuum. The Appendices~\ref{sec:l10}, \ref{sec:l11} and \ref{sec:int.inv}
develop the calculation of the in-medium integrals needed for the 
evaluations performed in other sections.

\section{In-medium chiral power counting}
\def\theequation{\arabic{section}.\arabic{equation}}
\setcounter{equation}{0}
\label{sec:pw}

We briefly review the chiral power counting for nuclear matter 
developed in ref.~\cite{nlou}. Let us start by introducing the concept 
of an ``in-medium generalized vertex'' (IGV) given in ref.~\cite{prcoller}.  
Such type of  vertices arises because one can connect several bilinear vacuum 
vertices through the exchange of baryon propagators with the flow through the 
loop of one unit of baryon number, contributed by the nucleon Fermi-seas.  
At least one Fermi-sea insertion is needed because otherwise we would have a 
closed vacuum nucleon loop that in a low-energy effective field theory  is 
completely decoupled.  It is also stressed in ref.~\cite{annp} that within 
a nuclear environment a nucleon propagator could have a ``standard'' or 
``non-standard'' chiral counting. To see this note that a soft momentum $Q
\sim p$, related to pions or external sources can be associated to any of the
 vertices. Denoting by $k$ the on-shell four-momentum associated with one
 Fermi-sea insertion in the IGV, the four-momentum
 running through the $j^{th}$ nucleon propagator can be written as $p_j=k+Q_j$, so that 
 \begin{align}
i\frac{\barr{k}+\barr{Q}_j+m}{(k+Q_j)^2-m^2+i\epsilon}=
 i \frac{\barr{k}+\barr{Q}_j+m}{Q_j^2+2 Q_j^0 E(\vk) -2 {\mathbf{Q}}_j
 \mathbf{k}+i\epsilon}~,
\label{pro.1}
\end{align}
 where  $E(\vk)=\vk^2/2m$, with $m$ the physical nucleon mass (not the bare
one), and $Q_j^0$ is the temporal component of $Q_j$. We have just shown in the previous equation the free part of an in-medium nucleon propagator because this is enough for our present discussion.  
Two different situations occur depending on the value of $Q_j^0$. If $Q_j^0={\cal O}(m_\pi)={\cal O}(p)$  one has the standard counting so that the chiral expansion of the propagator in eq.~(\ref{pro.1}) is
\begin{align}
i \frac{\barr{k}+\barr{Q}_j+m}{2 Q^0_j m+i\epsilon} 
 \left(1-\frac{Q_j^2-2 {\mathbf{Q}}_j \cdot \mathbf{k}}{2Q_j^0 m}+\Opd \right)~.
\end{align}
Thus, the baryon propagator  counts as a quantity of ${\cal O}(p^{-1})$. But it could also
occur that $Q_j^0$ is  of the order of a kinetic nucleon energy in the nuclear
medium or that it even vanishes.\footnote{An explicit example is shown in section 6 of ref.~\cite{annp}} The dominant term in eq.~(\ref{pro.1}) is then
\begin{align}
-i\frac{\barr{k}+\barr{Q}_j+m}{\mathbf{Q}^2_j+2\mathbf{Q}_j\cdot\vk-i
\epsilon} ~,
\end{align}
and the nucleon propagator should be counted as ${\cal O}(p^{-2})$, instead
of the previous ${\cal O}(p^{-1})$. This is referred to as the ``non-standard'' 
case in ref.~\cite{annp}. We should  stress that this situation also occurs 
already in the vacuum when considering the two-nucleon reducible diagrams in 
nucleon-nucleon scattering. This is indeed the reason advocated in
ref.~\cite{wein1} 
for solving a Lippmann-Schwinger equation with the nucleon-nucleon potential 
given by the two-nucleon irreducible diagrams. 

  In order to treat  chiral Lagrangians
with an arbitrary number of baryon fields (bilinear, quartic, etc)
ref.~\cite{nlou} considered firstly
bilinear vertices like in refs.~\cite{prcoller,annp},  but now  the additional 
exchanges of  heavy meson fields of any type are allowed. The latter should be
considered as
merely auxiliary fields that allow one to find a tractable
representation of the multi-nucleon interactions that result when  the
masses of the heavy mesons tend to infinity. Note that such methods are
also used in the so-called nuclear lattice simulations, see e.g.
\cite{Borasoy:2006qn}.
 These heavy meson fields are  denoted in the following by $H$, 
and a heavy meson propagator is counted as ${\cal O}(p^0)$ due to the large
 meson mass. On the other hand, ref.~\cite{nlou} takes  the non-standard counting
case from the start
and  any nucleon propagator is considered as  ${\cal O}(p^{-2})$.
In this way, no diagram, whose chiral order is actually lower
than expected if the nucleon propagators were counted assuming the standard
rules, is lost.  
In the following $m_\pi\sim k_F\sim {\cal O}(p)$ are taken of the same chiral
order,  and are considered  much smaller than a hadronic scale $\Lambda_\chi$
of several hundreds of MeV that results by integrating out all other particle
types, including nucleons with larger three-momentum, heavy mesons and nucleon
and delta resonances \cite{wein2}. The formula obtained in ref.~\cite{nlou} for the chiral order $\nu$ of a given diagram is
\be
\nu=4-E+\sum_{i=1}^{V_\pi}(n_i+\ell_i-4)+\sum_{i=1}^V(d_i+\omega_i-1)+
\sum_{i=1}^m(v_i-1)+\sum_{i=1}^{V_\rho} v_i~.
\label{fff}
\ee
where $E$ is the number of external pion lines, $n_i$ is the number of pion
lines attached to a vertex without baryons,  $\ell_i$ is the chiral order 
of the latter with   $V_\pi$ its total number. In addition, $d_i$ is the
chiral order of the $i^{th}$ vertex bilinear in the baryonic fields, $v_i$
is the number of mesonic lines attached to it, $\omega_i$ that of only the
heavy lines, $V$ is the total number of bilinear vertices, $V_\rho$ is the 
number of IGVs and $m$ is the total number of baryon propagators minus $V_\rho$,
$V=V_\rho+m$. The previous equation can be also written as
\be
\nu=4-E+\sum_{i=1}^{V_\pi}(n_i+\ell_i-4)+\sum_{i=1}^V(d_i+v_i+\omega_i-2)+
V_\rho~.
\label{ffg}
\ee

It is important to stress that $\nu$  is bounded from 
below as explicitly shown in ref.~\cite{nlou}. Because of the 
last term in eq.~(\ref{fff})  adding a new IGV to a connected diagram
increases the counting at least by  one unit because $v_i\geq 1$.
The number $\nu$ given in eq.~(\ref{fff}) represents a lower bound for the 
actual chiral power of a diagram, $\mu$, so that $\mu\geq \nu$. The
 actual
chiral order of a diagram might be
 higher than 
$\nu$ because 
the nucleon propagators are counted always as ${\cal O}(p^{-2})$ in
eq.~(\ref{fff}), while for some diagrams there could be propagators  
that follow the standard counting.  Eq.~(\ref{fff}) implies the following
conditions for
augmenting the number of lines in a diagram without increasing the chiral power
by adding i) pionic lines attached to lowest order mesonic vertices,
$\ell_i=n_i=2$, ii) pionic lines attached to lowest order meson-baryon
vertices, $d_i=v_i=1$ and iii) heavy mesonic lines attached to lowest order 
bilinear vertices, $d_i=0$, $\omega_i=1$.
One major difference between our counting, eq.~\eqref{ffg}, and Weinberg one
\cite{wein1,wein2} is that ours applies directly to the physical amplitudes
while the latter applies only to the potential.

We apply eq.~(\ref{fff})  by increasing step by step $V_\rho$ up to the order
considered. For each $V_\rho$ then we look for those diagrams that do not
further increase the order according to  the rules i)--iii). Some of these 
diagrams are indeed of higher order and one can refrain from 
 calculating them by  establishing which of the nucleon propagators 
scale as ${\cal O}(p^{-1})$. In this way, the  actual chiral order of the
diagrams is determined and one can select those diagrams that 
correspond to the precision required.

	It is worth realizing that eq.~\eqref{ffg} can be also applied in vacuum in order
to determine the relative weight of the different diagrams. In this case, the needed Fermi-sea insertion for each IGV is split in two external nucleon lines, both in- and out-going ones. For vacuum $V_\rho$ is constant because in the 
EFT there is no explicit closed nucleon  loops and baryon number is
conserved. As stressed above, the expressions between brackets in
eq.~\eqref{ffg} do not increase
 despite the diagrams become increasingly complicated.  As a result, one takes 
$V_\rho$ constant and determines the leading, next-to-leading, etc, contributions as 
indicated in the previous paragraph. In order to derive eq.~\eqref{ffg} in ref.~\cite{nlou}, a term $3V_\rho$ was summed because for each IGV there is a Fermi-sea insertion. Since in vacuum there is no sum over nucleons in the sea, one should subtract this contribution so that we would have $4-2V_\rho$ instead of $4+V_\rho$. However, this just modifies the absolute order of a diagram but not the relative one between contributions which remains invariant, and this is what matters for explicit calculations. 

\begin{figure}[t]
\psfrag{Vr=1}{{\small $V_\rho=1$}}
\psfrag{Vr=2}{{\small $V_\rho=2$}}
\psfrag{Op4}{{\small ${\cal O}(p^4)$}}
\psfrag{Op5}{{\small $ {\cal O}(p^5)$}}
\psfrag{i}{{\small $i$}}
\psfrag{j}{{\small $j$}}
\psfrag{q}{{\small $q$}}
\psfrag{Leading Order}{{\small Leading Order}}
\psfrag{Next to Leading Order}{Next-to-Leading Order}
\centerline{\fbox{\epsfig{file=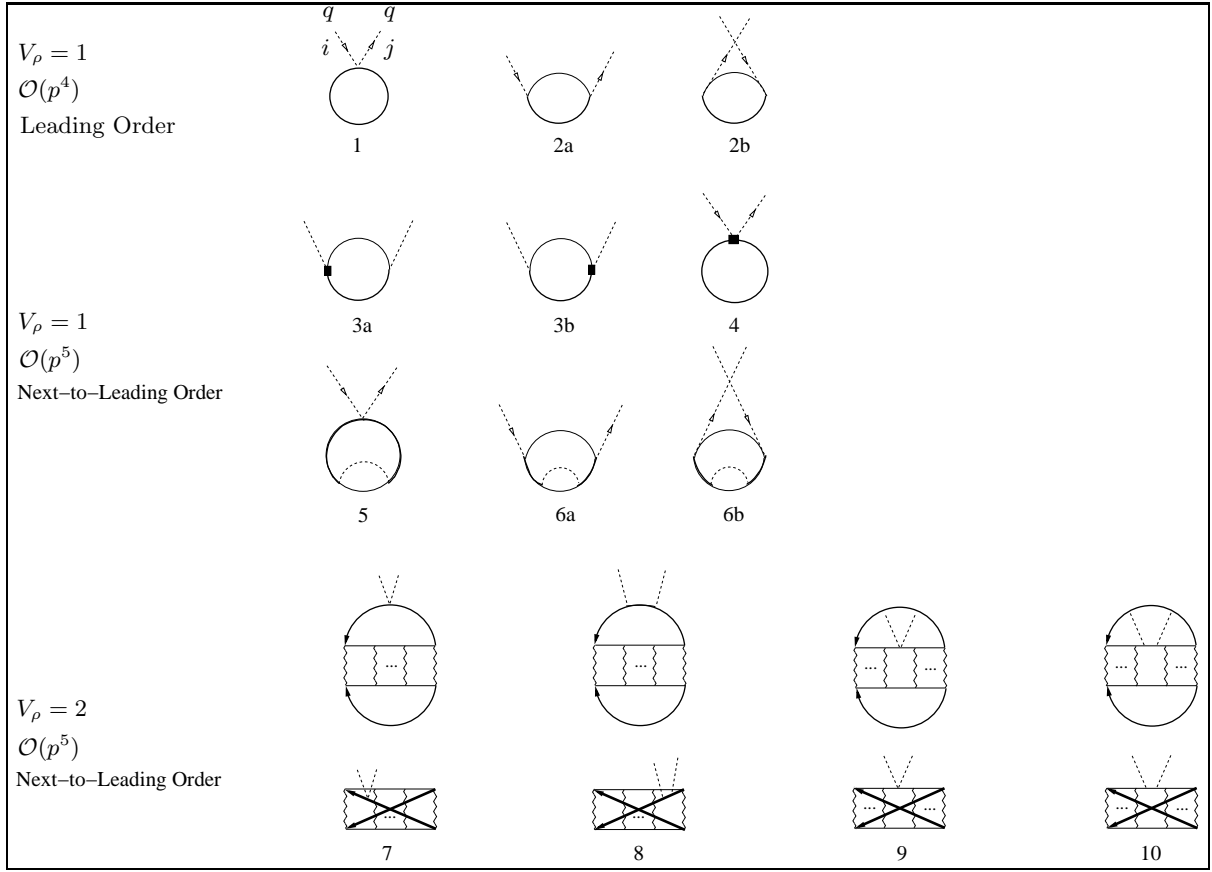,width=.9\textwidth,angle=0}}}
\vspace{0.2cm}
\caption[pilf]{\protect \small
Contributions to the in-medium pion self-energy $\Pi_i$ up
to NLO or ${\cal O}(p^5)$. The pions are indicated by the dashed lines and
the squares correspond to NLO pion-nucleon vertices. A wiggly line is the
nucleon-nucleon interaction kernel, given below in fig.~\ref{fig:wig}, 
which is iterated as indicated by the ellipsis. The thick lines correspond to
closed Fermi-sea insertions, while the thin lines represent in-medium nucleon
propagators, eq.~\eqref{nuc.pro.iso.2}.
The external pion lines in diagrams~3a, 3b, 8 and 10 should be understood as
leaving or entering the diagrams.
\label{fig:all}}
\end{figure}

\section{Meson-baryon contributions to the pion self-energy}
\def\theequation{\arabic{section}.\arabic{equation}}
\setcounter{equation}{0}
\label{sec.meson-baryon}

Here we start the application of the chiral counting in eq.~(\ref{fff}) to
calculate the pion self-energy in the nuclear medium up to NLO or ${\cal
  O}(p^5)$. The different contributions are denoted by $\Pi_i$ and are 
depicted in fig.~\ref{fig:all}.  In terms of the pion self-energy $\Pi$ 
the dressed pion propagators reads
\begin{align}
 \Delta (q) = \frac{1}{q^2-m_\pi^2+\Pi}~.
\end{align}

The in-medium nucleon propagator \cite{fetter}, $G_0(k)_{i_3}$, is 
\begin{align}
G_0(k)_{i_3}&=
\frac{\theta(\xi_{i_3}-|\vk|)}{k^0-E(\vk)-i\epsilon}+\frac{\theta(|\vk|-\xi_{i_3
})}{k^0-E(\vk)+i\epsilon}\nn\\
&=
\frac{1}{k^0-E(\vk)+i\epsilon}+i2\pi \theta(\xi_{i_3}-|\vk|)\delta(k^0-E(\vk))~.
\label{nuc.pro}
\end{align}
In this equation the subscript $i_3$ refers to the third component of
isospin of the nucleon, with $i_3=+1/2$ for the proton and $-1/2$ for the
neutron, and $\xi_{i_3}$ is the corresponding Fermi momentum. We consider 
that isospin symmetry is conserved so that all the nucleon and pion masses are 
equal. The first term on the right hand side (r.h.s.) of the first line of eq.~\eqref{nuc.pro} is the so-called hole contribution and the last term is the particle part. In the second line, the first term is the free-space part of the in-medium nucleon propagator and the last term is the density-dependent one (or a Fermi-sea insertion). 
The proton and neutron propagators can be combined in a common
expression 
\begin{align}
G_0(k)&=\left( \frac{1+\tau_3}{2}
\theta(\xi_p-\mk)+\frac{1-\tau_3}{2}\theta(\xi_n-\mk)\right)\frac{1}{
k^0-E(\vk)-i\epsilon}
\nn\\
&+\left( \frac{1+\tau_3}{2}
\theta(\mk-\xi_p)+\frac{1-\tau_3}{2}\theta(\mk-\xi_n)\right)\frac{1}{
k^0-E(\vk)+i\epsilon}~,
\label{nuc.pro.iso}
\end{align}
or in the equivalent form, 
\begin{align}
G_0(k)=\frac{1}{k^0-E(\vk)+i\epsilon}+ i(2\pi)\delta(k^0-E(\vk))
\left( \frac{1+\tau_3}{2}
\theta(\xi_p-\mk)+\frac{1-\tau_3}{2}\theta(\xi_n-\mk)\right)~.
\label{nuc.pro.iso.2}
\end{align}
In the following, $\sigma^i$ and $\tau^i$ correspond to the Pauli matrices in
the spin and isospin spaces, respectively. 

For the evaluation of the diagrams 1--6 we employ the ${\cal O}(p)$ and ${\cal
O}(p^2)$  Heavy Baryon CHPT (HBCHPT) Lagrangians \cite{ulfrev,peripheral}
\begin{align}
{\cal L}_{\pi N}^{(1)}&=\bar{N}\Bigl(i D_0-\frac{g_A}{2}\vec{\sigma}\cdot
\vec{u}\Bigl)N~,\nn\\
{\cal L}_{\pi N}^{(2)}&=\bar{N}\Biggl(
\frac{1}{2m}\vec{D}\cdot \vec{D}+i\frac{g_A}{4m}\left\{ \vec{\sigma}\cdot
\vec{D},u_0\right\}+2c_1
m_\pi^2(U+U^\dagger)+\left(c_2-\frac{g_A^2}{8m}\right)u_0^2+c_3 u_\mu u^\mu
\Biggr)N+\ldots
\label{lags}
\end{align}
where the ellipses represent terms that are not needed here.
In this equation, $N$ is the two component field of the nucleons, $g_A$ is the
axial-vector pion-nucleon coupling and $D_\mu=\partial_\mu+\Gamma_\mu$ the covariant
chiral derivative, with $\Gamma_\mu=\frac{1}{2}[u^\dagger, \partial_\mu u]$. The pion
fields $\vec{\pi}(x)$ enter in the matrix $u=\exp(i\vec{\tau}\cdot
\vec{\pi}/2f)$, in terms of which $u_\mu=i\left\{u^\dagger,\partial_\mu
u\right\}$ and $U= u^2$,  with $f$ the weak pion decay constant in the SU(2)
chiral limit. The $c_i$ are chiral low-energy constants whose values are fitted
from phenomenology \cite{ulfrev}. The in-medium pion self-energy from ${\cal
L}_{\pi N}$ in HBCHPT has been calculated up to two loops in
refs.~\cite{selfkaiser,korean}.

The diagrams 1--6 were calculated in ref.~\cite{nlou}. We give here more details
in their derivation.
The first diagram in fig.~\ref{fig:all} corresponds to
\begin{align}
\Pi_1&=\frac{-i q^0}{2f^2}\varepsilon_{i j 3}(\rho_p-\rho_n)~,
\label{eq:sig1}
\end{align}
and arises by closing the Weinberg-Tomozawa term (WT) in pion-nucleon
scattering. In the previous equation  $\rho_{p(n)}=\xi_{p(n)}^3/3\pi^2$  is the
proton(neutron) density.  
$\Pi_1$ is an S-wave isovector self-energy.

The diagram 2a in fig.~\ref{fig:all} is represented by $\Pi_{2a}$
and is obtained by closing the nucleon pole terms in pion-nucleon scattering,  
with the one-pion vertex from the lowest order meson-baryon chiral Lagrangian 
${\cal L}^{(1)}_{\pi N}$ \cite{ulfrev}, 
\begin{align}
\Pi_{2a}=-\frac{g_A^2}{4f^2}\int \frac{d^3 k}{(2\pi)^3}\hbox{ Tr}\left[
\left(\frac{1+\tau_3}{2}\theta(\xi_p-|\vk|)+\frac{1-\tau_3}{2}
\theta(\xi_n-|\vk|)\right)
\frac{\tau^i \tau^j \,\vec{\sigma}\cdot \vq \vec{\sigma}\cdot
\vq}{E(\vk)-q^0-E(\vk-\vq)+i\epsilon}
\right]~,
\label{eq.si1.1}
\end{align}
In eq.~(\ref{eq.si1.1})  we have not included the in-medium part of the
intermediate nucleon propagator 
because $q^0\simeq m_\pi\gg E(\vk)-E(\vk-\vq)$, so that the argument of the 
in-medium Dirac delta-function of eq.~(\ref{nuc.pro.iso.2}) cannot be
fulfilled. By the same token 
\be
\frac{1}{E(\vk)-E(\vk-\vq)-q^0}=-\frac{1}{q^0}-\frac{E(\vk)-E(\vk-\vq)}{(q^0)^2}
+{\cal O}(q)~,
\label{exp.nuc.pro}
\ee
and the ${\cal O}(q)$ terms contribute one order higher than NLO. On the other
hand, 
\be
E(\vk)-E(\vk-\vq)=-\frac{\vq^2-2 \vk\cdot \vq}{2 m}~,
\ee
and the $\vk\cdot \vq$ term, when included in eq.~(\ref{eq.si1.1}), does not 
contribute because of the angular integration. Then,  
\be
\Pi_{2a}=\frac{g_A^2}{4f^2q^0}\left(1-\frac{\vq^2}{2 m q^0}\right)
\int \frac{d^3 k}{(2\pi)^3}\hbox{ Tr}\left[
\left(\frac{1+\tau_3}{2}\theta(\xi_p-|\vk|)+\frac{1-\tau_3}{2}
\theta(\xi_n-|\vk|)\right)
 \tau^i \tau^j \,\vec{\sigma}\cdot \vq \vec{\sigma}\cdot \vq
\right]~.
\ee
The same procedure can be applied to the diagram 2b of fig.~\ref{fig:all}  
(which corresponds to the same expression as $\Pi_{2a}$ but 
with the exchanges $q^0\to -q^0$ and $i\leftrightarrow j$). Summing both, one has
\begin{align}
\Pi_2^{iv}&=\frac{ig_A^2\,\vq^2}{2f^2 q^0}\varepsilon_{i j
3}(\rho_p-\rho_n)~,\nn \\
\Pi_2^{is}&=\frac{-g_A^2}{4f^2}\frac{(\vq^2)^2}{m {q_0}^2}\delta_{i
j}(\rho_p+\rho_n)~.
\label{eq:sig2}
\end{align}
The superscripts $iv$ and $is$ refer to the isovector and isoscalar nature of the
corresponding contribution to $\Pi_2$, respectively. 
Both are P-wave self-energies but $\Pi_2^{is}$ is a recoil correction
of $\Pi_2^{iv}$ and it is suppressed by the inverse of the  nucleon mass.
 
 The rest of the diagrams in fig.~\ref{fig:all} are NLO. 
We now consider the sum of the diagrams~3a and 3b, where the squares indicate a
NLO one-pion vertex from ${\cal  L}^{(2)}_{\pi N}$, eq.~\eqref{lags}. It should
be understood that the pion lines can leave or enter these diagrams.  We also
employ the expansion of eq.~(\ref{exp.nuc.pro})
for the nucleon propagator, although  for this case it is only necessary to keep
the 
term $\pm 1/q^0$ because the diagram is already a NLO contribution. The
calculation yields
\be
\Pi_3=\frac{g_A^2 \vq^2}{2m f^2}(\rho_p+\rho_n)\delta_{ij}~.
\label{eq:sig3}
\ee
This is a P-wave isoscalar contribution. In this case the NLO pion-nucleon
vertex  is a recoil correction of the  LO one and this is why $\Pi_3$
is suppressed by the inverse of the nucleon mass. 
The diagram~4 in fig.~\ref{fig:all} is given by 
\be
\Pi_4=\frac{-2\delta_{i j}}{f^2}\left(
2 c_1\,m_\pi^2-q_0^2 (c_2+c_3-\frac{g_A^2}{8
m})+c_3\,\vq^2\right)(\rho_p+\rho_n)~.
\label{sigma4}
\ee
$\Pi_4$ is an isoscalar contribution in which  the term 
$-2\delta_{ij}c_3\vq^2(\rho_p+\rho_n)/f^2$ is P-wave and the rest is S-wave.
Indeed, the low-energy constant $c_3$ is known to be dominated by the contribution of
the $\Delta(1232)$ \cite{aspects}. For a Fermi momentum $\xi\simeq 2 m_\pi$,
corresponding to symmetric nuclear matter saturation, the Fermi energy of a
two-nucleon system is around 80~MeV, which is still significantly smaller than
the $\Delta$-nucleon mass difference. One
then expects that integrating out the $\Delta$-resonance  and parameterizing its
 effects in terms of the chiral counterterms is meaningful in the range of
energies we are considering. This is indeed an important  conclusion of
ref.~\cite{kreb1} where chiral EFTs with/without $\Delta$s are employed to
evaluate different orders of the two-nucleon and three-nucleon potentials. We
leave as a future improvement of our results to include explicitly the 
$\Delta$ resonances.

Let us consider the contributions to the pion self-energy due to the one-pion 
loop nucleon self-energy. This is represented by the diagrams~5, 6a and 6b in
fig.~\ref{fig:all}. These diagrams originate by dressing the in-medium nucleon 
propagator of the diagrams~1, 2a, 2b, in order, with the one-pion loop.
\begin{figure}[ht]
\psfrag{k}{$k$}
\psfrag{p}{$p$}
\psfrag{l}{$\ell$}
\psfrag{pi}{$\pi$}
\psfrag{r}{$k-\ell$}
\centerline{\epsfig{file=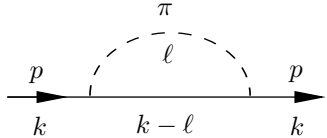,height=.8in,width=1.7in,angle=0}}
\vspace{0.2cm}
\caption[pilf]{\protect \small
One-pion loop contribution to the nucleon self-energy in the nuclear medium.
The four-momenta are indicated below the corresponding lines in the figure. 
\label{fig:piloop}}
\end{figure} 
  As
a preliminary result let us first evaluate the nucleon self-energy in the nuclear medium corresponding to fig.~\ref{fig:piloop}. First, we consider the case of a neutral pion. The results for the charged pion
contributions follow immediately from the $\pi^0$ case. In HBCHPT the proton
self-energy due to the one-$\pi^0$ loop is given by,
\begin{align}
\Sigma_p^{\pi^0}&=-i\frac{g_A^2}{f^2} S_\mu S_\nu\int \frac{d^D \ell}{(2\pi)^D}
 \frac{\ell^\mu \ell^\nu}{(\ell^2-m_\pi^2+i\epsilon)(v(k-\ell)+i\epsilon)} \nn  \\
&+2\pi \frac{g_A^2}{f^2} S_\mu S_\nu 
\int \frac{d^D\ell}{(2\pi)^D}\frac{\ell^\mu
\ell^\nu}{\ell^2-m_\pi^2+i\epsilon}\delta(v(k-\ell)) \theta(\xi_p-|\vk-\mathbf{l}|)~.
\label{nuc.self.1}
\end{align}
Here, $\mathbf{l}$ is the vector made up from the spatial components of $\ell$ and $v$ is the four-velocity normalized to unity ($v^2=1$), such that the
four-momentum of a nucleon is given by  
$p=m v +k$,  with $k$ a small residual momentum ($v\cdot k\ll m$).
In practical calculations we take  $v=(1,\mathbf{0})$ and $D\to 4$. Notice that the last integral in the previous equation is convergent because of the presence of the Dirac delta and Heaviside step functions.  
Instead of the full non-relativistic nucleon propagator eq~(\ref{nuc.pro}),
HBCHPT typically implies the so-called extreme non-relativistic limit in which
$E(\vk)\to 0$, see e.g. ref.~\cite{ulfrev}.  Given the properties of the
covariant spin operator $S_\mu$ 
\cite{ulfrev} it follows that
 \begin{align}
 S_\mu S_\nu \ell^\mu \ell^\nu=\frac{1}{2}\{S_\mu,S_\nu\}\ell^\mu \ell^\nu=\frac{1}{4}
 \left((v\cdot \ell)^2-\ell^2\right)~.
\label{s.com}
 \end{align}
For the vacuum part we then have the integral,
\be
\Sigma_{p,f}^{\pi^0}=-i\frac{g_A^2}{4 f^2} \int \frac{d^D \ell}{(2\pi)^D}
 \frac{(v \ell)^2-\ell^2}{(\ell^2-m_\pi^2+i\epsilon)(v(k-\ell)+i\epsilon)}~,
 \label{sigma.free}
 \ee
This can be evaluated straightforwardly in dimensional regularization. Adding
the contributions from the charged pions one has the free nucleon
self-energy due to a one-pion loop  \cite{ulfrev}, denoted in the following by
$\Sigma_f^\pi$,  
\be
\Sigma_f^\pi\equiv
\Sigma_{p,f}^{\pi}=\Sigma_{n,f}^{\pi}=\frac{3 g_A^2 b}{32 \pi^2 f^2}\left\{
-\omega+\sqrt{b}\left(i\log\frac{\omega+i\sqrt{b}}{-\omega+i\sqrt{b}}
+\pi\right) 
\right\}-\frac{3g_A^2 m_\pi^3}{32\pi f^2}~,
\label{sig.pi.fre}
\ee 
where $\omega=v\cdot k=k^0$ and $b=m_\pi^2-\omega^2-i\epsilon$.\footnote{In all the calculations that follow the square roots and logarithms have the cut along the negative real axis.} In the previous
expression we have subtracted the value of the one-pion loop nucleon self-energy
at $\omega=0$ since we are using the physical nucleon mass. 
 We also need below its derivative
 \be
 \frac{\partial\Sigma_f^\pi}{\partial w}=\frac{3 g_A^2}{32\pi^2
f^2}\left[
 m_\pi^2+\omega^2-3\omega \sqrt{b}\left(
 i\log\frac{\omega+i\sqrt{b}}{-\omega+i\sqrt{b}}+\pi
 \right)
 \right]~.
 \label{d.sif}
 \ee

The in-medium contribution to the proton self-energy due to the one-$\pi^0$
loop corresponds to the last line in eq.~\eqref{nuc.self.1}. Taking also into account
the charged pions in the loop we have for the in-medium part of the one-pion loop
contribution to the  proton and neutron self-energies, $\Sigma_{p,m}$ and
$\Sigma_{n,m}$, respectively, 
\begin{align}
\Sigma_{p,m}^\pi&=2\pi \frac{g_A^2}{f^2} S_\mu S_\nu \int
\frac{d^4\ell}{(2\pi)^4}\frac{\ell^\mu
\ell^\nu}{\ell^2-m_\pi^2+i\epsilon}\delta(v(k-\ell))\biggr[
\theta(\xi_p-|\vk-\mathbf{l}|)+2\theta(\xi_n-|\vk-\mathbf{l}|)\biggl]~,\nn\\
\Sigma_{n,m}^\pi&=2\pi \frac{g_A^2}{f^2} S_\mu S_\nu \int
\frac{d^4\ell}{(2\pi)^4}\frac{\ell^\mu
\ell^\nu}{\ell^2-m_\pi^2+i\epsilon}\delta(v(k-\ell))\biggr[
\theta(\xi_n-|\vk-\mathbf{l}|)+2\theta(\xi_p-|\vk-\mathbf{l}|)\biggl]~.
\label{1.pi.l.n.s.e}
\end{align}

$S_\mu S_\nu \ell^\mu \ell^\nu=\mathbf{l}^2/4$ for our choice of $v$  and the second
integral on the r.h.s. of eq.~(\ref{nuc.self.1}) reads
\begin{align}
I_m^{(1)}=2\pi \int\frac{d^4\ell}{(2\pi)^4}\frac{{\mathbf{l}}^2 \delta(k^0-\ell^0)
\theta(\xi_p-|\vk-\mathbf{l}|)}{\ell^2-m_\pi^2+i\epsilon}=\int \frac{d^3\ell}{(2\pi)^3}
\frac{{\mathbf{l}}^2 \theta(\xi_p-|\vk-\mathbf{l}|)}{k_0^2-\mathbf{l}^2-m_\pi^2+i\epsilon}~.
\label{im1.def}
\end{align}
The step function in the previous integral implies the requirement $\xi_p^2\geq
(\vk-\mathbf{l})^2=\vk^2+\mathbf{l}^2-2|\vk||\mathbf{l}|\cos\theta$~.  Then,
\be
\cos\theta\geq \frac{\vk^2+\mathbf{l}^2-\xi_p^2}{2 |\vk||\mathbf{l}|}=y_0~.
\ee 
It is necessary that $y_0\leq 1$, otherwise $\cos\theta$ would be larger than 1.
This implies that
\be
|\vk|-\xi_p\leq |\mathbf{l}| \leq |\vk|+\xi_p~.
\ee
On the other hand, if $|\mathbf{l}|\geq \xi_p-|\vk|$ then $y_0\geq -1$. Taking into
account these constraints, one has:
\begin{align}
& a)~|\vk|\geq \xi_p:~~|\mathbf{l}|\in \bigl[|\vk|-\xi_p,|\vk|+\xi_p\bigr] \hbox{ and
} 
\cos\theta\in \bigl[y_0,1\bigr] ~,\nn\\
& b)~\xi_p\geq |\vk|:~~|\mathbf{l}| \in \bigl[0,\xi_p-|\vk|\bigr] \hbox{ and }
\cos\theta\in \bigl[-1,1\bigr]~  ;~~
|\mathbf{l}| \in \bigl[\xi_p-|\vk|,\xi_p+|\vk|\bigr] \hbox{ and } \cos\theta\in
\bigl[y_0,1\bigr]~.
\end{align}

The same expression of $I_m^{(1)}$ results for the cases a) and b), 
\begin{align}
I_m^{(1)}(\xi_p)&=-\frac{\rho_p}{2}
+\frac{b}{4\pi^2}\left\{\xi_p-\sqrt{b}\hbox{ arctan}\frac{\xi_p-|\vk|}{\sqrt{b}}
-\sqrt{b}\hbox{ arctan}\frac{\xi_p+|\vk|}{\sqrt{b}}
-\frac{\vk^2-\xi_p^2-b}{4|\vk|}\log
\frac{(\xi_p+|\vk|)^2+b}{(\xi_p-|\vk|)^2+b}\right\}~.
\label{im1.exp}
\end{align}
 In terms of $I_m^{(1)}$ the in-medium part of the one-pion loop contribution to the nucleon self-energy, eq.~\eqref{1.pi.l.n.s.e}, reads
\begin{align}
\Sigma_{p,m}^\pi&= \Sigma_{p,m}^{\pi^0}+\Sigma_{p,m}^{\pi^+}=
\frac{g_A^2}{4f^2}\left(I_m^{(1)}(\xi_p)+2 I_m^{(1)}(\xi_n)\right)~,\nn\\
\Sigma_{n,m}^\pi&= \Sigma_{n,m}^{\pi^0}+\Sigma_{n,m}^{\pi^-}=
\frac{g_A^2}{4f^2}\left(I_m^{(1)}(\xi_n)+2 I_m^{(1)}(\xi_p)\right)~,
\label{sig.pi.med}
\end{align}
where the superscript refers to the pion species in the loop.  
The full in-medium nucleon self-energy is given by the sum of eqs.~\eqref{sig.pi.fre} and \eqref{sig.pi.med}.
In this way, the proton and neutron self-energies due to the one-pion loop are, in that order,
\begin{align}
\Sigma_p^\pi&=\Sigma_f^\pi+\Sigma_{p,m}^\pi~,\nn\\
\Sigma_n^\pi&=\Sigma_f^\pi+\Sigma_{n,m}^\pi~.
\label{sig.nuc.full}
\end{align}
The self-energies for both the proton and neutron can be joined together in $\Sigma^\pi$, given by
\begin{align}
\Sigma^\pi&=\frac{1+\tau_3}{2}\Sigma_p^\pi+\frac{1-\tau_3}{2}\Sigma_n^\pi~.
\end{align}

\begin{figure}[ht]
\psfrag{exact}{{\small exact}}
\psfrag{fact}{{\small fact}}
\psfrag{a}{a)}
\psfrag{b}{b)}
\psfrag{c}{c)}
\centerline{\epsfig{file=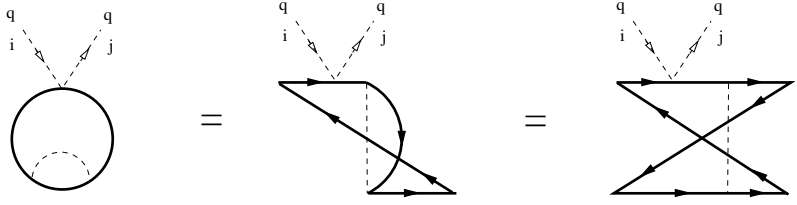,width=.6\textwidth,angle=0}}
\vspace{0.2cm}
\caption[pilf]{\protect \small The equivalence between the diagram~5 of fig.~\ref{fig:all}, when only density-dependent parts in the nucleon propagators are considered, and the 
one-pion exchange reduction of the diagram~7 is shown.
The second diagram from the left is an intermediate step in the continuous 
transformation of the diagram from the far left to that on the far right.
\label{fig:equiv}}
\end{figure}

The diagram~5 of fig.~\ref{fig:all} originates by dressing the in-medium nucleon propagator, 
eq.~\eqref{nuc.pro.iso.2}, with the in-medium one-pion loop nucleon self-energy.
Its contribution, $\Pi_5$, can then be written as 
\begin{align}
\Pi_5 &= \frac{q^0}{2f^2}\varepsilon_{ijk} \int\frac{d^4k}{(2\pi)^4}
e^{ik^0\eta} \hbox{ Tr} \left\{ \tau^k \, G_0(k) \, \Sigma^\pi(k) \, G_0(k)
\right\} ~,
\label{sigma5}
\end{align}
with the convergence factor $e^{ik^0\eta}$, $\eta\rightarrow0^+$, associated
with any closed loop made up by a single nucleon line \cite{fetter}.
The trace acts in the spin and isospin spaces and gives  the result
\begin{align}
\Pi_5&=\frac{q^0}{f^2}\varepsilon_{ijk}\int\frac{d^4k}{(2\pi)^4}e^{ik^0\eta}
\left(G_0(k)_p^2 \Sigma_p^\pi-G_0(k)_n^2 \Sigma_n^\pi\right)~.
\label{sigma5.b}
\end{align}
Next, we employ the identity
\begin{align}
G_0(k)^2_{i_3}=-\frac{\partial}{\partial k^0}G_0(k)_{i_3}~,
\label{pro.go2}
\end{align}
that follows from the r.h.s. of the first line of eq.~\eqref{nuc.pro}. A similar identity also holds at the matrix level
\begin{align}
G_0(k)^2=-\frac{\partial}{\partial k^0}G_0(k)~,
\end{align}
because of the orthogonality of the isospin projectors $(1+\tau_3)/2$ and $(1-\tau_3)/2$.
 Integrating by parts, as the convergence factor allows, we then have
\begin{align}
\Pi_5&=\frac{q^0}{f^2}\varepsilon_{ij3}\int\frac{d^4k}{(2\pi)^4}e^{ik^0\eta}\left(G_0(k)_p\frac{\partial \Sigma_p^\pi}{\partial k^0}-G_0(k)_n\frac{\partial \Sigma_n^\pi}{\partial k^0}\right)~.
\label{sigma5.c}
\end{align} 
We perform the integration over $k^0$ making use of the Cauchy-integration theorem. For that we close the integration contour along the upper $k^0$-complex plane with an infinite semicircle. Because of the convergence factor the integration over the infinite semicircle is zero as $\hbox{Im}k^0\to +\infty$ along it.  One should then study the positions of the poles and cuts in   $k^0$ for  $G_0(k)_{i_3}$ and $\Sigma_{i_3}^\pi$ in eq.~\eqref{sigma5.c}. First let us note that   $\Sigma_f$ has only singularities  for $\hbox{Im}k^0<0$, as follows  from eq.~\eqref{sigma.free}. This is also evident for the free part of $G_0(k)_{i_3}$, see eq.~\eqref{nuc.pro}. As a result, there is no contribution when the integrand in eq.~\eqref{sigma5.c} involves only free nucleon propagators. 
The contribution with only the density-dependent part both in $G_0(k)_{i_3}$ as well as in the nucleon propagator involved in the loop for $\Sigma_{i_3}^\pi$  is part of the  diagram~7 in fig.~\ref{fig:all}, corresponding to a $V_\rho=2$ contribution.
In fig.~\ref{fig:equiv} we depict such an equivalence for the diagrams~5 and 7
of fig.~\ref{fig:all}. An analogous result would hold for the diagrams~6 and
8.
The different $V_\rho=2$ contributions are evaluated in section
\ref{sec:sigma8}, so we skip them right now.

Consequently we consider in this section only the contributions where we have simultaneously one
free-space and one density-dependent part of the nucleon propagators involved in eq.~\eqref{sigma5.c} and in the calculation of $\Sigma^\pi$. Two contributions arise. The first one results by employing the density-part for $G_0(k)_{i_3}$ in the r.h.s. of eq.~\eqref{nuc.pro}. The integration over $k^0$ is trivial due to the Dirac-delta function, with the result
\begin{align}
\Pi_{5I}&=i\frac{q^0}{f^2}\varepsilon_{ij3}\int\frac{d^3k}{(2\pi)^3}(\theta_p^--\theta_n^-)\left.\frac{\partial \Sigma_f^\pi}{\partial k^0}\right|_{k^0=E(\vk)}~.
\label{fin.sig.5}
\end{align}
We have introduced the shorter notation $\theta(\xi_p-|\vk|)\equiv \theta_p^-$ 
and $\theta(\xi_n-|\vk|)\equiv \theta_n^-$~.
\begin{figure}[ht]
\psfrag{l}{$\ell$}
\psfrag{r}{$\ell+k$}
\psfrag{k}{$k$}
\centerline{\epsfig{file=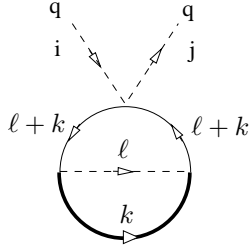,width=.15\textwidth,angle=0}}
\vspace{0.2cm}
\caption[pilf]{\protect \small This figure shows that the contribution  $\Pi_{5II}$, eq.~\eqref{sigma5.b.def}, involves two free nucleon propagators that follow the standard power counting. It is then of ${\cal O}(p^6)$.  
\label{fig:pi5b}}
\end{figure} 
The other contribution involves the free part of $G_0(k)_{i_3}$ and can be written as
\begin{align}
\Pi_{5II}&=\frac{q^0}{f^2}\varepsilon_{ij3}\int\frac{d^4k}{(2\pi)^4}\frac{e^{ik^0\eta}}{k^0-E(\vk)+i\epsilon}\left(\frac{\partial}{\partial k^0}\Sigma_{p,m}^\pi
-\frac{\partial}{\partial k^0}\Sigma_{n,m}^\pi\right)~.
\label{sigma5.b.def}
\end{align}
However, it is easily seen that this contribution, depicted in fig.~\ref{fig:pi5b}, is indeed of ${\cal O}(p^6)$. The reason is because there are  two free nucleon propagators of standard counting (those with four-momentum $\ell+k$ in the figure, $\ell={\cal O}(p)$), and each of them raises the counting with respect to $\nu$ given in eq.~\eqref{ffg} by one power of the small scale. As a result, we neglect in the following $\Pi_{5II}$.  The same reasoning is not applicable for $\Pi_{5I}$  because only one nucleon propagator, the one inside the pion-loop self-energy, follows the standard counting. Nonetheless, the  expression for $\Pi_{5I}$, eq.~\eqref{fin.sig.5}, explicitly shows that it is actually a contribution of 
${\cal O}(p^6)$.  This due to the fact that $\partial \Sigma_f^\pi/\partial k^0={\cal O}(p^2)$, as
follows directly from eq.~(\ref{d.sif}). We originally counted $\Pi_{5I}$ as ${\cal O}(p^5)$ because $\partial \Sigma_f^\pi/\partial k^0$ was taken  as 
${\cal O}(p)$, since $\Sigma_f^\pi={\cal O}(p^3)$ and
$k^0={\cal O}(p^2)$. However, this evaluation of the order of a 
derivative, based on dimensional analysis, represents indeed a lower bound and its actual order  might be higher, as it is the case here. This mismatch is  due to the presence of the variable $b$, defined after eq.~\eqref{sig.pi.fre}, in addition to $k^0$. The chiral order of the former is fixed by $m_\pi^2$ and not by $k_0^2$. We also recall that there is another contribution to  $\Pi_5$ that results by keeping the density-dependent parts both in $G_0(k)$ and $\Sigma_{p,m}^\pi$, $\Sigma_{n,m}^\pi$. It will be included in the evaluation of the $V_\rho=2$ contributions corresponding to the diagram 7 in fig.~\ref{fig:all},  section~\ref{sec:sigma8}. $\Pi_5$ is an isovector S-wave pion self-energy
contribution.

We consider now the diagrams~6 of fig.~\ref{fig:all}. The diagram~6a gives 
\begin{align}
 \Pi_{6}^a&=-i\frac{g_A^2}{4f^2}\int\frac{d^4k}{(2\pi)^4}\hbox{ Tr}\bigg\{
 \tau^i\vec{\sigma}\cdot\vq \, \frac{1}{k^0-q^0-E(\vk-\vq)+i\epsilon}\tau^j \,
\vec{\sigma}\cdot\vq
 \, G_0(k) \, \Sigma^\pi(k) \, G_0(k) \bigg\} e^{ik^0\eta}~,
\label{sigma6a}
\end{align}
where we have omitted the Fermi-sea insertion in the intermediate propagator,
following the discussion after eq.~\eqref{eq.si1.1}.
$\Pi_{6}^b$ is obtained from $\Pi_6^a$ by replacing $q^\mu\to -q^\mu$ and $i
\leftrightarrow j$ in the latter.
We now employ eq.~\eqref{pro.go2}, take into account that $\vec{\sigma}\cdot \vq\,\vec{\sigma}\cdot \vq=\vq^2$ and integrate by parts in $k^0$. In this way eq.~\eqref{sigma6a} becomes
\begin{align}
\Pi_6^a&=-i\frac{g_A^2 \vq^2}{4f^2}\int\frac{d^4k}{(2\pi)^4}\hbox{ Tr}\left[G_0(k)\tau^{i}\tau^{j}\Sigma^\pi\right]e^{ik^0\eta}\frac{\partial}{\partial k^0}\frac{1}{k^0-q^0-E(\vk-\vq)+i\epsilon}\nn\\
&-i\frac{g_A^2 \vq^2}{4f^2}\int\frac{d^4k}{(2\pi)^4}\hbox{ Tr}\left[
G_0(k)\tau^i\tau^j \frac{\partial \Sigma^\pi}{\partial k^0}
\right]e^{ik^0\eta}\frac{1}{k^0-q^0-E(\vk-\vq)+i\epsilon}~.
\label{pi6.cont.a}
\end{align}
Following an analogous procedure for $\Pi_6$ as the one given below eq.~\eqref{sigma5.c} for $\Pi_5$, the integration over $k^0$  
is performed first. As a result, for the previous equation we only take the contributions that simultaneously involve one free-space and one density-dependent part of the nucleon propagators in $G_0(k)$ and in the loop giving rise to $\Sigma^\pi$. The contribution with only free-space parts vanishes because the integration over $k^0$ along the upper half-plane. While that with only density-dependent parts is included in the evaluation of the diagram 8 in fig.~\ref{fig:all}, section~\ref{sec:sigma8}. On the other hand, applying here the argument in connection with fig.~\ref{fig:pi5b}, the contribution involving the free-space parts of $G_0(k)$ and the density-dependent one of $\Sigma^\pi$ in eq.~\eqref{sigma6a}  is ${\cal O}(p^6)$ because two free nucleon propagators (and not just one) are involved with standard counting. Hence, we neglect in the following this other contribution. In this way, we are left  with the contribution, denoted by $\Pi_{6I}^a$, that involves  $\Sigma_f^\pi$ and the density-dependent parts in $G_0(k)$. From eq.~\eqref{pi6.cont.a} it is given by
\begin{align}
\Pi_{6I}^a&=\frac{g_A^2 \vq^2}{4f^2}\int\frac{d^3k}{(2\pi)^3}\hbox{ Tr}\left[\left(\frac{1+\tau_3}{2}\theta_p^-+\frac{1-\tau_3}{2}\theta_n^-\right)\tau^{i}\tau^{j}\right]\Sigma^\pi_f\left.\frac{\partial}{\partial k^0}\frac{1}{k^0-q^0-E(\vk-\vq)+i\epsilon}\right|_{k^0=E(\vk)}\nn\\
&+\frac{g_A^2 \vq^2}{4f^2}\int\frac{d^3k}{(2\pi)^3}\hbox{ Tr}\left[
\left(\frac{1+\tau_3}{2}\theta_p^-+\frac{1-\tau_3}{2}\theta_n^-\right)\tau^i\tau^j\right]\left.\frac{1}{E(\vk)-q^0-E(\vk-\vq)+i\epsilon}\frac{\partial \Sigma^\pi_f}{\partial k^0}\right|_{k 0=E(\vk)}~.
\label{pit6.cont.aI}
\end{align}
Finally, taking into account the chiral expansion given in eq.~\eqref{exp.nuc.pro} and adding $\Pi_{6I}^b$, we have the new quantity $\Pi_{6I}$ given by
\begin{align}
\Pi_{6I}&=\frac{-i g_A^2}{f^2}\frac{\vq^2}{q^0}\varepsilon_{i j 3}
\int\frac{d^3k}{(2\pi)^3}(\theta_p^--\theta_n^-)\left.
\frac{\partial \Sigma_f}{\partial
k^0}\right|_{k^0=E(\vk)}
-\frac{g_A^2}{f^2}\frac{\vq^2}{{q_0}^2}\delta_{i j}
\int\frac{d^3k}{(2\pi)^3}(\theta_p^-+\theta_n^-)\left.\Sigma_f\right|_{k^0=E(\vk)}~.
\label{sig.6.f1}
\end{align}
$\Pi_{6I}$ is a P-wave self-energy contribution. However, while the first term
on the r.h.s. is isovector, $\Pi_{6I}^{iv}$, the second term is
isoscalar,  $\Pi_{6I}^{is}$. It is also the case, see \cite{nlou}, that
$\Pi_{6I}^{iv}$ is actually one order higher than expected, similarly  as for  
$\Sigma_{5I}$. For $ \Pi_{6I}^{is}$ this follows obviously from its explicit 
expression in eq.~(\ref{sig.6.f1}) as $\Sigma_f^\pi={\cal O}(p^3)$.
\\~\\
In this section we have undertaken the calculation of the diagrams in fig.~\ref{fig:all} that can be fully accounted for by pion-nucleon dynamics. All the contributions calculated in this sections, $\Pi_{1}$ to $\Pi_4$, as well as $\Pi_{5I}$ and $\Pi_{6I}$, are linear in density. We have shown that to NLO only the leading contributions, $\Pi_1$ and $\Pi_2$, and the NLO ones $\Pi_3$ and $\Pi_4$ have to be kept. $\Pi_{5I}$ and $\Pi_{6I}$ are finally one order higher.

\section{Nucleon-nucleon interactions}
\label{sec:nn-int}
\setcounter{equation}{0}

The inclusion of the nucleon-nucleon interactions for the calculation of the
pion self-energy takes place at NLO, because $V_\rho=2$ is required at least. 
 As a result, it is necessary to work out the nucleon-nucleon interactions only at the lowest chiral order,  ${\cal O}(p^0)$. 
These contributions correspond to the diagrams~7--10 in the last 
two rows of fig.~\ref{fig:all}.  First, we discuss these  interactions in  vacuum and then consider their extension
to the nuclear medium. For the vacuum case we also discuss  nucleon-nucleon 
scattering up to ${\cal O}(p)$.

\subsection{Free nucleon-nucleon interactions}
\label{sec:fnn}

\begin{figure}[ht]
\psfrag{k}{$k$}
\psfrag{p}{$p$}
\psfrag{l}{$\ell$}
\psfrag{pi}{$\pi$}
\psfrag{r}{$k-\ell$}
\centerline{\epsfig{file=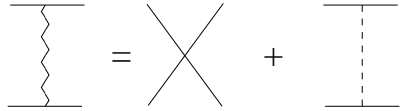,width=.3\textwidth,angle=0}}
\vspace{0.2cm}
\caption[pilf]{\protect \small
The exchange of a wiggly line between two nucleons corresponds to   
the sum of the local and the one-pion exchange contributions,  eqs.~\eqref{feynman} and \eqref{1pi.gen}, respectively.
\label{fig:wig}}
\end{figure}

The lowest order tree-level amplitudes for nucleon-nucleon scattering, ${\cal
O}(p^0$), 
are given by the one-pion exchange, with the lowest order pion-nucleon coupling, and local terms from the quartic nucleon Lagrangian  without quark masses or derivatives 
\be
{\cal L}_{NN}^{(0)}=-\frac{1}{2}C_S (\overline{N}N)(\overline{N}N)
-\frac{1}{2}C_T(\overline{N}\vec{\sigma} N)(\overline{N}\vec{\sigma} N)~.
\label{lnn}
\ee
 The fact that these are the leading tree-level contributions  is a consequence of our counting eq.~\eqref{ffg}, which determines that the 
 lowest order diagrams are those  with $(d_i=0,v_i=1,\omega_i=1)$ and 
$(d_i=1,v_i=1,\omega_i=0)$.
The former arises from the contact interaction Lagrangian,  
 eq.~\eqref{lnn}, and the latter corresponds to the lowest order one-pion exchange. 
The tree-level scattering amplitude for  $N_{s_1,i_1}(\vp_1) N_{s_2,i_2}(\vp_2) \to  N_{s_3,i_3}(\vp_3)
N_{s_4,i_4}(\vp_4)$ from eq.~\eqref{lnn} is
\begin{align}
T_{NN}^{c}=&-C_S\left(\delta_{s_3 s_1}\delta_{s_4 s_2}\,\delta_{i_3i_1}
\delta_{i_4 i_2}-\delta_{s_3 s_2}\delta_{s_4 s_1}\,\delta_{i_3i_2}
\delta_{i_4i_1}\right)\nn\\
&-C_T\left(\vec{\sigma}_{s_3 s_1}\cdot \vec{\sigma}_{s_4 s_2}\,
\delta_{i_3i_1}\delta_{i_4i_2}-\vec{\sigma}_{s_3 s_2}\cdot \vec{\sigma}_{s_4
s_1} \,\delta_{i_3i_2}\delta_{i_4i_1}\right)~,
\label{feynman}
\end{align}
 where $s_m$ is a spin label and $i_m$ an isospin one. 
Obviously, this amplitude only contributes to the nucleon-nucleon S-waves. 
The one-pion exchange tree-level amplitude is
\begin{align}
T_{NN}^{1\pi}&=\frac{g_A^2}{4f^2}\left[
\frac{
(\vec{\tau}_{i_3i_1}\cdot \vec{\tau}_{i_4i_2})
(\vec{\sigma}\cdot \vq)_{s_3s_1}(\vec{\sigma}\cdot
\vq)_{s_4s_2}}{\vq^2+m_\pi^2-i\epsilon}
-\frac{(\vec{\tau}_{i_4i_1}\cdot 
\vec{\tau}_{i_3i_2})(\vec{\sigma}\cdot \vq')_{s_4s_1}(\vec{\sigma}\cdot
\vq')_{s_3s_2}}{{\vq'}^2+m_\pi^2-i\epsilon}
\right]~,
\label{1pi.gen}
\end{align}
with $\vq=\vp_3-\vp_1$ and $\vq'=\vp_4-\vp_1$. 
The corresponding nucleon-nucleon partial waves due to one-pion exchange  
can be calculated using  eq.~(\ref{pw.exp.def}). Instead, we first take 
the one-pion exchange between nucleon-nucleon states with definite 
spin and isospin, so that eq.~(\ref{pw.exp.def})  simplifies to 
\begin{align}
N_{JI}^{1\pi}(\ell,\bar{\ell},S)&=\frac{Y_{\bar{\ell}}^0(\hat{\vz})}{2J+1}\sum_{
\sigma_i,\sigma_f=-S}^{S}
(0\sigma_i \sigma_i|\bar{\ell}S J)(m\sigma_f \sigma_i|\ell S J) \int d\hat{p}
\,T^{1\pi}_{\sigma_f\sigma_i}(S,I)Y_{\ell}^m(\hat{p})^*~,
\label{1pi.pw}
\end{align}
where $\ell$ and $\bar{\ell}$ are the final and initial orbital angular momentum in the two-nucleon rest frame, respectively. 
Explicit expressions for  $N_{JI}^{1\pi}(\ell,\bar{\ell},S)$ are given in
Appendix~D of~\cite{techrep}.
 The sum of the local vertex,
eq.~(\ref{feynman}), and the one-pion exchanges, eq.~(\ref{1pi.gen}), is
represented diagrammatically in the following by the exchange of a wiggly line, fig.~\ref{fig:wig}.

\begin{figure}[ht]
\psfrag{k}{$k$}
\psfrag{p}{$p$}
\psfrag{l}{$\ell$}
\psfrag{pi}{$\pi$}
\psfrag{r}{$k-\ell$}
\centerline{\epsfig{file=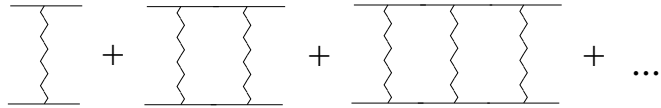,width=.5\textwidth,angle=0}}
\vspace{0.2cm}
\caption[pilf]{\protect \small
Resummation of the two-nucleon reducible diagrams. This is referred in the
text as a resummation of the right-hand cut or unitarity cut.
\label{fig:sum}}
\end{figure} 

\begin{figure}[ht]
\psfrag{deg}{{\small degrees}}
\psfrag{Tlab}{{\small $p_{cm}$ (MeV)}}
\psfrag{1S0}{$^1 S_0$}
\psfrag{3S1}{$^3S_1$}
\psfrag{3D1}{$^3D_1$}
\psfrag{epsilon_1}{$\epsilon_1$}
\centerline{\epsfig{file=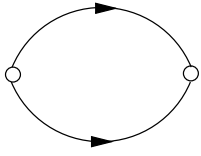,width=0.15\textwidth,angle=0}}
\vspace{0.2cm}
\caption[pilf]{\protect \small
Unitarity loop corresponding to the function $g(A)$,  eq.~(\ref{int.g}).
\label{fig:g}}
\end{figure}

Refs.~\cite{wein1,wein2} argued that  the
two-nucleon reducible diagrams 
should be resummed because they are infrared enhanced  (by large factors $\sim
m/|\vp_i|$) due to the large nucleon mass. This resummation,  depicted in
fig.~\ref{fig:sum},  is required by our power counting, eq.~\eqref{ffg}, 
  when the latter is applied to  the vacuum case as discussed at the end of section \ref{sec:pw}.
 Notice that every two-nucleon reducible loop in the string is connected by adding  ${\cal O}(p^0)$ local interactions and the exchange of pionic-lines at the lowest order.  As pointed out in the conditions  ii) and iii) of section~\ref{sec:pw} the counting does not increase then. 
Rephrasing the  discussion of this section to the present case, the nucleon propagators in a two-nucleon reducible loop follow the
non-standard counting and each of them is ${\cal O}(p^{-2})$, so that altogether are  ${\cal O}(p^{-4})$.  The leading wiggly line exchange is  ${\cal
 O}(p^0)$. When these two factors are  multiplied by the ${\cal
 O}(p^4)$ contribution from the measure of the loop integrals, associated with the running momenta of the wiggly lines, an ${\cal O}(p^0)$ contribution results. The latter
does not increase the chiral order and the series of diagrams in
fig.~\ref{fig:sum} must be resummed.\footnote{One could argue that if the nucleon propagator is taken as ${\cal O}(p^{-2})$ for the two-nucleon reducible loops, then the measure could be taken as ${\cal O}(p^5)$, counting $dp^0$ as ${\cal O}(p^2)$. If this counting  is followed,  a suppression by an extra power of $p$ seems to arise. However, this factor is multiplied by the large nucleon mass, 
so that  $m p$ finally results, which is then multiplied by local interactions. If the latter count as $1/m\Lambda$, with $\Lambda\sim m_\pi$, the resummation would be required as well within this point of view. We show below that this is the case in our approach.}  For this purpose, we follow the techniques of UCHPT \cite{nd,meis,npa} that  performs this resummation partial
wave by partial wave. Many recent nucleon-nucleon scattering analyses 
using CHPT \cite{ordo,kolck,entem,epe} follow refs.~\cite{wein1,wein2} and solve 
the Lippmann-Schwinger equation in order to accomplish such resummation. UCHPT has been applied with great 
success in meson-meson \cite{nd,report1,alba} and  meson-baryon scattering 
\cite{ramos,meis,prl1,prl2,Borasoy:2006sr,ojpa}. 
\begin{figure}[t]
\psfrag{inf}{{\small $\infty$}}
\psfrag{e}{{\small $\epsilon$}}
\psfrag{a}{{\small $-m_\pi^2/4$}}
\psfrag{C1}{{\small $C_I$}}
\psfrag{C2}{{\small $C_{II}$}}
\centerline{\epsfig{file=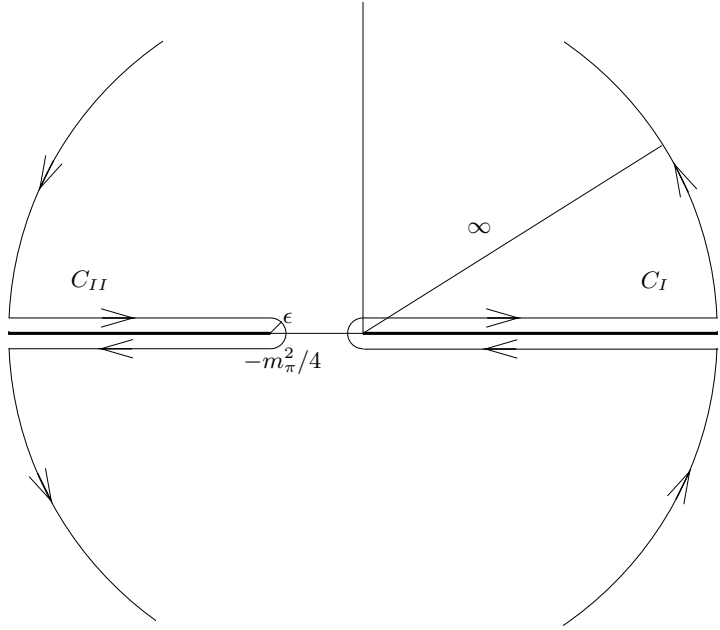,width=.55\textwidth,angle=0}}
\vspace{0.2cm}
\caption[pilf]{\protect \small
Right- and left-hand cuts of $T_{JI}(\ell,\bar{\ell},S)(\vp^2)$, for $\vp^2>0$ and $\vp^2<-m_\pi^2/4$, in order. We have also indicated the integration contours  $C_I$ and $C_{II}$ used for calculating $g(\vp^2)$ and $N_{JI}(\ell,\bar{\ell},S)(\vp^2)$, eqs.~\eqref{dis.rel.g} and \eqref{dis.nji}, respectively. The union of both contours $C_I\cup C_{II}$ is the one used  for $T_{JI}^{-1}(\ell,\ell,S)(\vp^2)$, eq.~\eqref{dis.inv.t}. In the calculation $\epsilon\to 0^+$.
\label{fig:cuts}}
\end{figure} 
The master equation for UCHPT is
\be
T_{JI}(\ell,\bar{\ell},S)=\left[I+ N_{JI}(\ell,\bar{\ell},S)\cdot 
g \right]^{-1}\cdot N_{JI}(\ell,\bar{\ell},S)~.
\label{master}
\ee

This equation, derived in detail in refs.~\cite{nd,meis,nn}, results by
performing  a once-subtracted dispersion relation of the inverse of a partial
wave amplitude.   The function $g$  is defined as follows. 
Let us denote by $\vp$ the center-of-mass (CM) three-momentum of the nucleon-nucleon system.
A nucleon-nucleon partial wave amplitude  has two cuts \cite{spearman}, the right hand-cut for $\infty>\vp^2>0$, due to unitarity, and the left-hand cut for $-\infty<\vp^2<-m_\pi^2/4$, due to the crossed channel dynamics. The upper limit for the latter interval is given by the one-pion exchange, as the pion is  the lightest particle that can be exchanged. These cuts are represented in fig.~\ref{fig:cuts}. 
Because of unitarity, a partial wave satisfies in the CM frame  that
\be
\hbox{ Im}T_{JI}(\ell,\bar{\ell},S)^{-1}_{\ell,\bar{\ell}}=-\frac{m |\vp|}{4
\pi}\delta_{\ell \bar{\ell}}~,
\label{unitarity}
\ee
  above the elastic threshold and below the pion production one. 
  The function $g$ in eq.~\eqref{master} only has a right-hand cut and 
its discontinuity along this cut is $2i$ times the right hand side of eq.~\eqref{unitarity}. 
  A once-subtracted dispersion relation  can be written down given the degree of divergence of eq.~\eqref{unitarity} for $\vp^2\to \infty$. The integration 
contour taken is a circle of infinite radius  
centered at the origin that engulfs the right-hand cut, as shown in fig.~\ref{fig:cuts} by $C_I$. In this way 
\begin{align}
g(A)=&g(D)-\frac{m(A-D)}{4\pi^2}\int_0^\infty
dk^2\frac{k}{(k^2-A-i\epsilon)(k^2-D-i\epsilon)}
\nn\\=&g(D)-\frac{i m}{4\pi}\left( \sqrt{A}-i\sqrt{|D|}\right)\nn\\
\equiv &
g_0-i\frac{m\sqrt{A}}{4\pi}~.
\label{dis.rel.g}
\end{align}
One subtraction has been taken at $D<0$ so that the integral is convergent.  
Note that the subtraction constant $g(D)$ is the value of $g(A)$ at $A=D$, in particular, $g_0=g(0)$. 
  Since $g(\vp^2)={\cal O}(p^0)$, as discussed above, it follows that
\begin{align}
g_0=g(0)={\cal O}(p^0)~.
\label{g0}
\end{align}

The function $g(A)$ corresponds to the divergent integral
\be
g(A)\to-m \int \frac{d^3 k}{(2\pi)^3}\frac{1}{k^2-A-i\epsilon}~.
\label{int.g}
\ee
The previous integral,  depicted in fig.~\ref{fig:g}, is linearly divergent 
although it shares the same analytical properties as eq.~(\ref{dis.rel.g}).
In dimensional regularization with $D\to 3$ one has, $g(A)=-i m\sqrt{A}/4\pi$.
 This result is purely imaginary above threshold, $A>0$, and it corresponds
 to the imaginary part of eq.~(\ref{dis.rel.g}). However, this is just a
 specific characteristic of the regularization method employed, since, as is 
explicitly shown in  eq.~(\ref{dis.rel.g}), there is an undetermined 
constant  $g_0$.  For $A=0$ 
the  integral in eq.~\eqref{int.g} is (infinitely-)negative, so that it is quite natural to assume that  $g_0<0$. Another more fundamental reason for taking $g_0<0$, required by the consistency of the approach, is given below. In the following, we regularize any 
two-nucleon reducible loop in terms of the subtraction constant $g_0$, taking into account 
eqs.~\eqref{dis.rel.g} and 
\eqref{g0}. The irreducible 
diagrams with respect to intermediate multi-nucleon states will be regularized employing 
dimensional regularization \cite{multikaiser}. This regularization method is shown up to NLO in the calculations performed in this work. For explicit calculations of loop integrals apart from $g(A)$ within this scheme see 
Appendix~\ref{sec:l11} and the calculation of the energy per nucleon, $E/A$ in section~\ref{sec:ener}.

Next, we consider how to fix  $N_{JI}(\ell,\bar{\ell},S)$ in
eq.~(\ref{master}). This function has only a left-hand cut, due to the exchange of pions in the chiral EFT (of course, in a meson-exchange calculation it would include further exchanges of other heavier mesons like $\rho$, $\omega$, etc). It has no right-hand cut since the latter is fully incorporated in the function $g(A)$ by construction. As a result, $N_{JI}$ should not be infrared enhanced since the effects of the large nucleon mass, associated with the 
two-nucleon reducible diagrams that give rise to the unitarity cut,  are taken into account by eq.~(\ref{master}). 
 Note that the latter results by integrating over the two-nucleon intermediate
states at the level of the inverse of a partial wave, eq.~\eqref{dis.rel.g}. 
In a plain perturbative chiral calculation of a nucleon-nucleon partial wave
the right-hand cut is not resummed and the convergence of the perturbative series is spoilt due to the infrared 
 enhancement of the two-nucleon reducible loops. However, since the latter are resummed in eq.~\eqref{master}, the idea is to match this general equation with a perturbative calculation within  CHPT up to the same number of two-nucleon reducible loops. The number must be the same to guarantee that $N_{JI}$ is real along the physical region and fulfills the requirement of not having  right-hand cut.  We can make use 
of the  geometric series in powers of $g$ of $T_{JI}$, eq.~\eqref{master}, 
\begin{align}
T_{JI}(\ell,\bar{\ell},S)&=
N_{JI}(\ell,\bar{\ell},S)
-N_{JI}(\ell,\bar{\ell},S)\cdot g \cdot N_{JI}(\ell,\bar{\ell},S)\nn\\
&+N_{JI}(\ell,\bar{\ell},S)\cdot g\cdot N_{JI}(\ell,\bar{\ell},S)\cdot g \cdot
N_{JI}(\ell,\bar{\ell},S) +\ldots
\label{geo.ser}
\end{align}
where we have used the matrix notation $N_{JI}\cdot g$ for the case with
coupled channels. Here, $g$ just corresponds to the identity matrix times eq.~(\ref{dis.rel.g}), because 
the latter is the same for all  partial
waves. Together with the previous geometric series one also has the standard chiral expansion 
\begin{align}
N_{JI}=\sum_{m=0}^n N_{JI}^{(m)}~,
\label{sum.ord}
\end{align} 
with the chiral order indicated by the superscript. Now, for the determination of the different $N_{JI}^{(m)}$, $m\leq n$, the matching  between eq.~\eqref{geo.ser} is performed with a perturbative chiral calculation for which the reducible part of every two-nucleon reducible (or unitarity) loop  is counted as ${\cal O}(p)$. It is important to stress that this counting is applied  for calculating $N_{JI}$  not $T_{JI}$, for the latter each two-nucleon reducible loop counts as ${\cal O}(p^0)$, eq.~\eqref{ffg}.  In this way, the matching up to a chiral order $n$ automatically comprises at most $n$ two-nucleon unitarity loops. In addition, the chiral order of the vertices employed will also make that no spurious imaginary parts are left since one is handling in the matching  with perturbative unitarity  up to order $n$. 

A few examples will clarify this process of matching and 
 why it makes sense to take as ${\cal O}(p)$ the reducible part of a two-nucleon reducible loop for calculating $N_{JI}$ within UCHPT.  At lowest order, $n=0$,  there are no two-nucleon reducible loops and  $N_{JI}^{(0)}(\ell,\bar{\ell},S)=L_{JI}^{(0)}(\ell,\bar{\ell},S)$, where the 
latter is the tree-level calculation in CHPT at ${\cal O}(p^0)$ given by the sum of  $T_{NN}^c$, eq.~\eqref{feynman}, and $T_{NN}^{1\pi}$, eq.~\eqref{1pi.gen},  projected in the appropriate partial wave. This is the wiggly line at the far left of fig.~\ref{fig:sum}. 
At ${\cal O}(p)$, $n=1$, the only new contribution is the two-nucleon reducible part of the 
second diagram in fig.~\ref{fig:sum}, denoted
by $L^{(1)}_{JI}(\ell,\bar{\ell},S)$ for a given partial wave.  Writing
$N_{JI}=N_{JI}^{(0)}+N_{JI}^{(1)}+{\cal O}(p^2)$, and matching
eq.~(\ref{geo.ser})  with the sum of the first two diagrams  
of fig.~\ref{fig:sum} one has
\begin{align}
N_{JI}^{(0)}+N_{JI}^{(1)}-N_{JI}^{(0)}\cdot g 
\cdot N_{JI}^{(0)}+{\cal O}(p^2)=L_{JI}^{(0)}+L^{(1)}_{JI}+{\cal O}(p^2)~,
\label{eq.a10}
\end{align}
 with the result  
 \begin{align}
N_{JI}^{(1)}=L^{(1)}_{JI}+N_{JI}^{(0)}\cdot g \cdot N_{JI}^{(0)}~.
\label{eq.a1} 
\end{align}
Notice that in the expansion of eq.~(\ref{geo.ser}) each factor of the kernel 
$N_{JI}(\ell,\bar{\ell},S)$ multiplies the loop function $g$ with its value on
shell.  This is why in eq.~(\ref{eq.a10}) we have $-N_{JI}^{(0)}\cdot g \cdot
N_{JI}^{(0)}$ for one iteration of $g$, which is then subtracted from the
function $L^{(1)}_{JI}$ in eq.~(\ref{eq.a1}).
 This equation  shows explicitly  that the simultaneous expansion in chiral powers and number of loops for fixing
$N_{JI}^{(n)}$ implies that UCHPT really takes as ${\cal O}(p)$  the difference
between a full calculation of {\it one} two-nucleon reducible loop and the result
obtained by factorizing the vertices on-shell, eq.~(\ref{geo.ser}). Ultimately this 
relies on the fact that the difference has no right-hand cut, which is the one associated with the infrared enhanced two-nucleon reducible loops, and it has only left-hand cut. The latter is incorporated perturbatively in the interaction kernel
$N_{JI}(\ell,\bar{\ell},S)$, which is  improved order by order. This is 
the reason why we have treated
 the expansion in two-nucleon reducible loops on the 
same foot as the chiral expansion. This procedure is iterated up to any desired order. 
E.g. at ${\cal O}(p^2)$ new contributions would arise that require the
calculation of the irreducible part of the box diagram in fig.~\ref{fig:sum} 
and the reducible parts of the last diagram of fig.~\ref{fig:sum} with the
wiggly line exchange iterated twice \cite{iteratedk}. In addition, there are also local  
interaction terms from the quartic nucleon Lagrangian and two-nucleon 
irreducible pion loops \cite{kaiser,epe,entem,multikaiser}. If we denote all these new 
contributions projected onto the
corresponding partial wave by $L_{JI}^{(2)}(\ell,\bar{\ell},S)$, the following equation results 
\begin{align}
N_{JI}^{(2)}=L_{JI}^{(2)}+N_{JI}^{(1)}\cdot g\cdot
N_{JI}^{(0)}+N_{JI}^{(0)}\cdot 
g \cdot N_{JI}^{(1)}-N_{JI}^{(0)}\cdot g\cdot N_{JI}^{(0)}\cdot g \cdot
N_{JI}^{(0)}~.
\label{eq.a2}
\end{align}
 The  $N_{JI}$ calculated up to some given order in the expansion  eq.~(\ref{sum.ord}) is
 then substituted in eq.~(\ref{master}). On the other hand, one can match
 formally eq.~(\ref{master}) with a perturbative chiral calculation of
 $T_{JI}(\ell,\bar{\ell},S)$ for any value of  the S-wave nucleon-nucleon
 scattering  lengths because the latter enter parametrically in the
 calculation. This procedure gives rise to values of the low-energy constants 
$C_S$ and $C_T$ that are consistent with their ascribed ${\cal O}(p^0)$
scaling, see eq.~(\ref{cs.ct}) below. It is worth pointing out that  
eq.~(\ref{master})  is algebraic, so that the numerical burden for in-medium 
calculations is reduced tremendously.

\begin{figure}[ht]
\psfrag{k}{$k$}
\psfrag{p}{$p$}
\psfrag{l}{$\ell$}
\psfrag{pi}{$\pi$}
\psfrag{r}{$k-\ell$}
\centerline{\epsfig{file=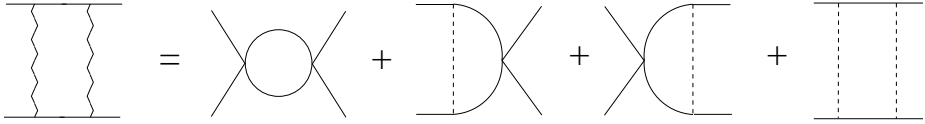,width=.7\textwidth,angle=0}}
\vspace{0.2cm}
\caption[pilf]{\protect \small
Box diagram, $L^{(1)}_{JI}$, originating  from the first iteration of a wiggly line. It
consists of the diagrams shown on the right-hand side of the figure with zero, 
one or two local vertices and/or one-pion exchange amplitudes. 
\label{fig:box}}
\end{figure}

The dependence on the parameter $g_0$
takes places because of the infrared enhanced  two-nucleon reducible loops. 
This has made necessary to resum the right-hand cut, which  requires the 
presence of one subtraction constant, $g_0$, eq.~\eqref{dis.rel.g}. 
 Indeed, for a fixed chiral order, according to the application of eq.~\eqref{ffg} to nucleon-nucleon scattering in vacuum, the dependence on $g_0$ becomes smaller as higher powers of $g$ are considered for calculating $N_{JI}$, eq.~\eqref{sum.ord}. 
To show this, we need to take advantage of the analytical properties of $N_{JI}$, eq.~\eqref{master}.  As discussed above, this quantity has only the left-hand cut, shown in fig.~\ref{fig:cuts}. 
 For the following discussion we take the case of one uncoupled channel to simplify the writing. 
  Its generalization to coupled channels is straightforward employing a matrix notation.  The imaginary  part of $N_{JI}$ along the left-hand cut is given by,
\begin{align}
\hbox{Im}N_{JI}&=\frac{|N_{JI}|^2}{|T_{JI}|^2}\hbox{Im}T_{JI}=|1+g N_{JI}|^2 
\hbox{Im}T_{JI}~,~|\vp|^2<-\frac{m_\pi^2}{4}~.
\label{disc.nji}
\end{align}
Note that $g$ is real along the left-hand cut. 
 We employ this result to write down a  once-subtracted dispersion
relation for $N_{JI}$. The integration contour is shown in fig.~\ref{fig:cuts} as $C_{II}$ and consists of a circle of infinite radius  
centered at the origin that engulfs the left-hand cut.
\begin{align}
N_{JI}(A)&=N_{JI}(D)+\frac{A-D}{\pi}\int_{-\infty}^{-m_\pi^2/4} dk^2\frac{ \hbox{Im}T_{JI}(k^2)\,|1+g(k^2) N_{JI}(k^2)|^2}{(k^2-A-i\epsilon)(k^2-D)}~,
\label{dis.nji}
\end{align}
  We have taken one subtraction in
the dispersion relation because the one-pion exchange amplitude, eq.~\eqref{1pi.gen},  tends to a constant for $\vk^2\to \infty$. 
 Then, we have for $T_{JI}$, eq.~\eqref{master},
\begin{align}
T_{JI}(A)=\left(\left[N_{JI}(D)+\frac{A-D}{\pi}\int_{-\infty}^{-m_\pi^2/4} dk^2\frac{\hbox{Im}T_{JI}\, |1+g N_{JI}|^2}{(k^2-A-i\epsilon)(k^2-D)}\right]^{-1}+g(A)\right)^{-1}~.
\label{t.nji}
\end{align}

In order to solve eq.~\eqref{dis.nji} one needs $\hbox{Im}T_{JI}$ as input  along the left-hand cut. CHPT could be used, since this imaginary part is due to  multi-pion exchanges. As a result one could afford its calculation perturbatively because  the infrared enhancements associated with the right-hand cut are absent in the discontinuity along the left-hand cut. The reason is because this discontinuity, according to Cutkosky's theorem \cite{landau,cutkosky}, implies to put on-shell pionic lines so that within  loops the pion poles are picked up making that the energy along nucleon propagators now is of ${\cal O}(p)$, instead of a nucleon kinetic energy. In this way, the order of the diagram rises compared to that of the reducible parts and it becomes a perturbation. E.g., let us take as illustration the last diagram on the r.h.s. of fig.~\ref{fig:box}, corresponding to  the twice iterated one-pion exchange. Its reducible part is infrared enhanced, which has been calculated by us in the presence of the nuclear medium, in agreement with ref.~\cite{peripheral} when reduced to the vacuum case. However, its discontinuity across the left-hand cut arises by putting on-shell the two intermediate pion lines. Its leading
 contribution to $\hbox{Im} T$ in a $1/m$ expansion in the $t$-channel CM frame is given by
\begin{align}
\hbox{Im}T=-{\cal N}\frac{(t/4-m_\pi^2)^{5/2}}{4\pi t^{3/2}}~,~t> 4m_\pi^2~,
\label{disc.tl}
\end{align}
with $t=-2\vp^2(1-\cos\theta),~\cos\theta\in[-1,1]$ and ${\cal N}$ is a numerical factor due to the spin algebra. No factor $m$ appears in the numerator and it follows the standard chiral counting. In this way, the leading contribution to $\hbox{Im}T_{JI}$ along the left-hand cut is given by the one-pion exchange. The latter can then be inserted in eq.~\eqref{dis.nji}, once projected in a given partial wave. The solution of this equation would correspond to the leading result for $N_{JI}$ in the chiral expansion of eq.~\eqref{ffg}, 
 without involving the expansion in the number of two-nucleon reducible loops. This interesting exercise will be left for future consideration. 

Pion exchange amplitudes  are treated perturbatively in the Kaplan-Savage-Wise (KSW) power counting \cite{kaplan,kswnnlo}. This is done  for any energy region and, in particular, along both the right- and left-hand cuts. On the other hand, the dispersive treatment offered here only needs as input  the discontinuity (imaginary part) of a nucleon-nucleon partial wave along the left-hand cut, see eq.~\eqref{t.nji}. This discontinuity arises due to pion exchanges which, as discussed in the previous paragraph, could  be calculated perturbatively in CHPT. Differences with respect to KSW arise due to the resummation of the right-hand-cut in eq.~\eqref{t.nji}, including both local and pion-exchange contributions. This would correspond to higher orders in KSW power counting \cite{kswnnlo}.  Notice also that while KSW is a strict perturbation theory calculation in quantum field theory (QFT) ours merges inputs from  perturbative QFT and S-matrix theory, see e.g. ref.~\cite{barton} for a pedagogical account of first applications of the similar N/D method to nucleon-nucleon scattering.

Two subtraction constants appear in eq.~\eqref{t.nji}, $N_{JI}(D)$ from  eq.~\eqref{dis.nji} and $g_0$ from the function $g(A)$, eq.~\eqref{dis.rel.g}. We are going to show that they are not independent, however. The two constants have appeared  due to the splitting between the functions $N_{JI}$ and $g$ when expressing   $T_{JI}^{-1}=N_{JI}^{-1}+g$,  eq.~\eqref{master}. This is analogous to the standard fact that in any renormalization scheme there is an exchange of
 contributions between  local parts in loops and local counterterms.
 In order to proceed with the demonstration that the resulting $T_{JI}$, eq.~\eqref{t.nji}, does not depend on the subtraction constant $g(D)$,  let us  write  directly a dispersion relation for $T_{JI}^{-1}$ taking the contour $C_I\cup C_{II}$ in fig.~\ref{fig:cuts}
\begin{align}
T_{JI}^{-1}(A)&=T_{JI}^{-1}(D)-\frac{m(A-D)}{4\pi^2}\int_0^\infty dk^2\frac{k}{(k^2-A-i\epsilon)(k^2-D)}
-\frac{(A-D)}{\pi}\int_{-\infty}^{-m_\pi^2/4} dk^2 \frac{\hbox{Im}T_{JI}/|T_{JI}|^2}{(k^2-A-\epsilon)(k^2-D)}\nn\\
&-\left.\frac{A-D}{(\ell-1)!}\frac{d^{\ell-1}}{d(k^2)^{\ell-1}}\frac{f_{JI}(k^2)}{(k^2-A)(k^2-D)}\right|_{k^2=0}~,
\label{dis.inv.t}
\end{align}
where $f_{JI}(\vp^2)=|\vp|^{2\ell} T_{JI}^{-1}(\vp^2)$~. The last term in the previous equation  gives contribution for $\ell\geq 1$ and arises due to the behaviour at threshold of a partial wave, vanishing as $|\vp|^{2\ell}$.
 Two-body unitarity is assumed all the way  along  the right-hand cut in the first integral. This is not essential for the discussion that follows and we could have written directly Im$T_{JI}^{-1}$ along the right-hand cut, as done for the left-hand one.\footnote{
In eq.~\eqref{dis.inv.t}  we could use different subtraction points for the two integrals, e.g. $B$ and $D$, respectively. One then has
\begin{align}
T_{JI}^{-1}(A)&=T_{JI}^{-1}(D)
+\frac{m(D-B)}{4\pi^2}\int_0^\infty dk^2\frac{k}{(k^2-D)(k^2-B)}
-\frac{m(A-B)}{4\pi^2}\int_0^\infty dk^2\frac{k}{(k^2-A-i\epsilon)(k^2-B)}\nn\\
&-\frac{A-D}{\pi}\int_{-\infty}^{-m_\pi^2/4}dk^2\frac{\hbox{Im}T_{JI}/|T_{JI}|^2}{(k^2-A-i\epsilon)(k^2-D)}-\left.\frac{A-D}{(\ell-1)!}\frac{d^{\ell-1}}{d(k^2)^{\ell-1}}\frac{f_{JI}(k^2)}{(k^2-A)(k^2-D)}\right|_{k^2=0}~.
\end{align}
}
As discussed above the input for solving $N_{JI}$ in  eq.~\eqref{dis.nji} is  $\hbox{Im}T_{JI}$ along the left-hand cut. This can also be shown explicitly from eq.~\eqref{dis.inv.t} by writing $1/|T_{JI}|^2=|N_{JI}^{-1}+g|^2$, as follows from eq.~\eqref{master}. Subtracting $g$ from $T_{JI}^{-1}$ we  arrive to the following equation for $N_{JI}^{-1}$,
\begin{align}
N_{JI}^{-1}(A)&=T_{JI}^{-1}(D)-g(D)-\frac{(A-D)}{\pi}\int_{-\infty}^{-m_\pi^2/4}dk^2\frac{\hbox{Im}T_{JI}\,|N_{JI}^{-1}+g|^2}{(k^2-A-\epsilon)(k^2-D)}\nn\\
&-\left.\frac{A-D}{(\ell-1)!}\frac{d^{\ell-1}}{d(k^2)^{\ell-1}}\frac{f_{JI}(k^2)}{(k^2-A)(k^2-D)}\right|_{k^2=0}~,
\label{dis.nji.inv}
\end{align}
 In the following we omit the last term in the previous equation for simplicity, since it does not depend on $g(D)$. The reader could include it straightforwardly if  desired.
If eq.~\eqref{dis.nji.inv} is solved by iteration, it is straightforward to show that $T_{JI}$ does not depend on $g_0$ at any order in the iteration. The zeroth iterated solution is $N_{JI;0}^{-1}=T_{JI}^{-1}(D)-g(D)$, which   yields $T_{JI;0}^{-1}(D)=T_{JI}^{-1}(D)-g(D)+g(A)$. Obviously, the sum $-g(D)+g(A)$ is independent of $g(D)$. For the first iterated solution one has
\begin{align}
N_{JI;1}^{-1}(A)&=T_{JI}^{-1}(D)-g(D)-\frac{A-D}{\pi}\int_{-\infty}^{-m_\pi^2/4} dk^2 \frac{\hbox{Im}T_{JI}\,|T^{-1}(D)-g(D)+g(k^2)|^2}{(k^2-A-i\epsilon)(k^2-D)}~.
\label{1st.iterated}
\end{align}
Notice that only the combination $-g(D)+g(k^2)$ appears in the integral, which is independent of $g(D)$. However, $N_{JI;1}^{-1}$ depends explicitly on $g(D)$ due to the term before the integral. Nevertheless, given that $T_{JI;1}^{-1}(A)=N_{JI;1}^{-1}(A)+g(A)$,  the first $g(D)$ on the r.h.s. of eq.~\eqref{1st.iterated} is accompanied again with $g(A)$ so that no dependence on $g(D)$ is left.  This process can be straightforwardly generalized to any order. For the $j^{th}$ iteration the combination $T^{-1}(D)-g(D)$  that  appears in $N_{JI;j}^{-1}(A)$  before the integral is added to $g(A)$ for calculating  $T_{JI;j}(A)$, so that no dependence on  $g(D)$ arises from this fact. In addition, under the integration sign we have repeatedly $j$ times the same term  $T_{JI}^{-1}(D)-g(D)+g(k^2)$, which does not depend on  $g(D)$.

From the previous discussion, one concludes quite confidently that no $g(D)$-dependence is  left because this was the case for $T_{JI}(A)$ evaluated at any order in the iterative solution of $N^{-1}_{JI}(A)$, eq.~\eqref{dis.nji.inv}.  
In this way, it is clear that one could interpret the constant $g(D)$,  eq.~\eqref{dis.rel.g}, and the subtraction point $D$ in close analogy with renormalization theory.   The latter corresponds to the ``renormalization scale" and the former fixes the ``renormalization scheme".  For a given $g(D)$ then $N_{JI}(D)$ is fixed so as to reproduce $T_{JI}(D)$ at  the point $|\vp^2|=D$. The dependence on $g(D)$ is then transmuted into the experimental input $T_{JI}(D)$. The final result should be independent of $g(D)$, which in turn, by taking the derivative of $T_{JI}^{-1}$,  eq.~\eqref{t.nji}, with respect to this parameter  implies the equation
\begin{align}
\frac{\partial N_{JI}(D)}{\partial g(D)}&=
N_{JI}^2
-\frac{2(A-D)}{\pi}\int_{-\infty}^{-m_\pi^2/4}dk^2
\frac{\hbox{Im}T_{JI}\,\hbox{Re}\bigl[(g \partial N_{JI}/\partial g(D)+N_{JI})(1+N_{JI}^* g)\bigr]}{(k^2-A-i\epsilon)(k^2-D)}~.
\end{align}
This discussion also shows that one always has the freedom to take $g_0$ to be the same for all the partial waves, as we have done.\footnote{There is  an infinity of solutions of eq.~\eqref{dis.inv.t} differing between each other in the number of zeros of $T_{JI}$. Each of these zeros is a pole of $T_{JI}^{-1}$ so that it brings altogether as free parameters the position of the pole and its residue. They are the so-called Castillejo-Dalitz-Dyson (CDD) poles \cite{castillejo}. The CDD poles are typically associated with resonances \cite{mandelstam,castillejo}. Notice that a pole in $T_{JI}^{-1}$ typically makes its real part to vanish if the remnant is a smooth function of energy around the pole. In low energy S-wave meson-meson scattering the Adler zeros  correspond to CDD poles \cite{nd}. However, for nucleon-nucleon scattering there is no evidence for a low energy zero in the partial waves (apart from the trivial one at threshold for $\ell\geq 1$.) In the pionless EFT for nucleon-nucleon interactions the third integration on the r.h.s. of eq.~\eqref{dis.inv.t} is absent. The infinity tower of chiral counterterms in this EFT can be accounted for by adding CDD poles, see ref.~\cite{nd} where this is shown explicitly for a similar problem.}

For higher partial waves it is convenient to derive the dispersion relation for $N_{JI}/|\vp|^{2\ell}$ instead of eq.~\eqref{t.nji}. In this way,  the low energy behaviour of a partial wave as $|\vp|^{2\ell}$ for $|\vp|\to 0$ is ensured, independently of the approximation for $\hbox{Im}T_{JI}$ \cite{goldberger}. The resulting expression is
\begin{align}
N_{JI}(A)&=\frac{A^\ell}{D^\ell} N_{JI}(D)+\frac{A^\ell(A-D)}{\pi }\int_{-\infty}^{-m_\pi^2/4} dk^2\frac{ \hbox{Im}T_{JI}(k^2)\,|1+g(k^2) N_{JI}(k^2)|^2}{k^{2\ell}(k^2-A-i\epsilon)(k^2-D)}~.
\label{dis.nji.k2l}
\end{align} 
Note also that for $\ell\geq 1$ no subtraction is needed if $\hbox{Im}T\sim const.$(mod $log$) for $k^2\to \infty$, as in the one-pion exchange. Then, one could also rewrite the previous equation for  $ \ell\geq 1$ as
\begin{align}
N_{JI}(A)&=\frac{A^\ell}{\pi }\int_{-\infty}^{-m_\pi^2/4} dk^2\frac{ \hbox{Im}T_{JI}(k^2)\,|1+g(k^2) N_{JI}(k^2)|^2}{k^{2\ell}(k^2-A-i\epsilon)}~.
\label{dis.nji.k2l.b}
\end{align} 
	The degree of divergence of $\hbox{Im}T_{JI}$ for $|\vp|\to \infty$ increases by including higher order loop contributions, see e.g. eq.~\eqref{disc.tl}. As a result, more subtractions should be taken and the resulting subtraction constants could be related with higher order chiral counterterms.

We have  proposed to consider $g$ as ${\cal O}(p)$ in order to fix $N_{JI}$. Indeed, $g$ is suppressed along the left-hand cut, vanishing  in the low momentum region of the dispersive integral of eq.~\eqref{dis.nji}, which  dominates its final value for low energy nucleon-nucleon scattering. On the physical Riemann sheet $|\vk|=+i\kappa$, with $\kappa=\sqrt{-k^2}>0$, and since $g_0$ is negative and of natural size $\sim -m m_\pi/4\pi$, it tends to cancel with   $-i m|\vk|/4\pi=m\kappa/4\pi>0$ and becomes zero for $\kappa=-4\pi g_0/m\sim m_\pi$. This is an important reason for having taken $g_0<0$ above. Proceeding along these lines, so that $g$ is treated as relatively small along the left-hand cut, eq.~\eqref{dis.nji} would simplify at leading order. On the one hand, $|1+g N_{JI}|^2$ is replaced by 1 and, on the other,  $\hbox{Im}T_{JI}$ is given by the one-pion exchange. Hence, one obtains for $N_{JI}^{(0)}$ in S-wave the sum of a constant plus one-pion exchange (resulting from the dispersive integral), precisely the content of the wiggly lines, fig.~\ref{fig:wig}. For $\ell\geq 1$ 
let us take directly eq.~\eqref{dis.nji.k2l.b}. In this way, when neglecting $g N_{JI}$, the dispersive integral  just gives rise to  the one-pion  exchange, as was the case for our previously calculated $N_{JI}^{(0)}$.   
One could continue further in this way, and solve eq.~\eqref{dis.nji} in a power series expansion of $g$ along the left-hand cut  at each chiral order in the calculation of $\hbox{Im}T_{JI}$. The truncation of such expansion leaves a residual $g_0$ dependence. We have followed the same point of view in order to determine $N_{JI}$ through the matching process discussed above. Indeed, alternatively to performing the geometric series expansion of eq.~\eqref{geo.ser}, we could  consider directly the inverse of $T_{JI}$, similarly as done in order to obtain eq.~\eqref{disc.nji}. Then, it follows from eq.~\eqref{master}  that
\begin{align}
\frac{1}{T_{JI}}&=\frac{1}{N_{JI}}+g~,\nn\\
\frac{T_{JI}^*}{|T_{JI}|^2}&=\frac{N_{JI}^*}{|N_{JI}|^2}+g~,\nn\\
N_{JI}&=T_{JI} |1+g N_{JI}|^2-|N_{JI}|^2 g^*~.
\label{eq.nji}
\end{align}
 The first method discussed above for determining $N_{JI}$ is the perturbative solution of eq.~\eqref{eq.nji} in a chiral series of powers of $g$.  The solutions of eqs.~\eqref{eq.nji} and \eqref{dis.nji} employing  the perturbative method are equivalent because $N_{JI}$ from eq.~\eqref{eq.nji} has only a left-hand cut, being its imaginary part along this cut the same as eq.~\eqref{disc.nji}, and it is analytical, so that it satisfies 
 the perturbative version in power of $g$ of 
  the dispersion relation eq.~\eqref{dis.nji}.\footnote{We have shown this equivalence explicitly for $N_{JI}^{(0)}$. It is also straightforward to show it for $N_{JI}^{(1)}$.} The following remark is in order. The perturbative solution in the chiral expansion of powers of $g$  of eq.~\eqref{eq.nji}  has the advantages over solving eq.~\eqref{dis.nji} that it is algebraic and  the chiral counterterms in $T_{JI}$ are taken into account in the solution $N_{JI}$ in a straightforward manner. It is also very versatile, so that it can be extrapolated straightforwardly to correct by initial and final state interactions and to the nuclear medium.  Notice that eq.~\eqref{eq.nji} can only be solved perturbatively since the input $T_{JI}$ is calculated in CHPT and only  fulfills unitarity perturbatively. However, the exact solution of the integral equation eq.~\eqref{dis.nji} has the advantage of not requiring the expansion in powers of $g$ but just the chiral series on $\hbox{Im}T_{JI}$ along the left-hand cut, and the latter expansion rests in a sound basis as discussed above.

We now concentrate on fixing the constants $C_S$ and $C_T$ from the local
quartic nucleon Lagrangian, eq.~(\ref{lnn}). These constants and $g_0$,
eq.~\eqref{g0}, are the only free parameters that enter in the evaluation of 
the nucleon-nucleon scattering amplitudes from eq.~(\ref{master}) up to ${\cal
  O}(p)$. We first discuss the LO result and then the NLO one. $C_S$ and $C_T$
are fixed by considering  the S-wave nucleon-nucleon scattering lengths $a_t$
and $a_s$ for the triplet and singlet channels, respectively. At ${\cal
  O}(p^0)$ we have at threshold 
\begin{align}
T_{01}(0,0,0)&=\frac{-(C_S-3C_T)}{1-g_0(C_S-3C_T)}~,\nn\\
T_{10}(0,0,1)&=\frac{-(C_S+C_T)}{1-g_0(C_S+C_T)}~.
\label{t.th}
\end{align}
The triplet S-wave is elastic at this energy, without mixing with 
the $^3D_1$ partial wave,  because of the vanishing of the three-momentum. 
The resulting expressions for the  scattering lengths from eq.~(\ref{t.th})
imply that
\begin{align}
C_S&=\frac{m}{16\pi}\frac{16\pi g_0/m+3/a_s+1/a_t}{(g_0+m/(4\pi a_s))(g_0+m/(4\pi a_t))}~,
\nn\\
C_T&=\frac{m}{16\pi}\frac{1/a_s-1/a_t}{(g_0+m/(4\pi a_s))(g_0+m/(4\pi a_t))}~.
\label{cs.ct}
\end{align}
One of the benchmark characteristics of nucleon-nucleon scattering are the large
absolute values of the S-wave scattering lengths $a_s=-23.758\pm 0.04$~fm 
and $a_t=5.424\pm 0.004$~fm, so that $m_\pi\gg |1/a_s|$, $1/a_t$. Given the expression 
for the imaginary part of $g(A)$ above threshold in eq.~\eqref{dis.rel.g} one can estimate that $g_0\sim -m m_\pi/4\pi\sim -0.54~m_\pi^2$,\footnote{As explicitly shown in the second line of eq.~\eqref{dis.rel.g} one can trade between the subtraction constant and $-i m\sqrt{A}/4\pi$ just by changing the subtraction point. In a natural way, both should be taken of similar size for estimations.} which is then
much larger in absolute value than $m/(4\pi |a_s|)$ and  $m/(4\pi a_t)$, although there is a difference because $a_t$ is smaller by around a factor 4 than $|a_s|$. As a result,  it follows from eq.~\eqref{cs.ct}  that
 $|C_S|\sim 1/|g_0| 
\gg |C_T|={\cal O}(m/16\pi a_t g_0^2)$. In this way, the low-energy constants
$C_S$ and
$C_T$ do not diverge for $a_s$, $a_t\to \infty$ and after iteration it is still consistent to treat 
 ${\cal L}_{NN}^{(0)}$, eq.~\eqref{lnn}, as ${\cal O}(p^0)$. Notice as well that the one loop iteration of the contact terms compared in absolute value with the tree level goes like $-m |\vp| (C_S-(4S-1)C_T)/4 \pi$, 
 taking into account the expression for $g(A)$ given in eq.~\eqref{dis.rel.g}. The three-momentum is divided  by the scale $\sim -4\pi/m C_S\sim m_\pi$, considering the just given estimates for $C_S\sim 1/g_0$ and  $g_0\sim -m m_\pi/4\pi$. This justifies to iterate these diagrams for $|\vp|={\cal O}(p)$ as discussed above. For the case of the once-iterated pion exchange one would have   the factor $m |\vp| g_A^2/16\pi f_\pi^2$ as compared with the tree level one-pion exchange. Then $|\vp|$ is divided by the scale $16\pi f_\pi^2/m g_A^2\sim 2 m_\pi={\cal O}(p)$, and the one-pion exchange should be as well iterated together with the lowest order contact terms.  The issue of iterating potential pions is analyzed in detail in ref.~\cite{kswnnlo}, in order to understand the failure of the KSW power counting in some triplet channels, particularly, for the $^3S_1{-}^3D_1$  and $^3P_{0,2}$ channels. The authors of ref.~\cite{kswnnlo} conclude that for some spin triplet channels the summation of potential pion diagrams is necessary to reproduce observables, while for the singlet channels this iteration does not seem to be a significant improvement over treating pion exchanges perturbatively.

\begin{figure}[ht]
\psfrag{li}{{\small $\ell_i$}}
\psfrag{L}{{\small $\begin{array}{l} \\ g_0~(m_\pi^2) \end{array}$}}
\centerline{\epsfig{file=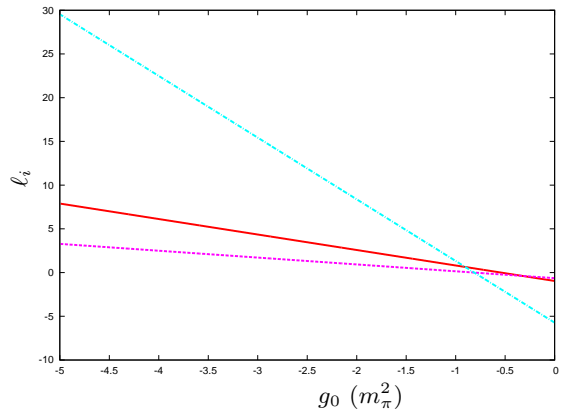,width=.3\textwidth,angle=-90}}
\vspace{0.2cm}
\caption[pilf]{\protect \small (Color online.) 
Values for $\ell_1$ (red solid line), $\ell_2(^1S_0)$ (magenta dashed line) 
and $\ell_2(^3S_1)$ (cyan dot-dashed line) as a function of $g_0$. 
$\ell_2$ is expressed in units of $m_\pi^{-2}$.    
\label{fig:rl112}}
\end{figure} 

Only local terms and one-pion exchange contributions 
enter in the calculation of $N^{(0)}_{JI}(\ell,\bar{\ell},S)$. 
This is rather simplistic in order to describe  properly the nucleon-nucleon
interactions as a function of energy  soon 
 above threshold.  
 Let us now consider eq.~(\ref{master}) with $N_{JI}$  up to ${\cal
  O}(p)$.
At this order, $\hbox{Im}T_{JI}$ along the left-hand cut is still given by the one-pion exchange, so that the exact solution of eq.~\eqref{dis.nji} for $N_{JI}$ would be the same. The differences observed in the results at ${\cal O}(p^0)$ and ${\cal O}(p)$ are then due to keep a one more factor $g$ in the perturbative solution of eq.~\eqref{dis.nji}. This discussion shows clearly the mixed nature of the chiral expansion in powers of $g$ for obtaining $N_{JI}$. 

We employ the ${^1S_0}$ and ${^3S_1}$ scattering lengths for evaluating  
 $C_S$ and $C_T$ at ${\cal O}(p)$. We denote by $a$ any of these scattering lengths and apply eq.~(\ref{master}) at threshold. We obtain  
\begin{align}
a=-\left.\frac{1}{k}\frac{\hbox{Im}T_{JI}}{\hbox{Re}T_{JI}}\right|_{k\to 0}=-\left.\frac{m}{4\pi}\frac{N_{JI
}}{1+g_0 N_{JI} }\right|_{k\to 0}~. 
\label{eq.sc1}
\end{align}
Taking eq.~(\ref{eq.a1}) at threshold we rewrite $N_{JI}^{(0)}=-C$ and  express
$L^{(1)}_{JI}\equiv -C^2 g_0+C\ell_1+\ell_2$ because the box diagram
$L^{(1)}_{JI}$,  fig.~\ref{fig:box}, consists of four contributions with two,
one and zero local  vertices. The first contribution is given by $-C^2 g_0$,
the second by $C\ell_1$ and the last one by $\ell_2$, respectively.  The
coefficients $\ell_1$ and $\ell_2$ are given in terms of $g_0$ and the known
parameters $m$, $g_A$ and $m_\pi$. $\ell_1$ is the same for the partial waves
 $^1S_0$ and ${^3S_1}$  while $\ell_2$ is different.  The values of $\ell_1$ and $\ell_2$ as a function of $g_0$ are shown in fig.~\ref{fig:rl112}. Substituting these expressions in eq.~(\ref{eq.sc1}) 
\begin{align}
C=\frac{C^{(0)}+\ell_2}{1-\ell_1}~,
\label{cs.nlo}
\end{align}
with $C^{(0)}=1/(\frac{m}{4\pi a}+g_0)$ 
the ${\cal O}(p^0)$ result.

\begin{figure}[ht]
\psfrag{deg}{{\small degrees}}
\psfrag{p}{{\small $|\vp|$ (MeV)}}
\psfrag{1S0}{$^1 S_0$}
\psfrag{3S1}{$^3S_1$}
\psfrag{3D1}{$^3D_1$}
\psfrag{epsilon_1}{$\epsilon_1$}
\centerline{\epsfig{file=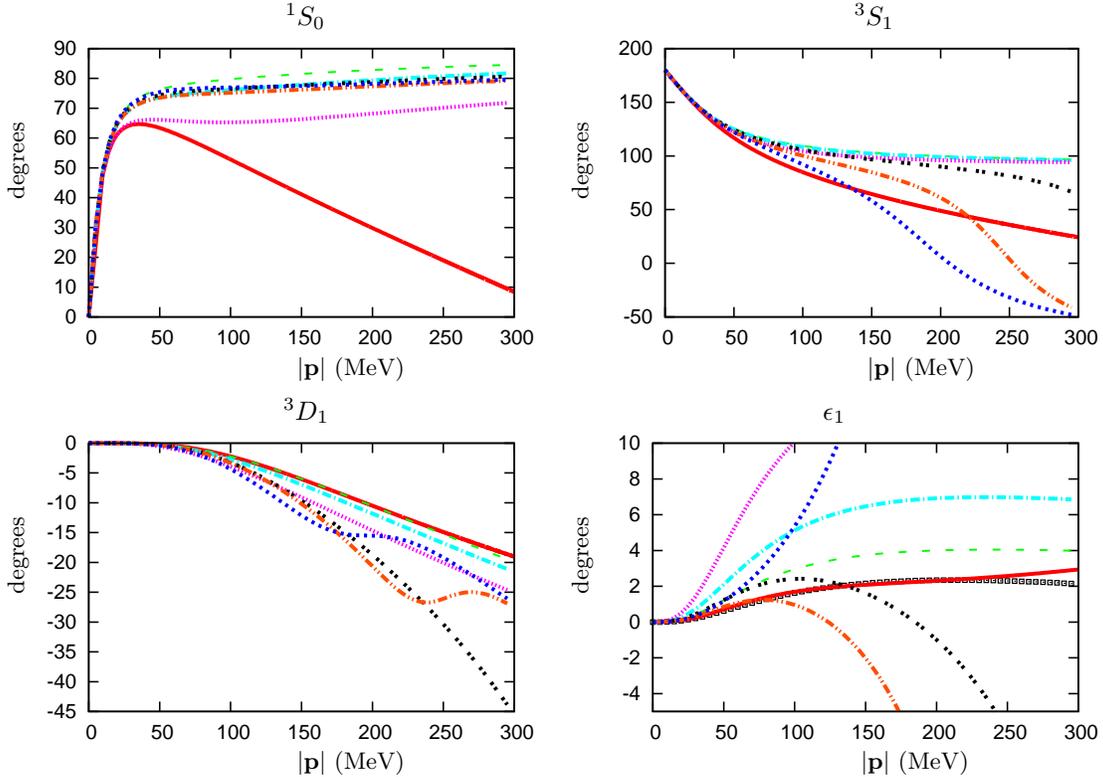,width=0.6\textwidth,angle=-90}}
\vspace{0.2cm}
\caption[pilf]{\protect \small (Color online.) 
$^1 S_0$, $^3S_1$, $^3D_1$ phase shifts and the mixing angle $\epsilon_1$ as a 
function of $|\vp|$. The (red) solid lines correspond to the Nijmegen data 
\cite{nijmegen,nijweb}. For the rest of the lines three values of $g_0=-1/4$,
$-1/3$ and $-1/2~m_\pi^2$ are employed. For LO these lines are the (green) dashed,  (cyan)
dot-dashed and (magenta) dotted lines, respectively. While to NLO these are the
(black) double-dotted, (orange) double-dot-dashed  and (blue) short-dashed
lines, in that order.  The squared points in the panel for $\epsilon_1$ corresponds to a NLO calculation with $g_0=-0.1$~$m_\pi^2$.
\label{fig:swave}}
\end{figure} 

\begin{figure}[ht]
\psfrag{deg}{{\small degrees}}
\psfrag{p}{{\small $|\vp|$ (MeV)}}
\psfrag{1P1}{$^1P_1$}
\psfrag{3P0}{$^3P_0$}
\psfrag{3P1}{$^3P_1$}
\psfrag{3P2}{$^3P_2$}
\psfrag{3F2}{$^3F_2$}
\psfrag{epsilon_2}{$\epsilon_2$}
\centerline{\epsfig{file=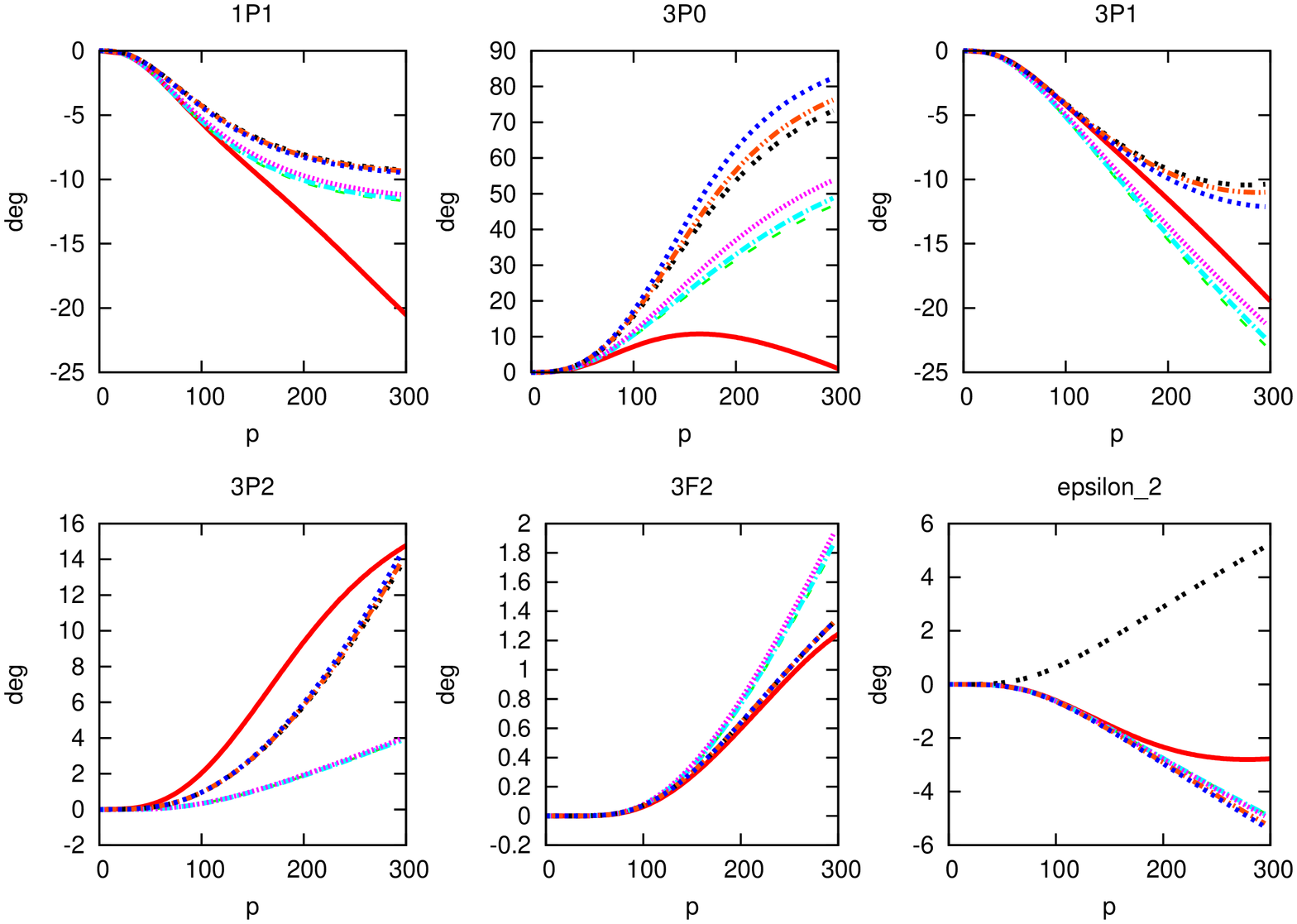,width=0.6\textwidth,angle=-90}}
\vspace{0.2cm}
\caption[pilf]{\protect \small (Color online.)
$^1 P_1$, $^3P_0$, $^3P_1$, $^3 P_2$, $^3 F_2$ phase shifts and the mixing
angle $\epsilon_2$ as a function of $|\vp|$. For notation, see
fig.~\ref{fig:swave}.
\label{fig:pwave}}
\end{figure} 

\begin{figure}[ht]
\psfrag{deg}{{\small degrees}}
\psfrag{p}{{\small $|\vp|$ (MeV)}}
\psfrag{1D2}{$^1D_2$}
\psfrag{3D2}{$^3D_2$}
\psfrag{3D3}{$^3D_3$}
\psfrag{3G3}{$^3G_3$}
\psfrag{epsilon_3}{$\epsilon_3$}
\centerline{\epsfig{file=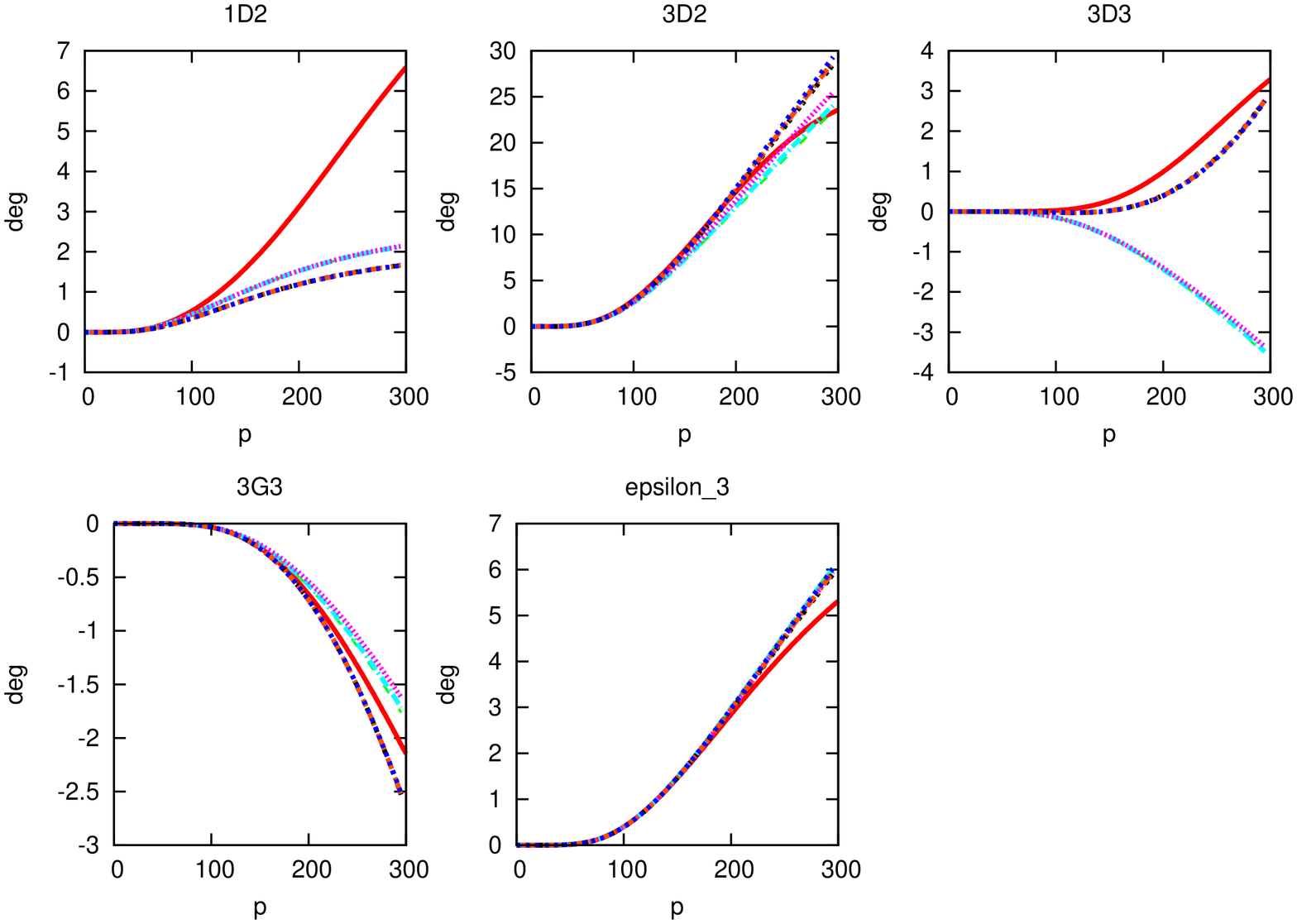,width=0.6\textwidth,angle=-90}}
\vspace{0.2cm}
\caption[pilf]{\protect \small
 (Color online.) $^1 D_2$, $^3D_2$, $^3D_3$, $^3 G_3$ phase shifts and the
mixing angle
$\epsilon_3$ as a function of $|\vp|$. For notation, see
fig.~\ref{fig:swave}.
\label{fig:dwave}}
\end{figure} 

In figs.~\ref{fig:swave}, \ref{fig:pwave} and
\ref{fig:dwave} we show the LO
and NLO results for the nucleon-nucleon scattering data (phase shifts and mixing angles)  up to $|\vp|=300~$MeV making use of eq.~(\ref{master}). Since $C_S$ at LO is close to 
$1/g_0$, as explained above, we show the results for the values  $g_0=-1/4~m_\pi^{2}$, $-1/3~m_\pi^{2}$ and $-1/2~m_\pi^{2}$ because its inverses are $-4~m_\pi^{-2}$, $-3~m_\pi^{-2}$ and $-2~m_\pi^{-2}$, respectively. In this way, the resulting $C_S$ at LO is of order $1~ m_\pi^{-2}$, a natural size. E.g. employing the estimation for $g_0\simeq -0.54~m_\pi^{-2}$, given below eq.~\eqref{cs.ct}, one would obtain $1/g_0\sim -1.8~m_\pi^{-2}$. On the other hand, let us recall that negative values for 
$g_0$, and not far from $-0.5~m_\pi^2$, are the required ones in order to optimize the perturbative solution of eq.~\eqref{dis.nji}. 
 For the LO  results the lines are the dashed, dot-dashed and dotted
lines, corresponding to $g_0=-1/4$, $-1/3$ and $-1/2~m_\pi^2$, respectively.
 While to NLO these are the double-dotted, double-dot-dashed and short-dashed lines, in the same order. 
For $|\vp|\simeq 360$~MeV the pion production threshold opens and it does not
make sense to compare with data above this point. An ${\cal O}(p^2)$ calculation, which  includes important new physical mechanisms, as non-reducible two-pion
exchanges between others, as indicated above before eq.~\eqref{eq.a2}, is
presumably needed to improve the agreement with data \cite{ordo,epe}. E.g., it is well known that for the $^1S_0$ partial wave an ${\cal O}(p^2)$ chiral counterterm, 
 in the standard chiral counting,\footnote{At ${\cal O}(p^0)$ in the KSW counting \cite{kaplan}.} is required in order to reproduce its relatively large effective range so that the agreement with data improves. This can be understood by considering the effective range expansion. For the $^1S_0$ partial wave $1/a_s$ is extremely small so that the contribution from the effective range $r_0 \vp^2/2$ rapidly overcomes $-1/a_s$ (the leading order contribution). Then, this problem is not so much related to the fact of having too large higher order corrections but more it arises because the leading order is anomalously small. The largest differences in absolute values between the LO and NLO results are
observed in the ${^3S_1}$-${^3D_1}$ and $^3P_0$ partial waves. These partial waves, as discussed in depth in ref.\cite{kswnnlo}, have large non-analytic corrections from two potential pion exchange.  For the $^3 P_1$, $^3 P_2$ and $^3D_3$ waves the difference in absolute terms is small, a few degrees, although relatively it can be large typically for $|\vp|\gtrsim 150~$MeV. For higher partial waves these differences are typically much smaller since the iteration of one-pion exchange becomes smaller  \cite{peripheral}. 
 Our ${\cal O}(p)$ results  are of comparable quality to those obtained at LO within the Weinberg's counting approach \cite{epe}. The $^3P_0$ phase shifts are also not well reproduced at this order in ref.~\cite{epe}. Both approaches share the same input for $\hbox{Im}T_{JI}$ along the left-hand cut, and at  ${\cal O}(p)$ we have already considered the iteration of one $g$ factor in determining $N_{JI}$, as discussed above. The main differences between our results and ref.~\cite{epe} at LO concern $\epsilon_1$ and the phase shifts for $^3 P_1$ and $^3D_3$.   For the latter our results are closer to experiment while for the two former observables the LO calculation of ref.~\cite{epe} is closer to data. 
 It is known that one-pion exchange has a too large tensor force which is reduced by higher order counterterms. In the meson exchange picture this
cancellation at short distances of the one-pion exchange tensor force is
produced by the exchange of  $\rho$-mesons \cite{brown}.   The mixing  ${^3S_1}$-${^3D_1}$
and the partial wave $^3P_0$ have large attractive matrix elements of the one-pion exchange tensor operator, as stressed in refs.~\cite{timm,majo}. These are the partial waves that depart more from data in absolute terms. The $^3P_0$ phase shifts were reproduced  accurately in ref.~\cite{timm} at LO for low energies. In this reference, a counterterm was promoted to LO in all the partial waves with attractive tensor interactions, and in particular to the $^3 P_0$ channel. 
Such free parameter is needed to fit the $^3 P_0$ data \cite{timm,epe}. 
 The results of ref.~\cite{timm} are then cut-off independent for high enough values of the employed cut-off. 
 
  As a result of the perturbative approach actually followed in this paper for determining $N_{JI}$ by solving eq.~\eqref{eq.nji} in an expansion in the number of two-nucleon reducible loops, a residual dependence on $g_0$ is left in the solution due to higher orders in this expansion (and not from the pure chiral one, eq.~\eqref{ffg}). As more orders are included  the exact solution of $T_{JI}$, obtained by solving  eq.~\eqref{dis.nji}, is better approached and any dependence on $g_0$ should tend to vanish.  From here one could also infer 
 that   contributions with one-pion exchange twice iterated in $N_{JI}$ are expected to be significant at least in those observables with a clear $g_0$ dependence in
figs.~\ref{fig:swave}-\ref{fig:dwave}.   It is also worth noticing that the dependence on $g_0$ in figs.~\ref{fig:swave}-\ref{fig:dwave} at LO/NLO is much smaller for the P- and higher partial waves than for the S-waves. 
 This should be expected because $N_{JI}$ for $\ell\geq 1$ vanishes at threshold as $|\vp|^{2\ell}$ so that both   $g$ and $N_{JI}$ are small in the low energy part of the left-hand cut. In this way the perturbative solution of eq.~\eqref{dis.nji} should typically converge faster for higher $\ell$. Conversely, the convergence of the S-waves should be slower, something that it is clear for the $^3S_1-^3D_1$ coupled channels from fig.~\ref{fig:swave}. Particularly noticeable is the dependence on $g_0$ of $\epsilon_1$, a fact that is in agreement with the results of Fleming, Mehen and Stewart \cite{kswnnlo}.  The squared points in the panel for $\epsilon_1$ in fig.~\ref{fig:swave} are obtained with $g_0=-0.1$~$m_\pi^2$. They agree closely with data \cite{nijmegen}, though such good agreement seems to be accidental.

\subsection{Nucleon-nucleon scattering in the nuclear medium}
\label{nn.medium}

When calculating a loop function in the nuclear medium we typically use the
notation $L_{ij}$, where $i$ indicates the number of two-nucleon states in the 
diagram (0 or 1) and $j$ the number of pion exchanges (0, 1 or 2). In
addition, we also use  $L_{ij,f}$, $L_{ij,m}$ and  $L_{ij,d}$, with the
subscripts $f$, $m$ and $d$ indicating zero, one or two Fermi-sea insertions 
from the nucleon propagators in the medium, respectively. In this way, 
the function $g=L_{10,f}$ and its in-medium counterpart is $L_{10}$, that is  
calculated in the Appendix \ref{sec:l10}.

The evaluation of the nucleon-nucleon scattering amplitudes in the nuclear
medium at lowest order can be easily obtained from our previous result in 
vacuum since the only modification without increasing the chiral order
corresponds to use the full in-medium nucleon propagators. This is  directly 
accomplished by replacing $g(A)$ by $L_{10}$ in eq.~\eqref{master}. At any order
for nucleon-nucleon scattering in
the nuclear medium, we use eq.~(\ref{master}) but now with the function
$g$ substituted by $L_{10}$ so that
 \be
T_{JI}^{i_3}(\ell,\bar{\ell},S)
=\left[I+ N^{i_3}_{JI}(\ell,\bar{\ell},S)\cdot L_{10}^{i_3}\right]^{-1}
\cdot N^{i_3}_{JI}(\ell,\bar{\ell},S)~.
\label{l10fa3}
\ee
The same process as previously discussed is followed to fix $N_{JI}$.  Note
that any other in-medium contribution requires $V_\rho=1$, which then
increases the order at least by one more unit, cf. eq.~(\ref{fff}).  This new 
in-medium generalized vertex must be associated with the nucleon-nucleon
scattering diagrams of leading order. The modification of the meson
propagators (both heavy and pionic ones) by the inclusion of an in-medium 
generalized vertex increases the chiral order by two units. However, the
modification of the enhanced nucleon propagators with one in-medium
generalized vertex only increases the order by one unit and these
contributions must be kept at NLO. It goes beyond the scope of this article 
to offer a complete study of the in-medium pion self-energy at N$^2$LO where
the full NLO in-medium nucleon-nucleon interactions are needed. What we do
here for illustration is to change the free nucleon propagators by the
in-medium ones in the calculation of the box diagram $L^{(1)}_{JI}$ that enter
in fixing $N_{JI}^{(1)}$, eq.~(\ref{eq.a1}), with $g$ also replaced by
$L_{10}$. 
In eq.~\eqref{l10fa3} we have included the superscript $i_3$, which
corresponds  to the third component of the total isospin of the two nucleons
involved in the scattering process, both in the partial wave
$T_{JI}^{i_3}(\ell,\bar{\ell},S)$ and in $L^{i_3}_{10}$, as the Fermi momentum
of the neutrons and protons are different for asymmetric nuclear matter.  
The function $L_{10}^{i_3}$  conserves total isospin $I$, because it is
symmetric under the exchange of the two nucleons, though it depends on the 
charge (or third component of the total isospin) of the intermediate state. 
This is a general rule, all the $i_3=0$ operators are symmetric under the 
exchange $p\leftrightarrow n$, so that they do not mix isospin representations 
with  different exchange symmetry properties.
\\~\\
In this section we have determined the vacuum nucleon-nucleon scattering at LO and NLO 
following the novel counting of eq.~\eqref{ffg} \cite{nlou}. For nuclear matter the LO 
nucleon-nucleon scattering amplitudes have been also obtained. The infrared enhancement of the two-nucleon reducible 
loops have made it necessary to resum the right-hand cut. This is  accomplished by 
a once-subtracted dispersion relation of the inverse of a partial wave giving rise to eq.~\eqref{master}.  
 The important function $g(A)$, eq.~\eqref{dis.rel.g}, which is defined in terms of a subtraction constant, $g(D)$ or $g_0$, is introduced. 
It has been argued that the subtraction constant is ${\cal O}(p^0)$, because by changing the 
subtraction point $B$ the subtraction constant  is modified  reshuffling  the form of the function $g(A)$, which is invariant.  The process for determining the interaction kernel $N_{JI}$, eq.~\eqref{master}, 
has been 
also discussed in detail. It was obtained that the subtraction point $D$  acts as a ``renormalization scale" where an experimental point is reproduced. The subtraction constant $g(D)$ just fixes the ``renormalization scheme" and the exact results should not depend on it. A natural value for $g_0\sim -m m_\pi/4\pi$ was argued to be adequate for obtaining  $N_{JI}$ as a perturbative solution of eq.~\eqref{dis.nji} in order to suppress the effects of the iterative factor $|1+g N_{JI}|^2$ in the equation. The couplings $C_S$ and $C_T$ from the local nucleon-nucleon Lagrangian, eq.~\eqref{lnn}, have been fixed in terms of $g_0$ at LO and NLO reproducing the S-wave nucleon-nucleon scattering lengths.  These couplings keep their estimated size 
of ${\cal O}(p^0)$ after the iteration, despite the well known fact that the nucleon-nucleon scattering lengths are much larger than $1/m_\pi$. The resulting phase shifts and mixing angles at LO and NLO are depicted in figs.~\ref{fig:swave}, \ref{fig:pwave} and \ref{fig:dwave}. It is argued that higher orders should be included in order to improve the reproduction of data. Particularly, a N$^2$LO analysis should be pursued since it would include the important two-pion irreducible exchange and new counterterms, in particular the one necessary to reproduce the effective range for the $^1S_0$ partial wave \cite{Epelbaum:2008ga}. This is left as a future task since our present main aim  is to work the results up to NLO and settle the formalism in detail.

\section{Contributions from the nucleon self-energy due to nuclear interactions}
\label{sec:sigma8}
\setcounter{equation}{0}
\label{sec:contri}

In this section we consider those diagrams in fig.~\ref{fig:all} that include the nucleon-nucleon contributions to the nucleon self-energy in the medium, diagrams 7 and 8. In turn, for each of these figures  the one on the top corresponds to the direct nucleon-nucleon interactions, while the 
exchange part gives rise to the diagram on the bottom (that includes the part of the diagrams 5 and 6 with all nucleon propagators corresponding to Fermi-sea insertions.) 

First, let us consider the evaluation of the diagrams~7 in fig.~\ref{fig:all},  denoted by $\Pi_7$. It is given by
\begin{align}
\Pi_7&=\frac{q^0}{2f^2}\varepsilon_{i j k}\int\frac{d^4k_1}{(2\pi)^4}
e^{ik_1^0\eta}\hbox{ Tr}\left\{
\tau^k G_0(k_1) \Sigma_{NN} G_0(k_1)  \right\}~,
\label{sel.nn}
\end{align}
where  
\be
\Sigma_{NN}=\frac{1+\tau_3}{2}\Sigma_{p,NN}+\frac{1-\tau_3}{2}\Sigma_{n,NN}~,
\label{sigmann}
\ee
with $\Sigma_{p,NN}$ and $\Sigma_{n,NN}$ the proton and neutron self-energies  due to 
the nucleon-nucleon interactions, in order. 
Performing the trace in isospin, 
\be
\Pi_7=\frac{q^0}{2f^2}\varepsilon_{i j 3}\sum_{\sigma_1}
 \int\frac{d^4k_1}{(2\pi)^4}e^{ik_1^0\eta}\left(
 G_0(k_1)_p^2  \Sigma_{p,NN}
 -G_0(k_1)_n^2  \Sigma_{n,NN}
 \right)~.
 \label{sigma.8}
\ee
Here $\sigma_1$ corresponds to the spin of the incident nucleon.
Taking into account the identity eq.~\eqref{pro.go2} we can integrate by parts
eq.~\eqref{sigma.8} with the result
\begin{align}
\Pi_7&=\frac{q^0}{2f^2}\varepsilon_{i j 3}\sum_{\sigma_1}
 \int\frac{d^4k_1}{(2\pi)^4}e^{ik_1^0\eta}\left(
 G_0(k_1)_p  \frac{\partial \Sigma_{p,NN}}{\partial k_1^0}
 -G_0(k_1)_n  \frac{\partial \Sigma_{n,NN}}{\partial k_1^0}
 \right)~.
 \label{sigma.7.2}
\end{align}

\begin{figure}[ht]
\psfrag{k}{$k$}
\psfrag{p}{$p$}
\psfrag{l}{$\ell$}
\psfrag{pi}{$\pi$}
\psfrag{r}{$k-\ell$}
\centerline{\epsfig{file=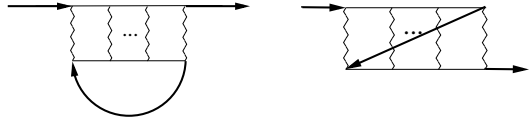,width=.4\textwidth,angle=0}}
\vspace{0.2cm}
\caption[pilf]{\protect \small
In-medium nucleon self-energy due to the nucleon-nucleon interactions with the
Fermi-seas.
\label{fig:nnself}}
\end{figure}
The nucleon 
self-energy due to the nucleon-nucleon interactions, represented in 
fig.~\ref{fig:nnself}, is given by the expression
 \be
\Sigma_{\alpha_1,NN}=-i\sum_{\alpha_2,\sigma_2}\int\frac{d^4
k_2}{(2\pi)^4}e^{ik_2^0\eta} G_0(k_2)_{\alpha_2}
T_{NN}(k_1\sigma_1\alpha_1,k_2\sigma_2\alpha_2|k_1\sigma_1\alpha_1,
k_2\sigma_2\alpha_2)
 \label{sel.n}
 \ee
where  $T_{NN}$ is the two-nucleon scattering operator between the nucleon states
characterized by the four-momentum $k_i$, spin $\sigma_i$ and third component
of isospin $\alpha_i$. We also use the variables 
\begin{align}
a=\frac{1}{2}(k_1+k_2)~,~p=\frac{1}{2}(k_1-k_2)~,
\label{change.variables}
\end{align} 
and 
\begin{align}
A&=2ma^0-\val^2~,
\label{def.a.1}
\end{align}
with $\val$ the three-momentum made up from $a^i$,~$i=1,2,3$. 
 We introduce the shorter notation
\begin{align}
T^{\sigma_1\sigma_2}_{\alpha_1\alpha_2}(\vp,\val;A)=T_{NN}(k_1
\sigma_1\alpha_1,k_2\sigma_2\alpha_2|k_1\sigma_1\alpha_1,k_2
\sigma_2\alpha_2)~,
\label{def.t.short}
\end{align}
that is more convenient for forward scattering than the notation followed in Appendix \ref{app.pwd}.
For on-shell scattering $A=\vp^2$. Eq.~\eqref{sigma.7.2}, after using  eq.~\eqref{sel.n},  becomes 
\begin{align}
\Pi_7=-i\frac{q_0}{2f^2}\varepsilon_{ij3}\sum_{\sigma_1,\sigma_2}\int\frac{d^4
k_1}{(2\pi)^4}\frac{d^4 k_2}{(2\pi)^4}e^{ik_1^0\eta}e^{ik_2^0\eta}&\Biggl(
G_0(k_1)_pG_0(k_2)_p\frac{\partial
T^{\sigma_1\sigma_2}_{pp}(\vp,\val;A)}{\partial k_1^0}\nn\\
&-
G_0(k_1)_nG_0(k_2)_n\frac{\partial
T^{\sigma_1\sigma_2}_{nn}(\vp,\val;A)}{\partial k_1^0}
\Biggr)~.
\label{sigma.77}
\end{align}

In order to obtain this result we have used that 
\begin{align}
\sum_{\sigma_1,\sigma_2}
 \int\frac{d^4k_1}{(2\pi)^4}\frac{d^4k_2}{(2\pi)^4} 
e^{ik_1^0\eta}e^{ik_2^0\eta} &\Biggl[G_0(k_1)_p G_0(k_2)_n
\frac{\partial}{\partial   k_1^0}T_{NN}(k_1 \sigma_1 p, k_2 \sigma_2 n|k_1
\sigma_1 p,
k_2\sigma_2 n)\nn\\
 &-G_0(k_1)_n  G_0(k_2)_p \frac{\partial}{\partial
   k_1^0}T_{NN}(k_1 \sigma_1  n, k_2 \sigma_2 p|k_1 \sigma_1 n, k_2\sigma_2
p)\Biggr]=0~,
   \label{sig.7.can}
\end{align}
which follows  for two reasons. First, let us notice that because of  Fermi-Dirac
statistics 
\begin{align}
T_{NN}(k_1 \sigma_1 p, k_2 \sigma_2 n|k_1 \sigma_1 p,k_2\sigma_2 n)=T_{NN}(k_2
\sigma_2 n, k_1 \sigma_1 p|k_2 \sigma_2 n,k_1\sigma_1 p)~.
\end{align}
Second, at LO the amplitude $T_{NN}$, as commented above, is given by the
iteration of the wiggly line in fig.~\ref{fig:sum}.  The latter does neither
depend on $k_1^0$ nor on $k_2^0$, see eqs.~\eqref{feynman} and \eqref{1pi.gen}.
Since  $L^{i_3}_{10}$  depends on $k_1^0$ and $k_2^0$ only through their sum,
$k_1^0+k_2^0$, then  $T_{NN}$ at LO  only depends on them in the same way and 
$\partial T_{NN}/ \partial k_1^0=\partial T_{NN}/\partial k_2^0$ holds. Taking
these two facts into account, as  $k_i$ and $\sigma_i$ are dummy variables,  
eq.~\eqref{sig.7.can}  is obtained.

It is convenient to give  the nucleon-nucleon scattering amplitude as an
expansion 
in partial waves, eq.~(\ref{pw.d2}).   The partial wave
decomposition of the nucleon-nucleon amplitudes is derived in detail in
Appendix~\ref{app.pwd}. A nucleon-nucleon partial wave is denoted by $T_{JI}^{i_3}(\ell',\ell,S)$, where 
$\vec{J}=\vec{\ell}+\vec{S}$ is the total angular momentum, $I$ is the total
isospin, $i_3=\alpha_1+\alpha_2$, $\ell'$ and   $\ell$ are the final and initial
orbital angular momenta, respectively, and $S$ is the total spin. The partial
wave 
 is a function of $\val^2$, $\vp^2$ and $A$ for our previously calculated
nucleon-nucleon amplitudes. Since for our present case,
eq.~\eqref{sigma.77}, $A\neq \vp^2$   an analytical extrapolation in $A$ of
$T_{JI}^{i_3}(\ell',\ell,S)$ is necessary. 
While eq.~(\ref{pw.d2}) is given in the CM of the two nucleons involved in the 
scattering process,  eqs.~(\ref{sel.nn}) and (\ref{sel.n}) are given in the
nuclear matter rest-frame. This implies that one must take into account the
boost from the
former frame to the latter in order to use eq.~(\ref{pw.d2}). However, as is
shown in Appendix~C of ref.~\cite{techrep}, the angle of the associated  
Wigner rotation is suppressed and it is ${\cal O}((p/m)^2$). Then, the leading and 
next-to-leading nucleon-nucleon
scattering amplitudes can be used as Lorentz invariants, similarly as for the meson-meson
ones, and  eq.~(\ref{pw.d2}) can be directly used  in eq.~(\ref{sel.n}). Let us 
recall that our calculation of the pion self-energy in nuclear matter is up to NLO, ${\cal O}(p^5)$, and these relativistic corrections are  of ${\cal O}(p^7)$. From eqs.~(\ref{sigma.8})
and (\ref{sel.n}) one has to sum over the spins $\sigma_1$ and $\sigma_2$. The
fact that both the initial and final nucleon-nucleon states are the same
implies a great simplification in the equations. First,
if we set $\si_1=\sigma'_1$ and $\si_2=\sigma'_2$ in eq.~\eqref{pw.d2} and sum,
  \be
 \sum_{\sigma_1, \sigma_2}(\sigma_1\sigma_2 s'_3|s_1 s_2 S')
(\sigma_1\sigma_2 s_3|s_1 s_2 S)=\delta_{s'_3s_3} \delta_{S'S}~.
  \ee
The sum over the third components of orbital angular momentum and $s_3$  
in the partial wave decomposition of eq.~(\ref{sel.n}) becomes
  \be
 \sum_{m',m,s_3}(m' s_3\mu|\ell' S J)(m s_3 \mu|\ell S J)
Y_{\ell'}^{m'}(\hat{\vp})
 {Y_{\ell}^{m}}(\hat{\vp})^*=\delta_{\ell'\ell}\frac{2J+1}{4\pi}~.
 \label{sum.ms} 
 \ee
Here we have made use of the symmetry properties of the Clebsch-Gordan 
coefficients and of the addition theorem for the spherical harmonics 
\cite{rosen},
  \begin{align}
 & (m's_3\mu|\ell' S J)
=(-1)^{s_3+S}\left(\frac{2J+1}{2\ell'+1}\right)^{1/2}(-s_3\mu m'|SJ\ell')~,\nn\\
& \frac{1}{2\ell+1}\sum_m \left|Y_{\ell}^m(\hat{\vp})\right|^2 =\frac{1}{4\pi}~.
 \end{align}
Whence, the sum of partial waves that matters for
eq.~(\ref{sel.nn}) can be expressed as
\be
\sum_{\sigma_1,\sigma_2} T^{\sigma_1\sigma_2}_{\alpha_1\alpha_2}(\vp,\val;A)=
\sum_{I,J,\ell,S}(2J+1)T_{JI}^{i_3}(\ell,\ell,S)\chi(S\ell I)^2 
(\alpha_1\alpha_2 i_3|I_1I_2 I)^2~,
\label{sum.mat}
\ee
 with $\chi(S\ell1)$ defined in eq.~(\ref{chi}).
Inserting the previous equation in eq.~(\ref{sigma.77}) the following expression for 
$\Pi_7$ results 
\begin{align}
\Pi_7&=-i\frac{q^0}{2f^2}\varepsilon_{i j 3}\int\frac{d^4k_1}{(2\pi)^4}\int
\frac{d^4k_2}{(2\pi)^4}e^{ik_1^0\eta}e^{ik_2^0\eta}\sum_{J,\ell,S} (2J+1) 
\chi(S\ell 1)^2
\left\{G_0(k_1)_pG_0(k_2)_p
\frac{\partial T_{J1}^{+1}(\ell, \ell, S)}{\partial k_1^0} \right.\nn\\
&-\left.G_0(k_1)_nG_0(k_2)_n
\frac{\partial T_{J1}^{-1}(\ell, \ell, S)}{\partial k_1^0}
\right\}~.
\label{sig.8.f1}
\end{align}
 $\Pi_7$ is  an
S-wave isovector self-energy contribution. This should be expected and it is due 
to the presence of the WT vertex for the coupling of the in- and out-going pions 
with a nucleon, see diagram 7 of fig.~\ref{fig:all} and eq.~\eqref{sigma.7.2}.

We now consider the diagrams~8 in fig.~\ref{fig:all}, that involve the Born terms of pion-nucleon 
scattering. They are similar to the
diagrams~6, though  the nucleon self-energy is now due to the in-medium
nucleon-nucleon interactions. Making use of  eq.~\eqref{pro.go2} and then
integrating by parts, we have
\begin{align}
\Pi_8&=\frac{-g_A^2}{2f^2}\frac{\vq^2}{q^0}\varepsilon_{i j 3}
\sum_{\sigma_1}\int\frac{d^4k_1}{(2\pi)^4}e^{ik_1^0\eta}\left(
 \frac{\partial \Sigma_{p,NN}}{\partial k_1^0}G_0(k_1)_p -
\frac{\partial \Sigma_{n,NN}}{\partial k_1^0}G_0(k_1)_n
\right)\nn\\
&+\frac{ig_A^2}{2f^2}\frac{\vq^2}{{q_0}^2}\delta_{i j}
\sum_{\sigma_1}\int\frac{d^4k_1}{(2\pi)^4}e^{ik_1^0\eta}\Bigl(
\Sigma_{p,NN}G_0(k_1)_p+\Sigma_{n,NN} G_0(k_1)_n\Bigr)~,
\label{sig.9}
\end{align}
where the first term on the r.h.s. of the previous expression is
isovector and the last one is isoscalar. The former is referred to as
$\Pi_8^{iv}$ and the latter as $\Pi_8^{is}$. Taking into account
eq.~(\ref{sum.mat})   
one is left with
\begin{align}
\Pi_8&=\frac{i g_A^2}{2f^2}\frac{\vq^2}{q^0}\varepsilon_{i j
3}\int\frac{d^4k_1}{(2\pi)^4}\int
\frac{d^4k_2}{(2\pi)^4}e^{ik_1^0\eta}e^{ik_2^0\eta}\sum_{J,\ell,S}(2J+1)\chi(S\ell 1)^2 \left\{
G_0(k_1)_pG_0(k_2)_p\frac{\partial
T_{J1}^{+1}(\ell,\ell,S)}{\partial k_1^0}-G_0(k_1)_nG_0(k_2)_n\right.\nn\\
&\left.\times\frac{\partial
T_{J1}^{-1}(\ell,\ell,S)}{\partial k_1^0}
\right\}+\frac{g_A^2}{2f^2}\frac{\vq^2}{{q^0}^2} \delta_{i
j}\int\frac{d^4k_1}{(2\pi)^4}\int \frac{d^4k_2}{(2\pi)^4}e^{ik_1^0\eta}e^{ik_2^0\eta}
\sum_{J,\ell,S} (2J+1)
\Biggl(
\chi(S\ell 1)^2 G_0(k_1)_pG_0(k_2)_p
T_{J1}^{+1}(\ell,\ell,S)\nn\\
&+ \chi(S\ell 1)^2 G_0(k_1)_nG_0(k_2)_n T_{J1}^{-1}(\ell,\ell,S)
+G_0(k_1)_pG_0(k_2)_n \Bigl[
\chi(S\ell 0)^2 T_{J0}^0(\ell,\ell,S)+\chi(S\ell 1)^2 T_{J1}^0(\ell,\ell,S)
\Bigr]\Biggr)~.
\label{sig.9.f1}
\end{align}
Eqs.~(\ref{sig.8.f1}) and (\ref{sig.9.f1}) involve the knowledge of the
derivative of the nucleon-nucleon partial wave amplitude with respect to the
energy $k_1^0$. 
Instead of the variable $k_1^0$ we use the variable $A$, eq.~(\ref{def.a.1}), 
which is also the argument of $L_{10}$ and use the relation
 \begin{align}
\frac{\partial f(a^0)}{\partial k_1^0}=\frac{\partial f(a^0)}{\partial
k_2^0}=m\frac{\partial f(a^0)}{\partial A}~,
 \end{align}
 with $f(a^0)$ an arbitrary function that depends on $k_1^0$ and $k_2^0$ only through 
 their sum.  Let us now obtain an expression for the derivative of $\partial T_{JI}/\partial A$. For that,  rewrite eq.~(\ref{master}) as 
\begin{align}
T_{JI}=N_{JI}-N_{JI}\cdot L_{10}\cdot T_{JI}~.
\end{align}
Taking the derivative on both sides of the previous equation and isolating 
$\partial T_{JI}/\partial A$, 
\begin{align}
\frac{\partial T_{JI}}{\partial A}=D_{JI}^{-1}\cdot \frac{\partial
N_{JI}}{\partial A}
-D_{JI}^{-1}\cdot \frac{\partial N_{JI}}{\partial A}\cdot L_{10}\cdot
D_{JI}^{-1}\cdot N_{JI}-
D_{JI}^{-1}\cdot N_{JI}\cdot \frac{\partial L_{10}}{\partial A}\cdot
D_{JI}^{-1}\cdot N_{JI}~,
\label{der.1st}
\end{align}
with
\begin{align}
D_{JI}^{i_3}(\ell,\bar{\ell},S)=I+N^{i_3}_{JI}(\ell,\bar{\ell},S)\cdot
L^{i_3}_{10}~,
\label{dmat}
\end{align}
the same matrix whose inverse is multiplying $N_{JI}(\ell,\bar{\ell},S)$ in
eq.~(\ref{l10fa3}). 
Eq.~(\ref{der.1st}) can be simplified by taking into account that $D_{JI}$ and
$N_{JI}$ commute so that 
\begin{align}
\frac{\partial T_{JI}}{\partial A}=D_{JI}^{-1}\cdot\left[
\frac{\partial N_{JI}}{\partial A}-N_{JI}^2\frac{\partial L_{10}}{\partial A}
\right]\cdot D_{JI}^{-1}~.
\label{dt.gen}
\end{align}
At LO and NLO the previous expression reduces to
\begin{align}
\left.\frac{\partial T_{JI}}{\partial A}\right|_{LO}&=
{D^{(0)}_{JI}}^{-1}
\cdot \left[
-{N^{(0)}_{JI}}^2
\frac{\partial L_{10}}{\partial A} 
\right] \cdot {D^{(0)}_{JI}}^{-1}~,\nn\\
\left.\frac{\partial T_{JI}}{\partial A}\right|_{NLO}&={D^{(1)}_{JI}}^{-1}\cdot
\left[
\frac{\partial L^{(1)}_{JI}}{\partial
A}-\left\{N^{(1)}_{JI},N^{(0)}_{JI}\right\}
\frac{\partial L_{10}}{\partial A}\right]\cdot {D^{(1)}_{JI}}^{-1}~.
\label{dta.lo}
\end{align}
with 
\begin{align}
D^{(0)}_{JI}&=I+N^{(0)}_{JI}\cdot L_{10}~,\nn\\
D^{(1)}_{JI}&=I+(N^{(0)}_{JI}+N^{(1)}_{JI})\cdot
L_{10}=D^{(0)}_{JI}+N^{(1)}_{JI}\cdot L_{10}~.
\end{align} 
Further, the standard notation  
$\left\{B,C\right\}=B\cdot C+C\cdot B$ has been used in eq.~\eqref{dta.lo}. 

Eqs.~(\ref{sig.8.f1}) and (\ref{sig.9.f1}) represent the contributions from diagrams 7 and 8 of fig.~\ref{fig:all} to the pion self-energy in the nucleon medium. Their contributions are  denoted by 
$\Pi_7$ and $\Pi_8$, respectively. The former is purely isovector while the
latter contains both an isovector and an isoscalar part, proportional to
$\ve_{ij3}$ and $\delta_{ij}$, in that order.  $\Pi_7$ and $\Pi_8^{iv}$
are given by the same expression except by the global factor, proportional  to
$q^0$ for the former and to $-g_A^2 \vq^2/q^0$ for the latter. This is just
a consequence of the chiral expansion eq.~\eqref{exp.nuc.pro} in the Born
terms. On the other hand, $\Pi_8^{is}$ is a N$^2$LO contribution because it
originates from the derivative with respect to $k_1^0$ of the
nucleon-propagator between the two pion lines. This propagator is not enhanced
so that one order higher
 results as compared with the isovector part.

\section{Other nucleon-nucleon contributions and the cancellation of the
isovector terms}
\setcounter{equation}{0}
\label{sec:sig.10}

We now consider  the calculation of those contributions that originate  
from the diagrams~9 and 10 of fig.~\ref{fig:all}, where a  pion scatters
inside a two-nucleon reducible loop. 
They are denoted by $\Pi_9$ and $\Pi_{10}$, in order. 
As usual the diagram on the top corresponds to 
the direct part of the nucleon-nucleon scattering while that on the bottom represents 
the exchange part. The loop with the pions  has to be corrected by initial 
(ISI) and final (FSI) state interactions, as denoted in the figure by the
ellipsis which represent iterated nucleon-nucleon interactions. This iteration 
is the same as occurs for the nucleon-nucleon scattering in the nuclear
medium, see
fig.~\ref{fig:sum}. The ``elementary'' nucleon-nucleon interaction $N_{JI}$  
is dressed by the iterative process which gives rise to eq.~(\ref{master}), 
with $N_{JI}$ multiplied by the inverse of the matrix $D_{JI}$. In this way, 
if we denote by $\xi_{JI}(\ell,\bar{\ell},S)$ the elementary partial wave for  
a generic ``production'' process, $F_{JI}(\ell,\bar{\ell},S)$, then the FSI
dress it so that   
\be
F_{JI}(\ell,\bar{\ell},S)=D_{JI}^{-1}(\ell,\bar{\ell},S)\cdot
\xi_{JI}(\ell,\bar{\ell},S)~.
\label{d1.mat}
\ee
The matrix $D_{JI}$, eq.~(\ref{dmat}), is already known from the study of the 
nucleon-nucleon interactions up to some order. On the other hand, $\xi_{JI}$
can be fixed following an analogous procedure to that used before for 
determining $N_{JI}$ in section~\ref{sec:fnn}. In this way, $\xi^{(n)}_{JI}$
is determined by expanding eq.~(\ref{d1.mat}) in powers of $L_{10}$ up to 
$(L_{10})^n$ and then comparing with a full CHPT calculation up to ${\cal
  O}(p^{m+n})$, with at most $n+1$ two-nucleon reducible diagrams. Note that
we have written $m+n$ and $n+1$ because for our present purposes the basic
process, made up by a two-nucleon reducible loop with the two pions attached 
to one nucleon propagator, starts at ${\cal O}(p^{-1})$, so that $m =-1$, and 
it implies already one two-nucleon reducible loop.  In addition, both ISI and FSI are involved 
in the diagrams~9 and 10 of fig.~\ref{fig:all}. 
Then, instead of eq.~(\ref{d1.mat}) we have
\begin{align}
H_{JI}(\ell,\bar{\ell},S)=\sum_{\ell',\ell''}D_{JI}^{-1}(\ell,\ell',S)\cdot
\xi_{JI}(\ell',\ell'',S)\cdot D_{JI}^{-1}(\ell'',\bar{\ell},S)~.
\label{dd.mat}
\end{align}
The LO result requires to employ $D^{(0)}_{JI}$ and to calculate the
two-nucleon reducible loop to which the two pions are attached by factorizing 
on-shell the nucleon-nucleon scattering amplitudes. We use the notation 
$D_{JI}^{(n);i_3}=I+N_{JI}^{(n);i_3}\cdot L_{10}^{i_3}$ with $n$ the chiral
order,
\begin{align}
\xi^{(0)}_{JI}&=-(N_{JI}^{(0)})^2 \cdot DL_{10}~,\nn\\
\left.H_{JI}\right|_{LO}&={D^{(0)}_{JI}}^{-1}\cdot \xi^{(0)}_{JI}\cdot
{D^{(0)}_{JI}}^{-1}~.
\label{eq.fix.xi0}
\end{align}
Explicit expressions for $DL_{10}$ are given below in eqs.~\eqref{fix.xi0.2} and
\eqref{fix.diss}. 
\begin{figure}[ht]
\psfrag{exact}{{\small exact}}
\psfrag{fact}{{\small fact}}
\psfrag{a}{a)}
\psfrag{b}{b)}
\psfrag{c}{c)}
\centerline{\epsfig{file=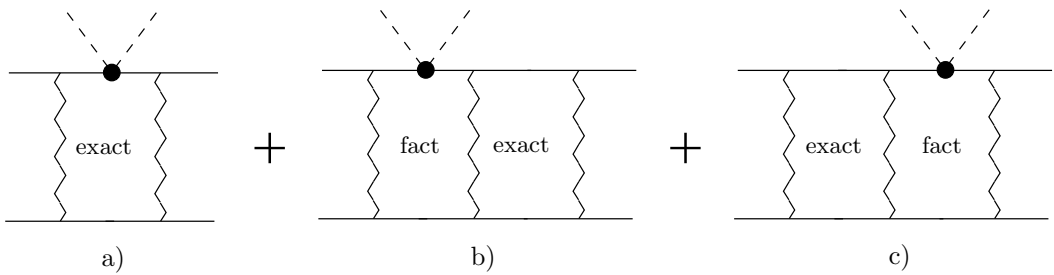,width=.8\textwidth,angle=0}}
\vspace{0.2cm}
\caption[pilf]{\protect \small
 Diagrams that contribute to the calculation of $\xi^{(1)}_{JI}$. Those
 two-nucleon  reducible loops that contain the label ``exact'' must be
 calculated exactly in the EFT, while those with the label ``fact'' must 
be calculated with the on-shell factorization  of the pertinent vertices. 
The filled circle in the figure indicates that the pion-nucleon scattering 
process contains both the WT and  Born terms.
\label{fig:prod}}
\end{figure}

\begin{figure}[ht]
\psfrag{k}{$k$}
\psfrag{p}{$p$}
\psfrag{l}{$\ell$}
\psfrag{pi}{$\pi$}
\psfrag{r}{$k-\ell$}
\psfrag{i}{$i$}
\psfrag{j}{$j$}
\centerline{\epsfig{file=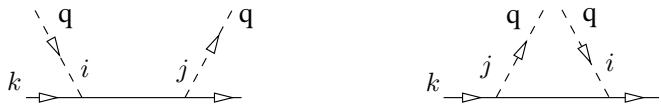,width=.5\textwidth,angle=0}}
\vspace{0.2cm}
\caption[pilf]{\protect \small
Born terms in pion-nucleon scattering. The vertices correspond to the lowest
order pion-nucleon vertex. 
\label{fig:effective}}
\end{figure}

At NLO one has an extra two-nucleon reducible loop. Expanding the
$D_{JI}^{-1}$ matrices in eq.~(\ref{dd.mat}) up to one $L_{10}$ and 
$\xi_{JI}$ up to ${\cal O}(p)$ we obtain
\begin{align}
\xi^{(0)}_{JI}+\xi^{(1)}_{JI}-2 N^{(0)}_{JI}\cdot L_{10}\cdot \xi^{(0)}_{JI}~.
\label{clearer0}
\end{align}
We now match the previous equation with the result of fig.~\ref{fig:prod}. 
In this  figure we have included inside each loop the labels
%
%
``exact'' or
``fact'' according to whether the loop is calculated exactly or by factorizing 
on-shell the nucleon-nucleon vertices. The filled circle refers to the
pion-nucleon scattering process that contains both the local and the Born
terms, fig.~\ref{fig:effective}. We denote by $L^{(1)}_{JI}$ the two-nucleon 
reducible loop without external pions calculated exactly in CHPT and that occurs
in 
figs.~\ref{fig:prod}b and \ref{fig:prod}c. There is also  the new contribution
of fig.~\ref{fig:prod}a  whose exact calculation is denoted by
$DL_{JI}^{(1)}$. The result is
\begin{align}
DL^{(1)}_{JI}- N_{JI}^{(0)}\cdot DL_{10}\cdot L_{JI}^{(1)}-L_{JI}^{(1)}\cdot
DL_{10}\cdot N_{JI}^{(0)}~.
\label{clearer1}
\end{align}
The equality of eqs.~\eqref{clearer0} and \eqref{clearer1}, taking into
account eq.~(\ref{eq.fix.xi0}) for $\xi_{JI}^{(0)}$, implies that
\begin{align}
\xi_{JI}^{(0)}+\xi_{JI}^{(1)}&=DL_{JI}^{(1)}-\left\{L_{JI}^{(1)}
+{N_{JI}^{(0)}}^2\cdot L_{10},\,N_{JI}^{(0)}\right\}\cdot DL_{10}~.
\label{fix.xi1}
\end{align}
In the last term we have the combination $L_{JI}^{(1)}+(N_{JI}^{(0)})^2\cdot
L_{10}$ which is ${\cal O}(p)$ in our counting because it corresponds to 
the difference between an exact calculation of a two-nucleon reducible loop
and that obtained by factorizing the vertices on-shell. The other contribution 
to $\xi_{JI}^{(1)}$ is given by $DL_{JI}^{(1)}-\xi_{JI}^{(0)}$, as follows
from eq.~(\ref{fix.xi1}), that is also ${\cal O}(p)$ by the same token. 
Finally, note that in the previous expression the two pions are attached to
the loops $DL_{JI}^{(1)}$ and $ DL_{10}$, while the remaining  terms
originate because of nucleon-nucleon scattering.

The nucleon
propagator before and after the filled circles in fig.~\ref{fig:prod} is  
the same so that it  appears squared. This is required as  the initial and final pion is also 
the same. We rewrite the nucleon propagator squared as 
\begin{align}
&\left[
\frac{\theta(\xi_\alpha-|\vp_1-\vk|)}{p_1^0-k_1^0-E(\vp_1-\vk)-i\epsilon}+
\frac{\theta(|\vp_1-\vk|-\xi_\alpha)}{p_1^0-k_1^0-E(\vp_1-\vk)+i\epsilon}
\right]^2\nn\\
&=-\frac{\partial}{\partial z}\left[
\frac{1}{p_1^0+z-k_1^0-E(\vp_1-\vk)+i\epsilon}
+i(2\pi)\delta(p_1^0+z-k_1^0-E(\vp_1-\vk))\theta(\xi_\alpha-|\vp_1-\vk|)\right]_
{z=0}~.
\label{pro.squ}
\end{align} 

The filled circles in fig.~\ref{fig:prod} consists of a WT pion-nucleon vertex  
and of the pion-nucleon scattering Born terms shown in
fig.~\ref{fig:effective}. 
Its sum is
\begin{align}
-\frac{iq^0}{2f^2}\ve_{ijk}\tau^k
-\left(\frac{g_A}{2f}\right)^2 \vq^2\left\{ 
\frac{\tau^j \tau^i}{{q}^0+p_1^0-k_1^0-E(\vp_1-\vk+\vq)+ i\epsilon}
+\frac{\tau^i \tau^j}{-{q}^0+p_1^0-k_1^0-E(\vp_1-\vk-\vq) + i\epsilon}
\right\}~.
\label{wt.mod}
\end{align}
We do not include the in-medium part of the nucleon propagator in the previous 
equation because for $q^0={\cal O}(m_\pi)$ the argument of the Dirac
delta-function in eq.~(\ref{nuc.pro}) is never satisfied as $m_\pi \gg {\cal
  O}$(nucleon kinetic energy). 
For the same reason, when performing the $k_1^0$-integration in the loop, 
the poles at $k_1^0=p_1^0\pm q^0-E(\vp_1-\vk\pm\vq)$, resulting from
eq.~(\ref{wt.mod}), are not considered because the nucleon propagators will 
not be any longer of ${\cal O}(p^{-2})$ but just of ${\cal O}(p^{-1})$
(standard counting). A contribution two orders higher would then result. 
Once the $k_1^0$-integration is done the latter acquires from
eq.~(\ref{pro.squ}) the value $z+p_1^0-E(\vp_1-\vk)$. The integration on 
$k_1^0$ for the evaluation of the two-nucleon reducible loop is analogous to 
the one performed in Appendix~\ref{sec:l10} for calculating the $L_{10}$
function. The point is that $L_{10}$ only depends on the energy of the
external legs through the variable $A=m(p_1^0+p_2^0)-\val^2$,
eq.~(\ref{def.a.1}), 
that in turn only depends on the total energy. As a result, when the
derivative with respect to $z$ acts on a baryon propagator not entering 
in eq.~(\ref{wt.mod}), one has
 \begin{align}
\frac{\partial L_{k\ell,r}^{ab\ldots}}{\partial z}
= \frac{\partial L_{k\ell,r}^{ab\ldots}}{\partial p_1^0}
=m\frac{\partial L_{k\ell,r}^{ab\ldots}}{\partial A}
=\frac{\partial L_{k\ell,r}^{ab\ldots}}{\partial p_2^0}~.
\label{da.l10}
\end{align}
Taking into account the chiral expansion of the nucleon propagators involved 
in eq.~(\ref{wt.mod}) for the pole terms and summing with the WT term, we have 
the leading contribution
\begin{align}
-\frac{iq^0}{2f^2}\left(1-g_A^2\frac{\vq^2}{{q_0}^2}\right)
\varepsilon_{ijk}\tau^k~.
\label{ver:wte}
\end{align}
The antisymmetric tensor in eq.~(\ref{ver:wte}) gives rise to 
the isospin factor $2i_3$ in the evaluation of the loops in fig.~\ref{fig:prod}.
Notice that in the loop the pions are attached to the propagators of the two
nucleons, the upper and the lower ones, and these 
two contributions sum symmetrically. We can then rewrite
eq.~(\ref{eq.fix.xi0}) for this case as
 \begin{align}
 \xi^{(0)}_{JI;iv}&=-i\frac{m q^0}{f^2}\left(1-g_A^2\frac{\vq^2}{q_0^2}\right)
i_3\ve_{ij3}\left[-(N_{JI}^{(0)})^2\frac{\partial L_{10}^{i_3}}{\partial
A}\right]~,\nn\\
DL_{10;iv}&=-i\frac{m
q^0}{f^2}\left(1-g_A^2\frac{\vq^2}{q_0^2}\right)i_3\ve_{ij3} \frac{\partial
L_{10}^{i_3}}{\partial A}~.
 \label{fix.xi0.2}
 \end{align}
In these equations we have included the subscript $iv$ given their isovector
character. In the same way for  $DL_{JI}^{(1)}$ one has 
 \begin{align}
DL_{JI;iv}^{(1)}=-i\frac{m q^0}{f^2}\left(1-g_A^2
\frac{\vq^2}{q_0^2}\right)i_3\ve_{ij3}\frac{\partial L^{(1);i_3}_{JI}}{\partial
A}~,
 \label{fix.dl2}
 \end{align}
 which corresponds to eq.~(\ref{fix.xi0.2}) but substituting the term between
 brackets, where the nucleon-nucleon vertices are on-shell, by its exact 
calculation. By applying eq.~\eqref{fix.xi1} we can fix $\xi_{JI;iv}^{(1)}$ in 
terms of eqs.~\eqref{fix.xi0.2} and \eqref{fix.dl2}.

 We now consider the case where the derivative with respect to $z$ from 
eq.~(\ref{pro.squ}) acts on the baryon propagators involved in the Born 
terms of eq.~(\ref{wt.mod}) with $k_1^0
=p_1^0+z-E(\vp_1-\vk)$. The term $E(\vp_1-\vk)-E(\vp_1-\vk\pm\vq)$ can be 
neglected when summed with $q^0$ for our calculation to NLO, so that the derivative with respect to $z$ 
of eq.~(\ref{wt.mod}) yields the isoscalar contribution
 \begin{align}
 -\frac{g_A^2}{2f^2}\frac{\vq^2}{q_0^2}\delta_{ij}~.
 \label{ver:iss}
 \end{align}
For any $i_3$ the isospin identity operator  gives rise to $+2$, instead of
the factor $2i_3$ of the isovector case. In this way, we can employ 
eqs.~(\ref{fix.xi0.2}) and \eqref{fix.dl2} substituting the vertex  
eq.~(\ref{ver:wte}) by eq.~(\ref{ver:iss}) and removing the action of the 
derivative $m\partial/\partial A$. Thus,
\begin{align}
\xi^{(0)}_{JI;is}&=-\frac{g_A^2\vq^2}{f^2q_0^2}\delta_{ij}\left[-(N_{JI}^{(0)}
)^2 L_{10}^{i_3}\right]~,\nn\\
DL_{10;is}&=-\frac{g_A^2 \vq^2}{f^2q_0^2} \delta_{ij}
L_{10}^{i_3}~,\nn\\
DL_{JI;is}^{(1)}&=-\frac{g_A^2\vq^2}{f^2q_0^2} \delta_{ij}
L^{(1);i_3}_{JI}~.
\label{fix.diss}
\end{align}
 Here the subscript $is$ is introduced given their isoscalar character. They give rise to $\Pi_{10}^{is}$. 
 Notice that both $DL_{10;is}$ and $DL_{JI;is}^{(1)}$ are one order higher than the analogous isovector terms in eqs.~\eqref{fix.xi0.2} and \eqref{fix.dl2}, respectively. This makes that  $\Pi_{10}^{is}$ starts to contribute at N$^2$LO. 
 
Next, we proceed to obtain the expressions for the pion self-energy
corresponding to the diagrams~9 and 10 of fig.~\ref{fig:all} as a sum over
partial waves. The leading contribution is obtained by using 
 $\xi_{JI;iv}^{(0)}$, eq.~(\ref{fix.xi0.2}), with the result
 \begin{align}
&\left.\left(\Pi_{9}+\Pi_{10}^{iv}\right)\right|_{NLO}=-i\frac{m q^0
\ve_{ij3}}{2f^2}\left(1-g_A^2\frac{\vq^2}{q_0^2}\right)\sum_{J,\ell,S}\chi(S
\ell 1)^2(2J+1)\int\frac{d^4k_1}{(2\pi)^4}\int\frac{d^4k_2}{(2\pi)^4} 
e^{ik_1^0\eta}e^{ik_2^0\eta}\Bigl(G_0(k_1)_p G_0(k_2)_p \nn\\
&\times 
\frac{\partial L_{10}^{+1}}{\partial A}[D^{(0);+1}_{J1}]^{-1}
\cdot{N_{J1}^{(0)}}^2\cdot [D^{(0);+1}_{J1}]^{-1}-G_0(k_1)_n G_0(k_2)_n
\frac{\partial L_{10}^{-1}}{\partial
A}[D^{(0);-1}_{J1}]^{-1}\cdot{N_{J1}^{(0)}}^2\cdot 
[D^{(0);-1}_{J1}]^{-1}
\Bigr)~,
\label{sig.10.isv.lo}
\end{align}
where the part corresponding to $\Pi_{10}^{iv}$ is the one proportional 
to $g_A^2$ in the previous equation. A symmetry factor 1/2 is included given the symmetry under the exchange of the two nucleonic external lines when they are finally closed.

In ref.~\cite{nlou} we established that at ${\cal O}(p^5)$ all the
contributions  to the pion self-energy involving nucleon-nucleon interactions 
 vanish ($V_\rho=2$). This implies that the contributions $\Pi_7$,
$\Pi_8^{iv}$, $\Pi_9$ and $\Pi_{10}^{iv}$  must cancel mutually at
this order. Recall that the isoscalar ones, $\Pi_8^{is}$ and $\Pi_{10}^{is}$, are ${\cal O}(p^6)$. The argument given in ref.~\cite{nlou} was a general one, without
assuming any specific procedure for resumming the  nucleon-nucleon
interactions. We now show that UCHPT fulfills this requirement. 
 When substituting  into 
 eqs.~\eqref{sig.8.f1} and (\ref{sig.9.f1})  the derivative $\partial T_{JI}/\partial
A$ eq.~(\ref{dta.lo}) at the lowest order, the following result is obtained
 \begin{align}
&\left.\left(\Pi_7+\Pi_8^{iv}\right)\right|_{NLO}=\frac{im
q^0\ve_{ij3}}{2f^2}\left(1-g_A^2\frac{\vq^2}{q_0^2}\right)
\sum_{J,\ell,S}\chi(S \ell 1)^2(2J+1)
\int\frac{d^4k_1}{(2\pi)^4}\int\frac{d^4k_2}{(2\pi)^4}
e^{ik_1^0\eta}e^{ik_2^0\eta}\Biggl( 
G_0(k_1)_pG_0(k_2)_p \nn\\
& \times [D_{J1}^{(0);+1}]^{-1}\cdot
{N_{J1}^{(0)}}^2\cdot
\frac{\partial L_{10}^{+1}}{\partial A}\cdot[D_{J1}^{(0);+1}]^{-1}
-G_0(k_1)_nG_0(k_2)_n[D_{J1}^{(0);-1}]^{-1}\cdot {N_{J1}^{(0)}}^2\cdot
\frac{\partial L_{10}^{-1}}{\partial A}\cdot[D_{J1}^{(0);-1}]^{-1}
\Biggr)~.
 \end{align}
This equation is the same as eq.~(\ref{sig.10.isv.lo}) but with opposite sign 
so that the cancellation  with $\Pi_{9}+\Pi_{10}^{iv}$ takes place. 

In the following of this section we work out several N$^2$LO contributions that comprise the isoscalar term $\Pi_{10}^{is}$ as well those that are obtained by employing 
 $\partial T/\partial A$ to NLO and $DL_{JI;iv}^{(1)}$ in   $\Pi_7+\Pi_8^{iv}$ and $\Pi_9+\Pi_{10}^{iv}$, respectively. 
 For the leading isoscalar contribution from
$\Pi_{10}$, which is already ${\cal O}(p^6)$ due to the same reason as for $\Pi_8^{is}$,  we obtain
\begin{align}
&\left.\Pi_{10}^{is}\right|_{N^2LO}=-\delta_{ij}\frac{g_A^2\vq^2}{2f^2 q_0^2}
\sum_{J,\ell,S}(2J+1)\int\frac{d^4k_1}{(2\pi)^4}\int\frac{d^4k_2}{(2\pi)^4}
e^{ik_1^0\eta}e^{ik_2^0\eta}\Biggl( 
\chi(S \ell 1)^2\Bigl\{G_0(k_1)_p G_0(k_2)_p
 L_{10}^{+1}[D^{(0);+1}_{J1}]^{-1}\cdot{N_{J1}^{(0)}}^2\nn\\
 &\cdot
[D^{(0);+1}_{J1}]^{-1}
+G_0(k_1)_n G_0(k_2)_n
L_{10}^{-1}[D^{(0);-1}_{J1}]^{-1}\cdot{N_{J1}^{(0)}}^2\cdot
[D^{(0);-1}_{J1}]^{-1}\Bigr\}+G_0(k_1)_pG_0(k_2)_nL_{10}^{0}\Bigl\{
\chi(S \ell 1)^2[D^{(0);0}_{J1}]^{-1}\nn\\
&\cdot{N_{J1}^{(0)}}^2\cdot[D^{(0);0}_{J1}]^{-1}
+\chi(S \ell
0)^2[D^{(0);0}_{J0}]^{-1}\cdot{N_{J0}^{(0)}}^2\cdot[D^{(0);0}_{J0}]^{-1}
\Bigr\}\Biggr)~.
\end{align}
 
 Including one more order in the calculation of  $\Pi_9$,
 $\Pi_{10}^{iv}$ and $\Pi_{10}^{is}$ requires the use of $DL_{JI}^{(1)}$,
 eq.~(\ref{fix.xi1}).  The input functions $DL_{JI}^{(1)}$ are given in 
eqs.~\eqref{fix.dl2} and \eqref{fix.diss} for the isovector and isoscalar
cases, respectively. In this way,
 \begin{align}
&\left.\left(\Pi_{9}+\Pi_{10}^{iv}\right)\right|_{N^2LO}=
i\frac{m q^0 \ve_{ij3}}{2f^2}\left(1-g_A^2\frac{\vq^2}{q_0^2}\right)
\sum_{J,\ell,S}\chi(S \ell 1)^2 (2J+1)
\int\frac{d^4k_1}{(2\pi)^4}\int\frac{d^4k_2}{(2\pi)^4} e^{ik_1^0\eta}e^{ik_2^0\eta}
\Biggl( G_0(k_1)_pG_0(k_2)_p
\nn\\
&\times [D^{(1);+1}_{J1}]^{-1}\cdot\left(
\frac{\partial L_{J1}^{(1);+1}}{\partial A}
-\left\{L_{J1}^{(1);+1}+{N_{J1}^{(0)}}^2\cdot L_{10}^{+1},\,
N_{J1}^{(0)}\right\}\cdot
\frac{\partial L_{10}^{+1}}{\partial A}\right)\cdot [D^{(1);+1}_{J1}]^{-1}
-G_0(k_1)_nG_0(k_2)_n[D^{(1);-1}_{J1}]^{-1}\nn\\
&\cdot \left(
\frac{\partial L_{J1}^{(1);-1}}{\partial A}
-\left\{L_{J1}^{(1);-1}+{N_{J1}^{(0)}}^2\cdot L_{10}^{-1},\,
N_{J1}^{(0)}\right\}\cdot
\frac{\partial L_{10}^{-1}}{\partial A}
\right)\cdot [D^{(1);-1}_{J1}]^{-1} \Biggl) ~.
\label{pi910.iv.nnlo}
\end{align}
 It is straightforward to  show that $\Pi_7+\Pi_8^{iv}$ calculated with $\partial T_{JI}/\partial A$ evaluated at NLO with eq.~(\ref{dta.lo}),  $\left.\Pi_7+\Pi_8^{iv}\right|_{N^2LO}$, cancels with   $\left.\left(\Pi_{9}+\Pi_{10}^{iv}\right)\right|_{N^2LO}$, eq.~\eqref{pi910.iv.nnlo}. One has to replace  $N_{JI}^{(1)}$ by its explicit expression in terms of $L_{JI}^{(1)}$, $N_{JI}^{(1)}=L_{JI}^{(1)}+N_{JI}^{(0)}\cdot L_{10}\cdot N_{JI}^{(0)}$.
Let us mention that there is an extra term for
$\Pi_{9}+\Pi_{10}^{iv}$  at ${\cal O}(p^6)$. It   
 stems from a one more term to eq.~(\ref{ver:wte}) in  the chiral
expansion 
of the baryon propagators eq.~(\ref{wt.mod}). This contribution is
suppressed by the inverse of the large nucleon mass and will be considered when 
a full N$^2$LO calculation of the pion self-energy in the nuclear medium is
given. In Appendix \ref{app:explicit}  we evaluate explicitly $DL_{JI}^{(1)}$ and
$L_{JI}^{(1)}$ needed for determining  $N_{JI}^{(1)}$ and $\xi_{JI}^{(1)}$,
eqs.~\eqref{eq.a1} and \eqref{fix.xi1}, in order. Some
steps introduced in the derivations of this and the previous section are also
calculated explicitly.
\\~\\
In summary, we have considered the calculation   of the diagrams 9 and 10 
of fig.~\ref{fig:all}, $\Pi_9$ and $\Pi_{10}$, in order. We have shown explicitly that up to and 
including NLO these contributions vanish with those evaluated in the previous section, $\Pi_7$ and $\Pi_8$. Indeed, the cancellation obtained between these contributions in ref.~\cite{nlou} is more general because it was shown that once the Born terms in $\Pi_8$ and $\Pi_{10}$ are reduced to their leading contribution and summed to WT, as in eq.~\eqref{ver:wte}, the cancellation occurs.  Some other contributions at N$^2$LO have been also calculated, though they do not exhaust a full calculation to this order of the in-medium pion self-energy. It is then shown that at N$^2$LO the just mentioned cancellation between the isovector contributions takes place as required by the findings of ref.~\cite{nlou}. Note that at this order the full calculation of the two-nucleon reducible loops takes place. This clearly shows that the cancellation is beyond the factorization approximation valid at NLO, as should be according to ref.~\cite{nlou}. 


\section{Nuclear matter energy density}
\label{sec:energy}
\def\theequation{\arabic{section}.\arabic{equation}}
\setcounter{equation}{0}
\label{sec:ener}

\begin{figure}[t]
\psfrag{Vr=1}{{\small $V_\rho=1$}}
\psfrag{Vr=2}{{\small $V_\rho=2$}}
\psfrag{Op6}{{\small ${\cal O}(p^6)$}}
\psfrag{Op5}{{\small $ {\cal O}(p^5)$}}
\psfrag{i}{{\small $i$}}
\psfrag{j}{{\small $j$}}
\psfrag{q}{{\small $q$}}
\psfrag{Leading Order}{Leading Order}
\psfrag{Next-to-Leading Order}{Next-to-Leading Order}
\psfrag{eq}{$\equiv$}
\centerline{\fbox{\epsfig{file=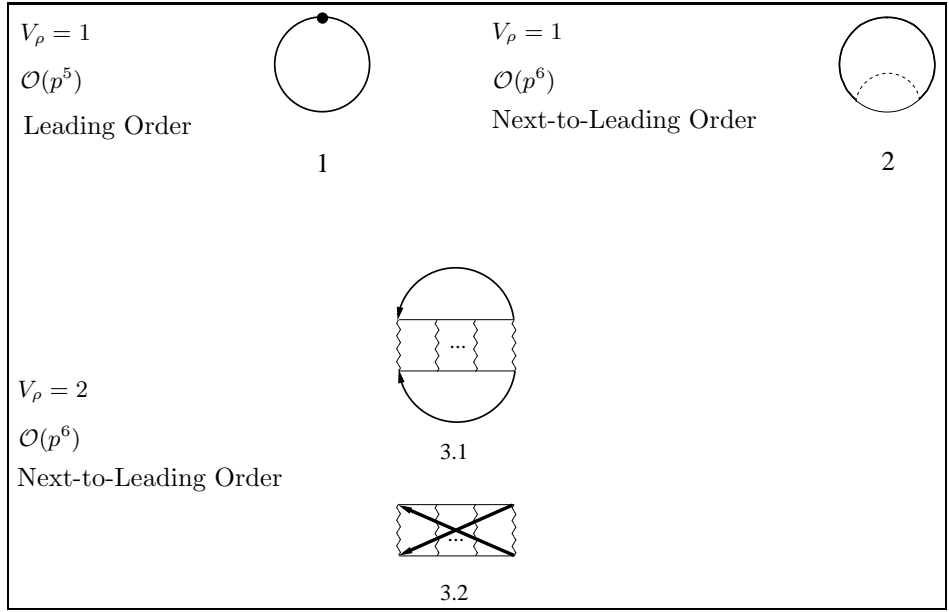,width=.7\textwidth,angle=0}}}
\vspace{0.2cm}
\caption[pilf]{\protect \small
Contributions to the nuclear matter energy up 
to  NLO or ${\cal O}(p^6)$. The lines have the same meaning as in fig.~\ref{fig:all}.
\label{fig:allE}}
\end{figure} 

We study now the problem of the nuclear matter equation of state by applying eq.~\eqref{ffg} up-to-and-including NLO. The diagrams required are shown in fig.~\ref{fig:allE}. The first diagram
 corresponds to the energy of a free Fermi-sea. Its
contribution,  ${\cal E}_1$, is given by
\begin{align}
{\cal E}_1&=\frac{3}{10 m}(\rho_p\,\xi_p^2+\rho_n\,\xi_n^2)~,
\label{cont.e1}
\end{align}
and is the only ${\cal O}(p^5)$ contribution. Nonetheless, since this is a
recoil correction originating from the first term of ${\cal L}_{\pi N}^{(2)}$,
eq.~\eqref{lags},  one expects it to be suppressed numerically (it
involves the inverse power of the hard scale 
$m$ instead of $\sim \sqrt{m m_\pi}$, the one in $1/g_0$.)  In addition, as shown below, there is further a dimensional suppression.  Hence, the NLO contributions involving the nucleon-nucleon interactions, suppressed by just one extra chiral order, could be of comparable size. This is of course an important remark for the possible saturation of nuclear matter.

The NLO or ${\cal O}(p^6)$ contributions  comprise the second and third
diagrams. The former is the
energy due to  the one-pion loop nucleon self-energy whose expression, including
a symmetry factor 1/2, is 
\begin{align}
i\int\frac{d^4k}{(2\pi)^4}
\sum_{i_3}G_0(k)_{i_3}\Sigma_{i_3}^\pi e^{i k^0\eta}&=
i\int\frac{d^4k}{(2\pi)^4}\frac{e^{i k^0\eta}}{k^0-E(\vk)+i\epsilon}
\left(\Sigma_{p}^\pi+\Sigma_{n}^\pi\right)\nn\\
&-\int\frac{d^3k}{(2\pi)^3}\Bigl[\theta(\xi_p-|\vk|)\Sigma_p^\pi
+\theta(\xi_n-|\vk|)\Sigma_n^\pi\Bigr]~.
\label{e.sigmapi}
\end{align}
The cut in $k^0$ from the free one-pion loop nucleon self-energy is restricted
to the lower half-plane of the $k^0$-complex plane, see eq.~\eqref{sigma.free}.
In this way,  by
closing  the first integral on the r.h.s. of the previous
equation along an infinite semicircle centered at the origin and on the upper
$k^0$-complex plane, the  contribution from the free part of $\Sigma_{i_3}^\pi$
cancels. Concerning the last term, an analogous reasoning to that given
previously in connection with fig.~\ref{fig:equiv} can be also applied here  for
the in-medium contribution $\Sigma_{i_3,m}^\pi$ of   $\Sigma_{i_3}^\pi$. This
part is accounted for by the diagram 3.2 of fig.~\ref{fig:allE}. Thus, the
contribution that we keep now for the
nuclear matter  energy density  from the second diagram of
fig.~\ref{fig:allE}, ${\cal E}_2$, is
\begin{align}
{\cal E}_2&=i\int\frac{d^4
k}{(2\pi)^4}\frac{e^{ik^0\eta}}{k^0-E(\vk)+i\epsilon}\bigl(\Sigma_{p,m}
^\pi+\Sigma_{n,m}^\pi\bigr)-\int\frac{d^3k}{(2\pi)^3}\bigl[\theta(\xi_p-|\vk|)+
\theta(\xi_n-|\vk|)\bigr]\Sigma_f^\pi~.
\label{eq.e2.1}
\end{align}
Let us show that up to ${\cal O}(p^6)$ these two integrals  give the same
result. Taking into account the expression for the
in-medium part of the one-pion loop nucleon self-energy,
eq.~\eqref{1.pi.l.n.s.e}, the first term on the r.h.s. of eq.~\eqref{eq.e2.1} 
yields the integrals
\begin{align}
&\int\frac{d^4
q}{(2\pi)^4}\frac{\vq^2}{q^2-m_\pi^2+i\ep}\int\frac{d^4k}{(2\pi)^4}\frac{1}{
k^0-E(\vk)+i\ep}(2\pi)\delta(k^0-q^0-E(\vk-\vq))\theta(\xi_{i_3}-|\vk-\vq|)~,
\label{eq.e2.2}
\end{align}
where the order of the integrations have been changed.  We perform the shift
$k\to k+q$ in the last integral, that is finite. In this way,
eq.~\eqref{eq.e2.2} can be rewritten as 
\begin{align}
i\int\frac{d^3k}{(2\pi)^3}\theta(\xi_{i_3}-|\vk|)(-i)\int\frac{d^4
q}{(2\pi)^4}\frac{\vq^2}{q^2-m_\pi^2+i\ep}\frac{1}{-q^0+E(\vk)-E(\vk-\vq)+i\ep}
~.
\end{align}
If the higher order corrections from the difference $E(\vk)-E(\vk-\vq)$  are  
neglected in the previous equation, the last integral is the one defining
$\Sigma_f^\pi$, compare with eq.~\eqref{sigma.free}. Then, after summing over
$i_3$, we have the same contribution as the last one in eq.~\eqref{eq.e2.1} 
up-to-and-including NLO. So finally we can write
\begin{align}
{\cal E}_2&=-2\int\frac{d^3
k}{(2\pi)^3}\Biggr(\theta(\xi_p-|\vk|)+\theta(\xi_n-|\vk|)
\Biggl)\Sigma_f^\pi~.
\label{eq.e2.3}
\end{align}

Let us evaluate now the diagrams 3.1 and 3.2 of fig.~\ref{fig:allE}, that
collectively are denoted as diagrams 3 in the following. These diagrams fully
involve the nucleon-nucleon scattering. Their contribution to the energy density, ${\cal E}_3$, is 
\begin{align}
{\cal
E}_3&=\frac{1}{2}\sum_{\sigma_1,\sigma_2}\sum_{\alpha_1,\alpha_2}\int\frac{d^4
k_1}{(2\pi)^4}\frac{d^4k_2}{(2\pi)^4}e^{ik_1^0\eta}e^{ik_2^0\eta}G_0(k_1)_{
\alpha_1}G_0(k_2)_{\alpha_2}T_{NN}(k_1
\sigma_1\alpha_1,k_2\sigma_2\alpha_2|k_1\sigma_1\alpha_1,k_2
\sigma_2\alpha_2)~.
\label{e3.1}
\end{align}
In this expression we have explicitly shown the symmetry factor 1/2 and the sum
over the spin ($\sigma_i$) and isospin ($\alpha_i$) labels, with $\eta\to 0^+$. 
 
The LO in-medium nucleon-nucleon scattering amplitude calculated in section
\ref{nn.medium} does not depend on $p^0$. The interaction kernel $N_{JI}$ only
depends on $\vp^2$ at this order, while the resummation over the two-nucleon
intermediate states, that gives rise to $L_{10}$, Appendix \ref{sec:l10},
depends on $A$ and   $|\val|$ as well. 
In the following we use as integration variables $a$ and $p$ introduced in 
 eq.~\eqref{change.variables}. The 
$p^0$-integration from eq.~\eqref{e3.1} can be readily performed, with the
result
\begin{align}
\int\frac{dp^0}{2\pi}G_0(a+p)_{\alpha_1}G_0(a-p)_{\alpha_2}=-i\Biggr[
\frac{\theta(|\val+\vp|-\xi_{\alpha_1})\theta(|\val-\vp|-\xi_{
\alpha_2})}{
2 a^0-E(\val+\vp)-E(\val-\vp)+i\ep}
-
\frac{\theta(\xi_{\alpha_1}-|\val+\vp|)\theta(\xi_{\alpha_2}-|\val-\vp|)}{
2 a^0-E(\val+\vp)-E(\val-\vp)-i\ep}
\Biggl]~.
\label{int.p0}
\end{align}
Here, we have made use of the fact that only those terms with poles located in
opposite halves of the $p^0$-complex 
plane survive after the $p^0$ integration.   Those terms with the two poles
located at the same half of the $p^0$-complex
 plane vanish, as it is clear by closing the integration contour with a
semicircle at infinite along 
the other half.  We insert eq.~\eqref{int.p0} into eq.~\eqref{e3.1}, 
and use the variable $A$, eq.~\eqref{def.a.1}, instead of $a^0$. Thus,  
\begin{align}
{\cal
E}_3&=-4i\sum_{\sigma_1,\sigma_2}\sum_{\alpha_1,\alpha_2}\int\frac{d^3a}{
(2\pi)^3}
\frac{d^3
p}{(2\pi)^3}\frac{dA}{2\pi}e^{iA\eta}\,T^{\sigma_1\sigma_2}_{\alpha_1\alpha_2}
(\vp,\val;A)\Biggr[
\frac{\theta(|\val+\vp|-\xi_{\alpha_1})\theta(|\val-\vp|-\xi_{
\alpha_2})}{A-\vp^2+i\ep} & \nn \\
  &- \frac{\theta(\xi_{\alpha_1}-|\val+\vp|)\theta(\xi_{\alpha_2}-|\val-\vp|)}{A-\vp^2-i\ep}
\Biggl] ~,
\label{e3.4}
\end{align}
where we made used that $k_1^0+k_2^0=(A+\val^2)/2m$. The resulting redefinition of $\eta$ is not indicated as it is not relevant for the
following manipulations, as well as the factor $\exp(i\eta \val^2)$ that is not shown. The first term between the square brackets on the r.h.s. of eq.~\eqref{e3.4}
corresponds to the
particle-particle part while the last one corresponds to the hole-hole part. 
Both of them originate by closing the nucleon lines in the diagrams 3 of
fig.~\ref{fig:allE}. 
Making use of
\begin{align}
&\theta(|\val+\vp|-\xi_{\alpha_1})\theta(|\val-\vp|-\xi_{
\alpha_2})=
\Bigr(1-\theta(\xi_{\alpha_1}-|\val+\vp|)\Bigl)\Bigr(1-\theta(\xi_{
\alpha_2}-|\val-\vp|)\Bigl)
\nn\\
&=1-\theta(\xi_{\alpha_1}-|\val+\vp|)-\theta(\xi_{
\alpha_2}-|\val-\vp|)
+\theta(\xi_{\alpha_1}-|\val+\vp|)\theta(\xi_{\alpha_2}-|\val
-\vp|)~,
\end{align}
eq.~\eqref{e3.4} becomes
\begin{align}
{\cal
E}_3&=-4i\sum_{\sigma_1,\sigma_2}\sum_{\alpha_1,\alpha_2}\int\frac{d^3a}{
(2\pi)^3}
\frac{d^3
p}{(2\pi)^3}\frac{dA}{2\pi}e^{iA\eta}\,T^{\sigma_1\sigma_2}_{\alpha_1\alpha_2}
(\vp,\val;A)
\Biggr[ \frac{1}{A-\vp^2+i\ep} \nn\\
&-\frac{\theta(\xi_{\alpha_1}-|\val+\vp|)
+\theta(\xi_{\alpha_2}-|\val-\vp|)}{A-\vp^2+i\ep}
-2\pi i \delta(A-\vp^2)
\theta(\xi_{\alpha_1}-|\val+\vp|)\theta(\xi_{\alpha_2}-|\val
-\vp|)
\Biggl] ~.
\label{e3.5}
\end{align}

 Let us discuss the calculation of the integral
 \begin{align}
\int_{-\infty}^{+\infty}\frac{dA}{2\pi}\frac{e^{iA\eta}}{A-\vp^2+i\epsilon}T^{
\sigma_1\sigma_2}_{\alpha_1\alpha_2}(\vp,\val;A)~.
\label{int.a}
 \end{align}
This integral
is involved in the first two terms of eq.~\eqref{e3.5}. As a preliminary result let us discuss  the particle-particle
and hole-hole parts in
$L_{10}$, since they represent the inclusion of two-nucleon intermediate states
in the medium. It is then convenient to express the function $L_{10}$ employing
the
first form of the nucleon propagator in eq.~\eqref{nuc.pro}. In this way
\begin{align}
L_{10}^{i_3}&=i\int\frac{d^4k}{(2\pi)^4}\left[
\frac{ \theta(\xi_1-|\val-\vk|)}
{a^0-k^0-E(\val-\vk)-i\ep}+\frac{\theta(|\val-\vk|-\xi_1)}{
a^0-k^0-E(\val-\vk)+i\ep}\right]\nn\\
&\times\left[\frac{\theta(\xi_2-|\val+\vk|)}{a^0+k^0-E(\val
+\vk)-i\ep}+\frac{\theta(|\val+\vk|-\xi_2)}{a^0+k^0-E(\val
+\vk)+i\ep}
\right]~.
\label{l10.b.f}
\end{align}

\begin{figure}[t]
\psfrag{f}{$\infty$}
\psfrag{A}{$A(\alpha)$}
\psfrag{B}{$B(\alpha)$}
\psfrag{ci}{$C_I$}
\psfrag{cip}{$C_{I'}$}
\psfrag{2ie}{$2i\ep$}
\centerline{\epsfig{file=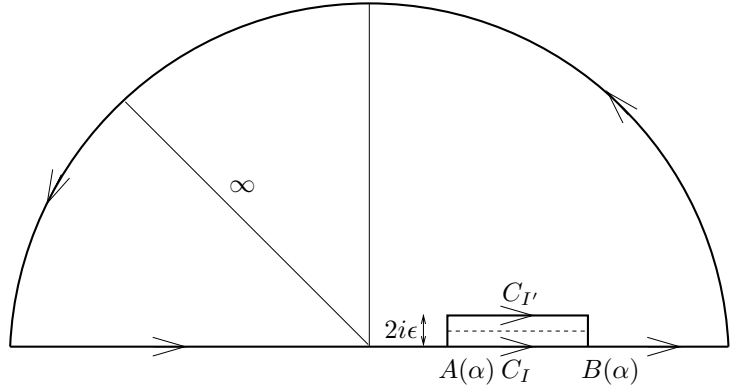,width=.55\textwidth,angle=0}}
\vspace{0.2cm}
\caption[pilf]{\protect \small
Contours of integration $C_I$ and $C_{I'}$ on the $A$-complex plane used to
perform the integral in eq.~\eqref{int.a}. The former contour runs below the
cut (dashed line) and the latter above it. The limits of the cut in $A$ due to
the hole-hole part of $L_{10}$, eq.~\eqref{l10.b.pp}, are
indicated by $A(\alpha)$ and $B(\alpha)$. 
\label{fig:ci}}
\end{figure}

Similarly as in eq.~\eqref{int.p0}, only those contributions in
eq.~\eqref{l10.b.f} with the two poles in $k^0$ lying on opposite halves of the
$k^0$-complex plane contribute. Then,
\begin{align}
L_{10}^{i_3}&=m\int\frac{d^3k}{(2\pi)^3}\left[
\frac{\theta(|\val-\vk|-\xi_1)\theta(|\val+\vk|-\xi_2)}{
A-\vk^2+i\ep}-\frac{\theta(\xi_1-|\val-\vk|)\theta(|\xi_2-|\val
+\vk|)}{A-\vk^2-i\ep}\right]~.
\label{l10.b.pp}
\end{align}
The first term is the particle-particle part and the last is the hole-hole
 one. Notice the different position of the cuts in $A$.
While for the particle-particle case $A$ has  a negative
imaginary part, $-i\ep$, for the hole-hole part the cut takes values with
 positive imaginary part, $+i\ep$. It is also worth mentioning that the extent
of
the cut in $A$ for the hole-hole part is finite. This cut in the last term of
eq.~\eqref{l10.b.pp} requires $\vk^2=A$, but $|\vk|$ is bounded so that the two
$\theta$-functions in the numerator are simultaneously satisfied. The extension
of this cut is
given in eq.~\eqref{l10d.exp}. We denote its lower limit by $A(\alpha)$ and its
upper one by $B(\alpha)$, corresponding to logarithmic branch points, where $\alpha=|\val|$.  This
observation is very useful for performing 
the integral of eq.~\eqref{int.a}. For its
 evaluation  we consider the contours $C_I$ and $C_{I'}$ of
fig.~\ref{fig:ci}. The dashed line in the figure represents the cut in
$T^{\sigma_1\sigma_2}_{\alpha_1\alpha_2}(\vp,\val;A)$ due to that in the
hole-hole part of $L_{10}$ for
$A(\alpha)<\hbox{Re}(A)<B(\alpha)$ and positive imaginary part $+i\ep$.
Physically it represents a real reshuffling of the occupied states by an
in-medium pair of baryons  respecting energy and three-momentum conservation. 
All the
contours of integration include  a semicircle at infinity centered at the origin
along the upper part of the $A$-complex plane.  While the contour $C_I$ runs
along the real axis, and then below the cut, the contour $C_{I'}$ runs above
it, 
with the imaginary part $+2i\ep$. In both cases the pole in the
denominator of eq.~\eqref{int.a} at $\vp^2-i\ep$ is outside the contours of
integration. Because of the convergent factor the integration along the
semicircle at infinity is zero so that we can write
 \begin{align}
\int_{-\infty}^{+\infty}\frac{dA}{2\pi}\frac{e^{iA\eta}}{A-\vp^2+i\epsilon}T^{
\sigma_1\sigma_2}_{\alpha_1\alpha_2}(\vp,\val;A)=\oint_{C_I}\frac{dA}{
2\pi}\frac{e^{iA\eta}}{
A-\vp^2+i\epsilon}T^{\sigma_1\sigma_2}_{\alpha_1\alpha_2}(\vp,\val;A)~.
\label{int.a.2}
 \end{align}
Since the cut is outside the contour $C_{I'}$, see fig.~\ref{fig:ci}, the
integration along this contour is zero
 \begin{align}
\oint_{C_{I'}}\frac{dA}{2\pi}\frac{e^{iA\eta}}{A-\vp^2+i\epsilon}T^{
\sigma_1\sigma_2}_{\alpha_1\alpha_2}(\vp,\val;A)=0~.
\label{int.a.3}
 \end{align}
Subtracting eq.~\eqref{int.a.3} to eq.~\eqref{int.a.2}  we are left with
\begin{align}
\oint_{C_I}\frac{dA}{2\pi}\frac{e^{iA\eta}}{A-\vp^2+i\epsilon}T^{
\sigma_1\sigma_2}_{\alpha_1\alpha_2}(\vp,\val;A)&-\oint_{C_{I'}}\frac{dA
}{2\pi}\frac{e^{iA\eta}}{
A-\vp^2+i\epsilon}T^{\sigma_1\sigma_2}_{\alpha_1\alpha_2}(\vp,\val
;A)\nn\\
&=\int_{A(\alpha)}^{B(\alpha)}\frac{dA}{2\pi}\frac{T^{\sigma_1\sigma_2}_{
\alpha_1\alpha_2}(\vp,\val;A)-T^{\sigma_1\sigma_2}_{\alpha_1\alpha_2}
(\vp,\val;A+2i\ep)}{A-\vp^2+i\ep
}~.
\label{int.a.4}
\end{align}
Since the branch points are
just of logarithmic type the integration along the vertical segments at
$A(\alpha)$ and $B(\alpha)$ in fig.~\ref{fig:ci} do not contribute in the limit
$\ep\to 0^+$. Notice as well that the limit $\eta\to 0^+$ is already taken 
in the last line of eq.~\eqref{int.a.4}. The amplitude
$T^{\sigma_1\sigma_2}_{\alpha_1\alpha_2}(\vp,\val,A)$  can be obtained
from the analytical extrapolation in $A$ of the partial waves amplitudes 
\begin{align}
T_{JI}^{i_3}(\ell',\ell,S;\vp^2,\val^2,A)=\left[N_{JI}^{i_3}(\ell',\ell,
S)^{-1}+ L_{10}^{i_3}(\val^2,A)\right]^{-1}~.
\label{master.2}
\end{align} 
At LO    $N_{JI}$ depends only on $\vp^2$, although for higher orders it could depend also on $i_3$, $A$ and $\val^2$ in addition to $\vp^2$. We can also apply here eq.~\eqref{sum.mat},  keeping explicitly the separation between the $\vp^2$ and $A$ variables,
\be
\sum_{\sigma_1,\sigma_2}
T_{\alpha_1\alpha_2}^{\sigma_1\sigma_2}(\vp^2,\val^2,A)=
\sum_{I,J,\ell,S}(2J+1)\chi(S\ell I)^2 
(\alpha_1\alpha_2 i_3|I_1I_2 I)^2 T_{JI}^{i_3}(\ell,\ell,S;\vp^2,\val^2,A)~.
\label{sum.mat.2}
\ee 
The analytical extrapolation in $A$ does not affect 
the expansion of the nucleon-nucleon scattering in spherical harmonics 
associated to the angular variables which are left intact.  From
eq.~\eqref{master.2} it follows that
\begin{align}
&T_{JI}^{i_3}(\vp^2,\val^2,A)-T_{JI}^{i_3}(\vp^2,\val^2,
A+2i\ep)=\left[{N_{JI}^{i_3}}^{-1}+L^{i_3}_{10}(\val^2,A)\right]^{-1}-
\left[{N_{JI}^{i_3}}^{-1}+L^{i_3}_{10}(\val^2,A+2i\ep)\right]^{-1}\nn\\
&=\left[{N_{JI}^{i_3}}^{-1}+L_{10}^{i_3}(\val^2,A)\right]^{-1}
\left[L_{10}^{i_3}(\val^2,A+2i\ep)-L^{i_3}_{10}(\val^2,A)\right]\left[{N_{JI}^{
i_3}}^{-1}+L^{i_3}_{10}(\val^2,A+2i\ep)\right]^{-1}~.
\end{align}
In this equation we have taken into account that although $N_{JI}$ could depend
on $A$ for higher orders, in the difference
$N_{JI}^{i_3}(\vp^2,\val^2,A)-N_{JI}^{i_3}(\vp^2,\val^2,
A+2i\ep)$ this dependence cancels. The point is that the discontinuity in
$T_{JI}^{i_3}$ due to the right-hand cut is fully taken into account by
multiplying the loop function $L_{10}$ by the kernel $N_{JI}$ on-shell, as in
eqs.~\eqref{master} and \eqref{l10fa3}. The right-hand cut
associated to the variable $A$ is then removed in the process of calculating $N_{JI}$
order by order,  as discussed in
sections \ref{sec:fnn} and \ref{nn.medium}.

As commented above, the difference
$L_{10}^{i_3}(\val^2,A+2i\ep)-L^{i_3}_{10}(\val^2,A)$ is due entirely to the
hole-hole part of
$L_{10}^{i_3}$, the last term on the r.h.s. of eq.~\eqref{l10.b.pp}.  From
eq.~\eqref{l10.b.pp} we have
\begin{align}
L_{10}^{i_3}(\val^2,A+2i\ep)-L_{10}^{i_3}(\val^2,A)&=-m\int\frac{d^3q}{(2\pi)^3}
\theta(\xi_{\alpha_1}-|\val+\vq|)
\theta(\xi_{\alpha_2}-|\val-\vq|)\left(
\frac{1}{A-\vq^2+i\ep}-\frac{1}{A-\vq^2-i\ep}\right)
\nn\\
&=
i2\pi m\int\frac{d^3q}{(2\pi)^3}\theta(\xi_{\alpha_1}-|\val+\vq|)
\theta(\xi_{\alpha_2}-|\val-\vq|)\delta(A-\vq^2)~.
\label{arriba}
\end{align}

Thanks to $\delta(A-\vq^2)$ the $A$-integration in eq.~\eqref{int.a.4} is
now trivial.\footnote{The values of $\vq^2$ that satisfy the two in-medium
$\theta$-functions in eq.~\eqref{arriba} for a given $|\val|$ are comprised in
the interval $[A(\alpha),B(\alpha)]$, which is the domain of the $A$-integration
in eq.~\eqref{int.a.4}.} As a result 
eq.~\eqref{e3.5} turns into
\begin{align}
&{\cal E}_3=-4\sum_{I,J,\ell,S}\sum_{i_3=-1}^1(2J+1)\chi(S\ell
I)^2\int\frac{d^3a}{(2\pi)^3}
\frac{d^3
q}{(2\pi)^3}\theta(\xi_{\alpha_1}-|\val+\vq|)\theta(\xi_{\alpha_2}-|\val-\vq|)\Biggl(
T_{JI}^{i_3}(\vq^2,\val^2,\vq^2)\nn\\
&+m\int\frac{d^3p}{(2\pi)^3}\frac{1-\theta(\xi_{\alpha_1}-|\val+\vp|)
-\theta(\xi_{\alpha_2}-|\val-\vp|)}{\vp^2-\vq^2-i\ep}
\nn\\
&\times
\left[{N_{JI}^{i_3}(\vp^2)}^{-1}+L_{10}^{i_3}(\val^2,\vq^2)\right]^{-1}
\cdot\left[{N_{JI}^{i_3}(\vp^2)}^{-1}+L_{10}^{i_3}(\val^2,\vq^2+2i\ep)\right]^{
-1}
\Biggr)_{(\ell,\ell,S)},
\label{e3.trans}
\end{align}
where we have indicated explicitly   the integration variables in the different
functions. 
 The isospin index $i_3=\alpha_1+\alpha_2$ and for $i_3=0$ one should take just 
one the two possible cases with $\alpha_1=-\alpha_2$, $|\alpha_1|=1/2$.
It is straightforward to show that ${\cal E}_3$ given in eq.~\eqref{e3.trans} is
purely real, as it should be. First, 
the two $\theta$-functions in the first line of eq.~\eqref{e3.trans} imply that
only the hole-hole part of $L_{10}$ 
can have imaginary part. It follows that 
$L_{10}^{i_3}(\vq^2+2i\ep)=L_{10}^{i_3}(\vq^2)^*$ and since, furthermore,
 $N_{JI}^{-1}(\vp^2)+L_{10}^{i_3}(\vq^2)$ is a symmetric matrix (for the $S=1$ 
and $J=\ell\pm 1$ partial waves) or 
just a number (for the rest of partial waves), the diagonal elements of the
product
\begin{align}
\left[{N_{JI}^{i_3}(\vp^2)}^{-1}+L_{10}^{i_3}(\vq^2)\right]^{-1}\cdot\left[{N_{
JI}^{i_3}(\vp^2)}^{-1}+L_{10}^{i_3}(\vq^2+2i\ep)\right]^{-1}~, 
\end{align}
are positive real numbers. In this way we have for the imaginary part of ${\cal
E}_3$, 
\begin{align}
&\hbox{Im}({\cal
E}_3)=-4\sum_{I,J,\ell,S}\sum_{i_3=-1}^1(2J+1)\chi(S\ell
I)^2\int\frac{d^3a}{(2\pi)^3}
\frac{d^3
q}{(2\pi)^3}\theta(\xi_{\alpha_1}-|\val+\vq|)\theta(\xi_{\alpha_2}-|\val-\vq|)\Biggl[\hbox{Im} T_{JI}^{i_3}(\vq^2,\val^2,\vq^2)\nn\\
&
+m\int\frac{d^3p}{(2\pi)^3}\Bigl\{1-\theta(\xi_{\alpha_1}-|\val+\vp|)
-\theta(\xi_{\alpha_2}-|\val-\vp|)\Bigr\}\pi\delta(\vp^2-\vq^2)
T_{JI}^{i_3}(\vp^2,\val^2,\vq^2)\cdot {T_{JI}^{i_3}(\vp^2,\val^2,\vq^2)}^*
\Biggr]_{(\ell,\ell,S)}.
\label{im.e3}
\end{align}
Taking into account eq.~\eqref{l10fa3} and the expression for $L_{10}$, eq.~\eqref{l10.b.pp}, the
imaginary part of $T_{JI}^{i_3}$ can also be calculated 
in terms of that of the  hole-hole part of $L_{10}$. Substituting the result in
eq.~\eqref{im.e3}, it follows that
\begin{align}
&\hbox{Im}({\cal
E}_3)=-4\sum_{I,J,\ell,S}\sum_{i_3=-1}^1(2J+1)\chi(S\ell
I)^2\int\frac{d^3a}{(2\pi)^3}
\frac{d^3
q}{(2\pi)^3}\theta(\xi_{\alpha_1}-|\val+\vq|)\theta(\xi_{\alpha_2}-|\val-\vq|) \, m\int\frac{d^3p}{(2\pi)^3}\pi\delta(\vp^2-\vq^2)\nn\\
&\times\Bigl\{1-\theta(\xi_{\alpha_1}-|\val+\vp|)
-\theta(\xi_{\alpha_2}-|\val-\vp|)+\theta(\xi_{\alpha_1}-|\val
+\vp|)
\theta(\xi_{\alpha_2}-|\val-\vp|)\Bigr\} 
T_{JI}^{i_3}\cdot {T_{JI}^{i_3}}^* \Biggr|_{(\ell,\ell,S)}.
\label{im.e3.2}
\end{align}
The quantity between curly brackets in the previous   equation is  
\begin{align}
\bigl[1-\theta(\xi_{\alpha_1}-|\val+\vp|)\bigr]
\bigl[1-\theta(\xi_{\alpha_2}-|\val-\vp|)\bigr]=
\theta(|\val+\vp|-\xi_{\alpha_1})
\theta(|\val-\vp|-\xi_{\alpha_2})~.
\label{pp.pp}
\end{align}
But given $|\val|$ and $|\vq|$ satisfying simultaneously the
$\theta$-functions in the first line, corresponding to two Fermi-sea 
insertions, it is not possible that they also satisfy simultaneously the two
$\theta$-functions in eq.~\eqref{pp.pp}, corresponding to 
the particle-particle part. Note that  due to the 
Dirac $\delta$-function in eq.~\eqref{im.e3.2} $|\vq|=|\vp|$. As as result, it
is clear that  $\hbox{Im}({\cal E}_3)=0$ once 
eq.~\eqref{pp.pp} is inserted in eq.~\eqref{im.e3.2}.

\begin{figure}[ht]
\psfrag{a}{$a$}
\psfrag{b}{$b$}
\psfrag{c}{$c$}
\psfrag{d}{$d$}
\centerline{\epsfig{file=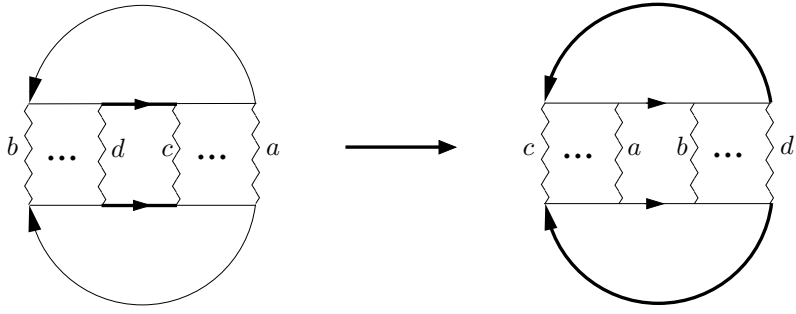,width=.6\textwidth,angle=0}}
\vspace{0.2cm}
\caption[pilf]{\protect \small
Cyclic permutation of the free nucleon propagators
 with Fermi-sea insertions. The labels on the potential lines are 
included to appreciate the permutation of lines in the graph. As usual the thick
lines in fig.~\ref{fig:cicle} 
refer to the insertion of a Fermi-sea. 
\label{fig:cicle}}
\end{figure}

Writing explicitly 
\begin{align}
\left.T_{JI}^{i_3}(\vp^2,\val^2,\vq^2)\cdot
{T_{JI}^{i_3}}(\vp^2,\val^2,\vq^2)^*\right|_{(\ell,\ell,S)}=
\sum_{\ell'}T_{JI}^{i_3}(\ell,\ell',S;\vp^2,\val^2,\vq^2)
T_{JI}^{i_3}(\ell',\ell,S;\vp^2,\val^2,\vq^2)^*,
\end{align}
we have for eq.~\eqref{e3.trans}
\begin{align}
&{\cal E}_3=-4\sum_{I,J,\ell,S}\sum_{i_3=-1}^{1}(2J+1)\chi(S\ell
I)^2\int\frac{d^3a}{(2\pi)^3}
\frac{d^3
q}{(2\pi)^3}\theta(\xi_{\alpha_1}-|\val+\vq|)\theta(\xi_{\alpha_2}-|\val
-\vq|)\Biggl[T_{JI}^{i_3}(\ell,\ell,S;\vq^2,\val^2,\vq^2)\nn\\
&+
m\int\frac{d^3p}{(2\pi)^3}\frac{1-\theta(\xi_{\alpha_1}-|\val+\vp|)
-\theta(\xi_{\alpha_2}-|\val-\vp|)}{\vp^2-\vq^2-i\ep}
\sum_{\ell'}T_{JI}^{i_3}(\ell,\ell',S;\vp^2,\val^2,\vq^2)
T_{JI}^{i_3}(\ell',\ell,S;\vp^2,\val^2,\vq^2)^*
\Biggr]~.
\label{e3.trans2}
\end{align}
The process followed from eq.~\eqref{e3.5} up to here can be schematically drawn
as in fig.~\ref{fig:cicle}. 
The particle-particle part that results by closing the external lines in the
diagrams 3 of fig.~\ref{fig:allE} is transferred to a reducible two-nucleon 
loop entering the in-medium nucleon-nucleon partial waves. This corresponds to
the integration in $\vp$ in eq.~\eqref{e3.trans2}, which is 
sandwiched between two partial waves amplitudes, one of them complex
conjugated. 
The integration on $\vp^2$ in eq.~\eqref{e3.trans2} is linearly divergent
because of the first integral on the second line. The  product 
\begin{align}
\sum_{\ell'}T_{JI}^{i_3}(\ell,\ell',S;\vp^2,\val^2,\vq^2)
T_{JI}^{i_3}(\ell',\ell,S;\vp^2,\val^2,\vq^2)^*
\end{align}
  tends  to a constant for  $\vp^2\to \infty$. Let us discuss how to regularize
this integral, in the same way as already done for the calculation of the 
$L_{ij}$ functions.
\begin{align}
&-m\int\frac{d^3p}{(2\pi)^3}\frac{\sum_{\ell'}T_{JI}^{i_3}(\ell,\ell',S;\vp^2,
\val^2,\vq^2)
T_{JI}^{i_3}(\ell',\ell,S;\vp^2,\val^2,\vq^2)^*}{\vp^2-\vq^2-i\ep}\nn\\
&=
-\frac{m}{2\pi^2}\int_0^\infty
dp\frac{\vp^2 }{\vp^2-\vq^2-i\ep}\sum_{\ell'}T_{JI}^{i_3}(\ell,\ell',S;\vp^2,
\val^2,\vq^2) T_{JI}^{i_3}(\ell',\ell,S;\vp^2,\val^2,\vq^2)^*~.
\label{int.div}
\end{align}
 Let us denote by $N_{JI;\infty}^{i_3}(\val^2,\vq^2)^{-1}$ the limit for
$\vp^2\to\infty$ of $N^{i_3}_{JI}(\vp^2,\val^2,\vq^2)^{-1}$. At LO  this limit
is a  constant for each partial wave. Then, we rewrite the previous integral as
\begin{align}
&-\frac{m}{2\pi^2}\int_0^\infty dp\frac{\vp^2}{\vp^2-\vq^2-i\ep}
\Biggl\{\left[{N_{JI}^{i_3}(\vp^2,\val^2,\vq^2)}^{-1}+L_{10}^{i_3}(\val^2,\vq^2)\right]^
{-1}\left[{N_{JI}^{i_3}(\vp^2,\val^2,\vq^2)}^{-1}+L_{10}^{i_3}(\val^2,
\vq^2+2i\ep)\right]^{-1}\nn\\
&\pm\left[{N_{JI,\infty}^{i_3}(\val^2,\vq^2)}^{-1}+L_{10}^{i_3}(\val^2,
\vq^2)\right]^{-1}\left[{N_{JI,\infty}^{i_3}(\val^2,\vq^2)}^{-1}+L_{10}^{i_3}
(\val^2,\vq^2+2i\ep)\right]^{-1}\Biggr\}~.
\label{e3.div.1}
\end{align}
 Now, since ${N_{JI}^{i_3}}^{-1}\to {N_{JI,\infty}^{i_3}}^{-1}+{\cal
O}(|\vp|^{-2})$ the integral 
\begin{align}
&-\frac{m}{2\pi^2}\int_0^\infty dp\frac{\vp^2}{\vp^2-\vq^2-i\ep}\Biggl\{
\left[{N_{JI}^{i_3}(\vp^2,\val^2,\vq^2)}^{-1}+L_{10}^{i_3}(\val^2,\vq^2)\right]^
{-1}\left[{N_{JI}^{i_3}(\vp^2,\val^2,\vq^2)}^{-1}+L_{10}^{i_3}(\val^2,
\vq^2+2i\ep)\right]^{-1}\nn\\
&-\left[{N_{JI,\infty}^{i_3}(\val^2,\vq^2)}^{-1}+L_{10}^{i_3}(\val^2,
\vq^2)\right]^{-1}\left[{N_{JI,\infty}^{i_3}(\val^2,\vq^2)}^{-1}+L_{10}^{i_3}
(\val^2,\vq^2+2i\ep)\right]^{-1}\Biggr\}
\end{align}
is convergent. The remaining integral in eq.~\eqref{e3.div.1} is expressed 
in terms of the function $g(\vq^2)$, eq.~\eqref{dis.rel.g}, 
\begin{align}
&-\frac{m}{2\pi^2}\int_0^\infty dp\frac{\vp^2}{\vp^2-\vq^2-i\ep}
\left[{N_{JI,\infty}^{i_3}(\val^2,\vq^2)}^{-1}+L_{10}^{i_3}(\val^2,
\vq^2)\right]^{-1}\left[{N_{JI,\infty}^{i_3}(\val^2,\vq^2)}^{-1}+L_{10}^{i_3}
(\val^2,\vq^2+2i\ep)\right]^{-1}\nn\\
&=g(\vq^2)\sum_{\ell'}T^{i_3}_{JI,\infty}(\ell,\ell',S;\val^2,\vq^2)T^{i_3}_{JI,\infty}
(\ell',\ell,S;\val^2,\vq^2)^* ~,
\label{ea.regu}
\end{align}
with 
\begin{align}
T_{JI,\infty}^{i_3}(\ell',\ell,S;\val^2,\vq^2)=\left[{N_{JI,\infty}^{i_3}(\val^2,\vq^2)}^{-1}+L_{10}^{i_3}(\val^2,
\vq^2)\right]^{-1}=\lim_{\vp^2\to\infty}
T_{JI}^{i_3}(\ell',\ell,S;\vp^2,\val^2,\vq^2)~.
\end{align}
To simplify the notation let us define the symbols
\begin{align}
\Sigma_{p\ell}&=\sum_{\ell'}T_{JI}^{i_3}(\ell,\ell',S;\vp^2,\val^2,\vq^2)T_{JI}^{i_3}(\ell',\ell,S;\vp^2,
\val^2,\vq^2)^*~,\nn\\
\Sigma_{\infty\ell}&=\sum_{\ell'}T_{JI,\infty}^{i_3}(\ell,\ell',S;\val^2,\vq^2)T_{JI,
\infty}^{i_3}(\ell',\ell,S;\val^2,\vq^2)^*~.
\label{sigma.symbols}
\end{align} 
The function $g(\vq^2)$ depends on the same subtraction constant $g_0$ already employed in 
the study of the vacuum nucleon-nucleon interactions. 
The final expression for ${\cal E}_3$ in eq.~\eqref{e3.trans2} is then
\begin{align}
& {\cal E}_3=4\sum_{I,J,\ell,S}\sum_{\alpha_1,\alpha_2}(2J+1)\chi(S\ell
I)^2 \int\frac{d^3a}{(2\pi)^3} \frac{d^3 q}{(2\pi)^3} \theta(\xi_{\alpha_1}-|\val+\vq|)\theta(\xi_{\alpha_2}-|\mathbf{a}-\vq|) 
\Bigl[
-T_{JI}^{i_3}(\ell,\ell,S;\vq^2,\val^2,\vq^2)\nn\\
& + g(\vq^2) \Sigma_{\infty\ell} +
m \int \frac{d^3p}{(2\pi)^3} 
\Bigl\{
\frac{\theta(\xi_{\alpha_1}-|\val+\vp|) + \theta(\xi_{\alpha_2}-|\val-\vp|)}{\vp^2-\vq^2-i\ep}
\Sigma_{p\ell}-\frac{
\Sigma_{p\ell}-\Sigma_{\infty\ell}}{\vp^2-\vq^2-i\ep}\Bigl\}
\Bigr]\nn\\
&=4\sum_{I,J,\ell,S}\sum_{\alpha_1,\alpha_2}(2J+1)\chi(S\ell
I)^2 \int\frac{d^3a}{(2\pi)^3}  \frac{d^3 q}{(2\pi)^3} \theta(\xi_{\alpha_1}-|\val+\vq|)\theta(\xi_{\alpha_2}-|\mathbf{a}-\vq|) 
\Bigl[
-T_{JI}^{i_3}(\ell,\ell,S;\vq^2,\val^2,\vq^2)\nn\\
& + g _0 \Sigma_{\infty\ell} -
m \int \frac{d^3p}{(2\pi)^3} 
\Bigl\{
\frac{1-\theta(\xi_{\alpha_1}-|\val+\vp|) - \theta(\xi_{\alpha_2}-|\val-\vp|)}{\vp^2-\vq^2-i\ep}
\Sigma_{p\ell}-\frac{1}{\vp^2}\Sigma_{\infty\ell}\Bigr\}
\Bigr]~.
\label{e3.reg}
\end{align}

 \begin{figure}[ht]
\psfrag{rho}{{\small $\begin{array}{c}
\\ \rho~(\hbox{fm}^{-3})
\end{array}$}}
\psfrag{xi}{$\xi$}
\psfrag{E/A (MeV)}{{\small $E/A$~(MeV)}}
\centerline{\epsfig{file=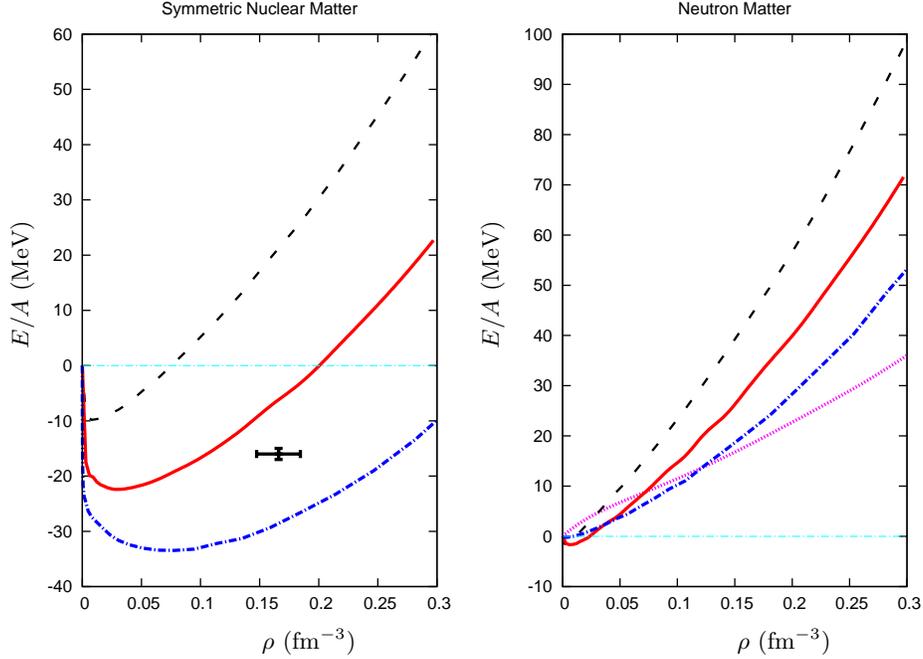,width=.5\textwidth,angle=-90}}
\vspace{0.2cm}
\caption[pilf]{\protect \small (Color online.) 
Energy per  nucleon, ${\cal E}/\rho$, for symmetric nuclear matter, left
panel, and for neutron matter, right panel. The (magenta) dotted line in the
right panel  is the result from the many-body calculation of ref.~\cite{urbana}
using realistic nucleon-nucleon potentials. The rest of the lines  from top to
bottom correspond to different values of $g_0=-0.25$, $-0.37$ and $-0.5~m_\pi^2$, respectively. The point on the left panel is the empirical
 saturation one for nuclear matter \cite{majo2}.
\label{fig:en.lo}}
\end{figure}

The sum of eqs.~\eqref{cont.e1}, \eqref{eq.e2.3} and \eqref{e3.reg} gives our
result for the energy density, ${\cal E}$, in nuclear 
matter at NLO, 
\begin{align}
{\cal E}={\cal E}_1+{\cal E}_2+{\cal E}_3~.
\label{energy}
\end{align}
We evaluate eq.~\eqref{e3.reg}  using the in-medium nucleon-nucleon partial
waves determined at LO in section \ref{nn.medium}. The sum over partial waves
shows good convergence already for maximum  $J=4$ and we sum up to $J=5$. The
results for the energy per nucleon, $E/A={\cal E}/\rho$, are shown in
fig.~\ref{fig:en.lo} for symmetric nuclear matter, left panel, and for neutron
matter, right panel.  The inserted point with errors on the left panel of
fig.~\ref{fig:en.lo} corresponds to the experimental values for the saturation of
nuclear matter $E/A=(-16\pm 1)$~MeV and $\rho=(0.166\pm 0.018)$~fm$^{-3}$ quoted
in ref.~\cite{majo2}.   The  dotted line in the right panel is the
result for neutron matter from the many-body calculation of the Urbana group
\cite{urbana}. It employs  realistic nucleon-nucleon potentials and a fitted
density dependent three-nucleon force in order to reproduce the experimental
saturation point for nuclear matter.  The rest of the curves, from top to bottom
in both panels, correspond to the values of $g_0=-0.25$, $-0.37$ and $-0.5~m_\pi^2$,  in order.
 \begin{figure}[ht]
\psfrag{rho}{{\small $\begin{array}{c}
\\{\rho~(\hbox{fm}^{-3})}
\end{array}$}}
\psfrag{xi}{$\xi$}
\psfrag{E/A (MeV)}{{\small $E/A$~(MeV)}}
\centerline{\epsfig{file=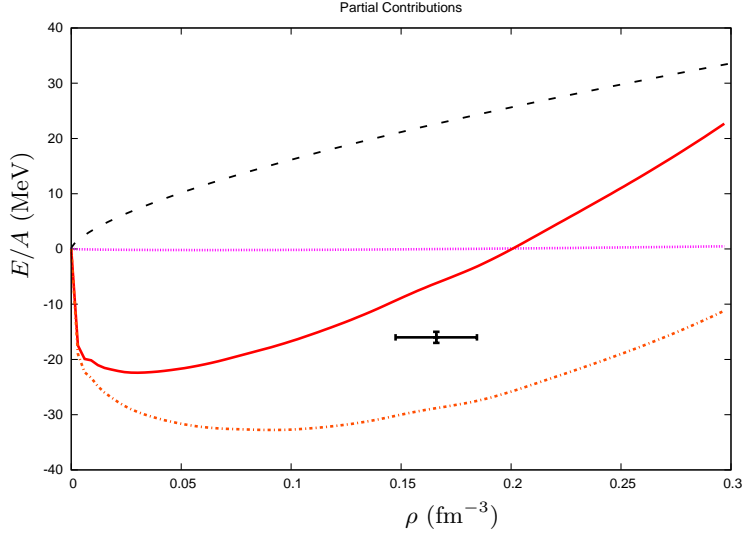,width=.4\textwidth,angle=-90}}
\vspace{0.2cm}
\caption[pilf]{\protect \small (Color online.) The partial contributions to
  ${\cal E}/\rho$, for symmetric nuclear matter and $g_0=-0.37~m_\pi^2$ are
  shown. The (black) dashed,
 (magenta) dotted, (orange) dot-dashed and solid lines are for ${\cal E}_1/\rho$, ${\cal E}_2/\rho$, ${\cal E}_3/\rho$ and their sum, respectively. The point  is the saturation one for nuclear matter \cite{majo2}.
\label{fig:en.lo.partial}}
\end{figure}
We observe that our curves for symmetric nuclear matter does have a minimum with
a value in agreement with the experimental one, $-16\pm 1$~MeV, for $g_0\simeq -0.30$. However,
the position is displaced towards too low values of $\xi=\xi_n=\xi_p\sim
150~$MeV, too small by a factor 1.7 compared with the value $\xi \simeq 266{\pm 10}~$MeV \cite{majo2}. For the case of neutron matter, the curves are repulsive and are
larger than the calculation of the Urbana group \cite{urbana} above some density.  It is clear from
fig.~\ref{fig:swave} and \ref{fig:pwave} that we are employing for the
calculation of eq.~\eqref{e3.reg} nucleon-nucleon partial waves that do not 
reproduce closely the Nijmegen data in several partial waves, see figs.~\ref{fig:swave}-\ref{fig:dwave}. As commented above, we are just employing the
iteration of the one-pion exchange and two four-nucleon local vertices. One needs
more elaborate nucleon-nucleon partial waves. Indeed, there are many mutual
cancellations involved in the case of symmetric nuclear matter, between the
purely kinetic energy term, ${\cal E}_1$, and between the S- and P-waves in
${\cal E}_3$, with ${\cal E}_2$  negligible small. Nevertheless,  we find rather
encouraging that our curves in fig.~\ref{fig:en.lo} can reproduce the main trends
of ${\cal E}/\rho$ both for symmetric nuclear and neutron  matter despite they are
obtained employing in-medium nucleon-nucleon amplitudes calculated only at LO.
 We already pointed out in section~\ref{sec:fnn} that the one-pion exchange has a too large tensor force which is
reduced by higher order counterterms (in the meson exchange approach this reduction is achieved  by the exchange of  $\rho$-mesons \cite{brown}.) In ref.~\cite{majo}
this point is emphasized in its study of nuclear binding because a large tensor
force leads to less binding energy. Indeed, the partial waves ${^3S_1}$-${^3D_1}$
and $^3P_0$ have large matrix elements of the one-pion exchange tensor operator
\cite{majo} and these partial waves are not well reproduced in our study at LO. 
 We show in fig.~\ref{fig:en.lo.partial} the different contributions to ${\cal E}/\rho$ for symmetric nuclear matter and 
 $g_0=-0.37~m_\pi^2$, that corresponds to the solid lines in fig.~\ref{fig:en.lo}. Namely, ${\cal E}_1/\rho$, ${\cal E}_2/\rho$ and ${\cal E}_3/\rho$ are given by the 
 dashed, dotted and dot-dashed lines, 
 with the sum corresponding to the full curve. As expected, the contribution from ${\cal E}_2$, eq.~\eqref{eq.e2.3}, is very small since the derivative of $\Sigma_f$ with respect to $k^0$ is ${\cal O}(p^2)$, as discussed in section~\ref{sec.meson-baryon}.\footnote{Notice that $\Sigma_f(k^0)=0$ for $k^0=0$ and $E(\vk)$ is very small.} The other terms, ${\cal E}_1$ and ${\cal E}_3$ have similar size, though the former is ${\cal O}(p^5)$ and the latter ${\cal O}(p^6)$. 
 This is due to the fact that the kinetic energy term  is a recoil correction stemming from 
 ${\cal L}_{\pi N}^{(2)}$, eq.~\eqref{lags}, being suppressed numerically by the inverse of the large nucleon mass $m$. Notice that ${\cal E}_3/\rho$ scales like $ \rho C$, which introduces, compared with ${\cal E}_1/\rho$, the additional power of $\xi$ times $1/2\pi^2$ from the density and $4\pi/ p$ from $m C\sim m/g_0$. Both contributions have the same order of magnitude as the resulting  factor $2\xi/\pi p\sim 1$. Additionally, there is also an extra suppression of the kinetic term contribution because of the dimensionality of space. The point is that ${\cal E}_1$ contains the integral of $\int_0^\xi d|\vk||\vk|^4=\xi^5/5$, while the extra factor of density in ${\cal E}_3$ goes like $\xi^3/3$, so that a numerical factor $3/5=0.6$ is suppressing ${\cal E}_1$. 
 Then, the cancellation between the kinetic energy term and the one due to the nucleon-nucleon interactions is a consequence of keeping the natural size for the chiral counterterms, of similar size than the one-pion exchange as seen in section~\ref{sec:fnn}. Of course, the precise value resulting from such cancellations depends on $g_0$ as shown in fig.~\ref{fig:en.lo}. 
 Additionally,  the presence of such cancellation enhanced this dependence. E.g. for the neutron matter case the kinetic energy dominates the energy per baryon and the dependence on $g_0$ is  smaller indeed.  
  ${\cal E}_3$ depends implicitly on $g_0$ through the nucleon-nucleon partial waves. Additionally, there is  an explicit  dependence  from the first term on the last line of eq.~\eqref{e3.reg}, $g _0 \Sigma_{\infty\ell}$.    The implicit  dependence is  due to the truncated solution of eq.~\eqref{dis.nji}, as discussed in detail in section~\ref{sec:fnn}. The explicit one should be also related to this truncation given the close similarity between them. To make this clear let us notice that the partial wave $-T_{JI}^{i_3}(\ell,\ell,S;\vq^2,\val^2,\vq^2)$ appearing in eq.~\eqref{e3.reg} can also be written from eq.~\eqref{l10fa3} as:
 \begin{align}
&-T_{JI}^{i_3}(\ell,\ell,S;\vq^2,\val^2,\vq^2)
=-\left[{N_{JI}^{i_3}}^{-1}(\vq^2)+L_{10}^{i_3}(\val^2,\vq^2) \right]^{-1}\nn\\
 &=
 -\sum_{\ell',\ell''}  T_{JI}^{i_3}(\ell,\ell',S;\vq^2,\val^2,\vq^2)T_{JI}^{i_3}(\ell',\ell'',S;\vq^2,\val^2,\vq^2)^*\left({N_{JI}^{i_3}}^{-1}(\ell'',\ell,S;\vq^2)+L_{10}^{i_3}(\val^2,\vq^2)^*\delta_{\ell'' \ell}\right)~.
\label{bla1}
 \end{align}
 From the previous equation the term  proportional to $L_{10}$ is
\begin{align}
-L_{10}^{i_3}(\val^2,\vq^2)^*\sum_{\ell'}T_{JI}^{i_3}(\ell,\ell',S;\vq^2,\val^2,\vq^2)T_{JI}^{i_3}(\ell',\ell,S;\vq^2,
\val^2,\vq^2)^*=\left(-\hbox{Re}L_{10}^{i_3}+i\, \hbox{Im}L_{10}^{i_3}\right)\Sigma_{q\ell}~.
\label{bla2}
\end{align}
Then, a similar dependence on $g_0$ as that of $g_0\Sigma_{\infty\ell}$ results  from eq.~\eqref{bla2} as $-g_0\Sigma_{q\ell}$. The sum of both is $-g_0(\Sigma_{q\ell}-\Sigma_{\infty\ell})$.  Thus, as a higher order calculation should dismiss the dependence on $-g_0\Sigma_{q\ell}$, by analogy,   we expect this to be the case also for $g_0\Sigma_{\infty \ell}$.

Another way of considering our power counting in eq.~\eqref{ffg} is to use it
for correcting order by order  nucleon-nucleon amplitudes  determined in vacuum.
In this way, one can use better nucleon-nucleon partial waves, e.g. calculated
at higher orders in momentum, and use in-medium corrections (whose calculation
is always more cumbersome than diagrams in the vacuum)  at lower orders.  
Another interesting issue left for further work is to explore the three-nucleon
force influence on ${\cal E}$. This requires to consider the calculation of the
energy density in the nucleon medium one order higher or ${\cal O}(p^7)$, since
$V_\rho=3$.

\subsection{Some phenomenology}
\label{phenomena}

In this section, we will give up the strict power counting scheme employed so
far and try to analyse the possible effects of higher orders by some
phenomenologically guided parameter fine-tuning. This will allow us to 
better understand the results obtained for nuclear and neutron matter
in comparison to other recent studies, as e.g. in refs.~\cite{lutz,kai1,kai3}.

\begin{figure}[ht]
\psfrag{rho}{$\begin{array}{c}
\\{\small \rho~(\hbox{fm}^{-3})}
\end{array}$}
\centerline{\epsfig{file=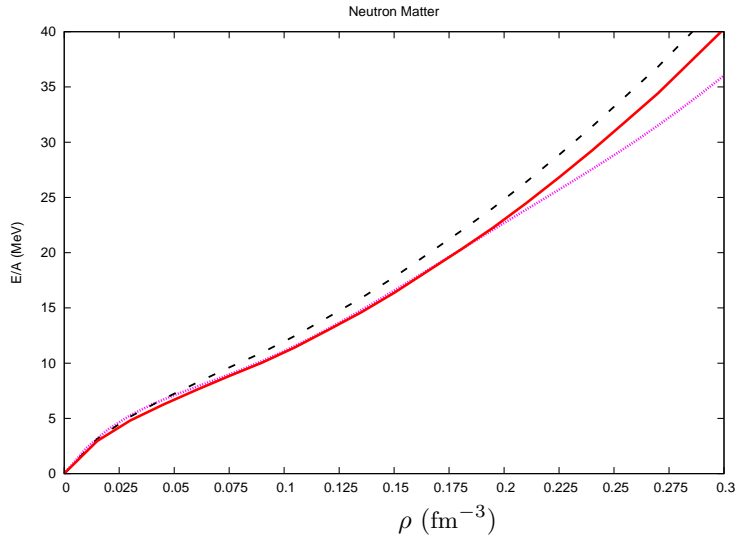,width=.4\textwidth,angle=-90}}
\vspace{0.2cm}
\caption[pilf]{\protect \small (Color online.) ${\cal E}/\rho$ for pure neutron matter. The (magenta) dotted line corresponds to the result of 
  ref.~\cite{urbana}. The (black) dashed line is obtained from
  eq.~\eqref{e3.reg}  with $g_0=-0.62~m_\pi^2$. The (red) solid line 
  represents eq.~\eqref{e3.reg.2} with $g_0=-0.62~m_\pi^2$ and 
  $\widetilde{g}_0=-0.65~m_\pi^2$. See the text for further details.
\label{fig:neu.com}}
\end{figure}

As a first exercise, let us vary the parameter $g_0$ in order to improve the
description of ${\cal E}/\rho$ for the case of neutron matter, so that our results 
agree better with the dotted line in fig.~\ref{fig:en.lo} corresponding to the 
sophisticated many-body calculation of ref.~\cite{urbana}.  The dashed line in 
fig.~\ref{fig:neu.com} is obtained employing $g_0=-0.62~m_\pi^2$. The so
obtained fine-tuned curve is  very close to the results of ref.~\cite{urbana}, 
even up to rather high densities (the deviation is then less than 10$\%$.) 
Note as well that this result is obtained with a value of $g_0$ still of natural 
size, in the expected range around $-0.55~m_\pi^2$. However, if we employ the
same  $g_0$ for evaluating  ${\cal E}$ for symmetric nuclear matter the
resulting curve has the minimum at its right position, $\rho\simeq 0.16$~fm$^{-3}$, but the value of the energy per baryon  is around
$-42~$MeV, which is an over-binding by a factor 2.5.

To further analyse the density dependence of nuclear matter, we rewrite the
contribution from the $NN$ interactions introducing by hand a new parameter, so that 
\begin{align}
{\cal E}_3&=4\sum_{I,J,\ell,S}\sum_{\alpha_1,\alpha_2}(2J+1)\chi(S\ell
I)^2 \int\frac{d^3a}{(2\pi)^3}  \frac{d^3 q}{(2\pi)^3} \theta(\xi_{\alpha_1}-|\val+\vq|)\theta(\xi_{\alpha_2}-|\mathbf{a}-\vq|) 
\Bigl[
-T_{JI}^{i_3}(\ell,\ell,S;\vq^2,\val^2,\vq^2)\nn\\
& + \widetilde{g}_0 \Sigma_{\infty \ell} -
m \int \frac{d^3p}{(2\pi)^3} 
\Bigl\{
\frac{1-\theta(\xi_{\alpha_1}-|\val+\vp|) - \theta(\xi_{\alpha_2}-|\val-\vp|)}{\vp^2-\vq^2-i\ep}
\Sigma_{p\ell}-\frac{1}{\vp^2}\Sigma_{\infty\ell}\Bigr\}
\Bigr]~.
\label{e3.reg.2}
\end{align}
Here we have distinguished between $\widetilde{g}_0$, that corresponds to the
parameter $g_0$ that appears explicitly in ${\cal E}_3$ because of the diverging integral in  
eq.~\eqref{int.div}, and $g_0$,
 on which ${\cal E}_3$ depends implicitly through the dependence on the
 nucleon-nucleon partial waves.  The idea is to mock up higher order effects  
by varying independently $g_0$ and $\widetilde{g}_0$ and exploit the
phenomenological  implications of such a procedure. While $g_0$ affects  
nucleon-nucleon scattering  in vacuum, as discussed at length in
section~\ref{sec:fnn},  $\widetilde{g}_0$ affects only the nuclear matter 
equation of state. E.g. employing $\widetilde{g}_0=-0.67~m_\pi^2$, which
implies a change of $7\%$  with respect to  $g_0=-0.62~m_\pi^2$  used above, 
and that we also keep here, the solid curve in fig.~\ref{fig:neu.com}
is obtained.  The agreement for $\rho\lesssim 0.2$~fm$^{-3}$ between our results and ref.~\cite{urbana} is almost perfect.  

 \begin{figure}[ht]
\psfrag{rho}{$\begin{array}{c}
\\{\small \rho~(\hbox{fm}^{-3})}
\end{array}$}
\centerline{\epsfig{file=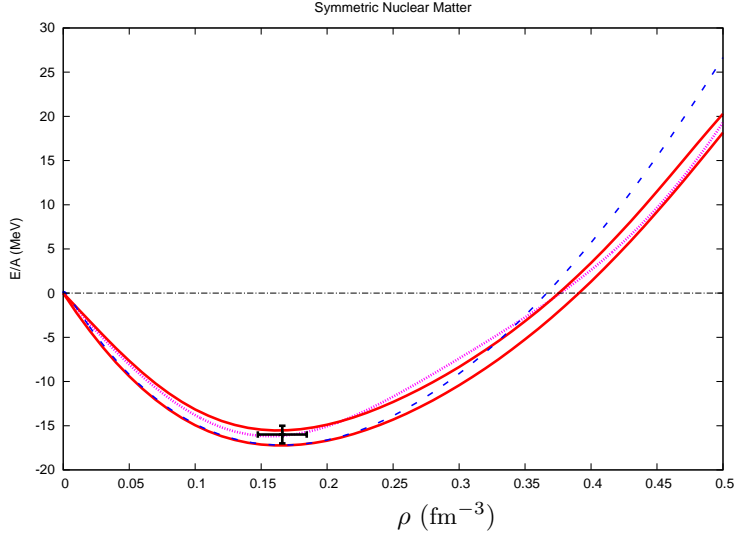,width=.4\textwidth,angle=-90}}
\vspace{0.2cm}
\caption[pilf]{\protect \small (Color online.) ${\cal E}/\rho$ for symmetric
  nuclear matter. The two (red) solid lines correspond from top to bottom to 
 $(g_0,\widetilde{g}_0)=(-0.977,-0.512)~m_\pi^2$ and $(-0.967,-0.525)~m_\pi^2$, in order.  The (blue) dashed line is obtained from eq.~\eqref{param.kw}  adjusting to  the minimum position and  value of the lowest of our curves. The (magenta) dotted line is the result  of ref.~\cite{urbana}.
\label{fig:sym.com}}
\end{figure}

For the case of symmetric nuclear matter the best results are obtained with  
$\widetilde{g}_0= -0.52~m_\pi^2$, and $g_0=-0.97~m_\pi^2$. Again the
magnitude of both numbers is of natural size. This is shown in
fig.~\ref{fig:sym.com} by the two solid lines which have  
$(g_0 , \widetilde{g}_0)   = (-0.977,-0.512)~m_\pi^2$ and  $(-0.967,-0.525)~m_\pi^2$, in order
from top to bottom in the figure. The corresponding value for $E/A$ at the 
saturation point is $-15.4$ and $-17.1$~MeV, respectively. The experimental
value given by the cross corresponds to $-16\pm 1~$MeV. The position of the
minima for the same  lines is $\rho=0.169$ and $\rho=0.168$~fm$^{-3}$, compared
to the empirical value $\rho=0.166\pm 0.019$~fm$^{-3}$. In addition, the
dotted  line is the result from the many-body calculation of
ref.~\cite{urbana} employing realistic nucleon-nucleon interactions, that includes a free parameter to fix the three-nucleon 
interaction in the nuclear medium  to reproduce the saturation point. We observe that our results reproduce very well the saturation point and agree 
closely with the calculation of ref.~\cite{urbana}.

As done in ref.~\cite{kai1} it is illustrative to compare our curves in 
fig.~\ref{fig:sym.com} with the following simple parameterization for 
the energy per baryon in symmetric nuclear matter 
\begin{align}
\frac{{\cal E}}{\rho}&=\frac{3\xi^2}{10 m}-\alpha\frac{\xi^3}{m^2}+\beta\frac{\xi^4}{m^3}~.
\label{param.kw}
\end{align}
Interestingly, the nuclear matter incompressibility \cite{globo1,globo2} 
\begin{align}
K=\left.\xi^2\frac{\partial^2} {\partial \xi^2}\frac{\cal E}{\rho}\right|_{\xi_0}
\label{k.exp}
\end{align}
is correctly given once $\alpha$ and $\beta$ are known by adjusting the
empirical nuclear matter saturation point. In this equation $\xi_0$ is the 
Fermi momentum at the saturation point. For the central values of the point 
in fig.~\ref{fig:sym.com}, $\rho=0.166$~fm$^{-3}$ and $E/A=-16~$MeV, the 
resulting nuclear matter incompressibility is $K=259~$MeV, which is compatible 
with the experimental value $K=250\pm 25~$MeV \cite{globo1}. Our curves can be 
also described rather accurately by eq.~\eqref{param.kw}, as shown by the
dashed curve in fig.~\ref{fig:sym.com} obtained by adjusting  the minimum position and  value of the lowest of solid curves. Both curves run very close to  each other and start to deviate for densities  above $\rho\simeq 0.25$~fm$^{-3}$. The
resulting nuclear matter incompressibility calculated from our results is 
$K=254$ and $233~$MeV, for the upper and lower solid curves, respectively. 
These values are compatible with the experimental value. The close agreement 
between our results and eq.~\eqref{k.exp}  shows that, to a good approximation, 
the former admits an expansion in powers of the Fermi momentum as in 
eq.~\eqref{param.kw}  for low $\xi$. 

In Table~\ref{tab:cont} we show the contributions to ${\cal E}_3$~(MeV) in nuclear matter for the different partial waves considered at the saturation point $\rho=0.16$~fm$^{-3}$. The first column to  the right of every partial wave corresponds to  $g_0=-0.977$~$m_\pi^2$ and the second one to $g_0=-0.521$~$m_\pi^2$, and $\widetilde{g}_0$ is fixed to $-0.512$~$m_\pi^2$ in the two cases. The former case is the top solid line in fig.~\ref{fig:sym.com} and the latter is nearly the dot-dashed line in the left panel of fig.~\ref{fig:en.lo}. The kinetic energy contributes with $22.11$~MeV per nucleon. One observes clearly the dominant role of the S-, P- and  $^3D_1$ 
 waves. It is remarkable the large influence of the medium on the $^3 P_0$ partial wave that despite being attractive in vacuum  gives a repulsive contribution in the medium for $(g_0,\widetilde{g}_0)=(-0.977,-0.512)~m_\pi^2$, and only slightly attractive for $g_0=\widetilde{g}_0=-0.512$~$m_\pi^2$.  The change of the  $^1S_0$ contribution with $g_0$ is due to the fact that for $g_0=-0.977$~$m_\pi^2$ the phase shifts decrease with energy (they are 30 and 20 degrees at $|\vp|\simeq 100$ and $300~$MeV, respectively), instead of stabilizing at around 60 degrees like happens for $g_0=-0.512~m_\pi^2$, see fig.~\ref{fig:swave}. In this way, its contribution is  less attractive. We also remark that the non-elastic partial waves, e.g. the $^3S_1{\to} ^3D_1$, also  contribute to ${\cal E}_3$ from $\Sigma_{p\ell}$ and $\Sigma_{\infty\ell}$ and  are included in the elastic ones. For a given partial wave  they correspond to one term in the sum of  two terms on $\ell'$ in eq.~(\ref{sigma.symbols}).

\begin{table}
\begin{center}
\begin{tabular}{|l|r|r|l|r|r|l|r|r|l|r|r|}
\hline
$^1 S_0$ & $-8.80$ & $-31.03$ & $^1 D_2$ & $-1.56$ & $-1.54$ & $^3S_1$ & $-42.85$& $-38.39$ & $^3G_3$ & $0.47$ & $0.62$ \\  
$^3 P_0$ & $9.37$  & $-1.42$  & $^3 F_3$ & $1.71$  & $1.76$  & $^3D_1$ & $-9.31$ & $2.94$   & $^1F_3$ & $0.60$ & $0.61$ \\
$^3 P_1$ & $10.20$ & $12.30$  & $^3 F_4$ & $-0.04$ & $-0.05$ & $^1P_1$ & $ 2.48$ & $2.78$   & $^3G_4$ & $0.29$ & $0.15$ \\
$^3 P_2$ & $-1.13$ & $-1.13$  & $^3 H_4$ & $0.03$  & $0.028$ & $^3D_2$ & $0.07$  & $-1.65$  & $^3G_5$ & $0.32$ & $0.35$ \\
$^3 F_2$ & $-0.28$ & $-0.28$  & $^1 G_4$ & $-0.38$ & $-0.38$ & $^3D_3$ & $0.91$  & $1.11$   & $^3I_5$ & $0.23$ & $0.26$ \\
\hline
\end{tabular}
\caption{Contributions to $E/A$~(MeV) in nuclear matter for the different partial waves considered at $\rho=0.16$~fm$^{-3}$. The first column to  the right of every partial wave corresponds to  $g_0=-0.977$~$m_\pi^2$ and the second one to $g_0=-0.521$~$m_\pi^2$, with $\widetilde{g}_0$ fixed to $-0.512$~$m_\pi^2$. The former is the top solid line in fig.~\ref{fig:sym.com} and the latter is nearly the dot-dashed line in the right panel of fig.~\ref{fig:en.lo}.  
\label{tab:cont}}
\end{center}
\end{table}

Thus, we are able to reproduce the equation of state of symmetric and neutron 
matter in terms of one fine-tuned  free parameter, the value of
$g_0\simeq -0.97~m_\pi^2$ for the symmetric nuclear matter case. The parameter 
$\widetilde{g}_0$ comes out always with a value close to the expected one, 
$-m m_\pi/4\pi=-0.54~m_\pi^2$. For  neutron matter one has $\widetilde{g}_0\simeq
-0.62~m_\pi^2$ and for symmetric nuclear matter $\widetilde{g}_0\simeq -0.52~m_\pi^2$.  
In addition, it is worth stressing that the resulting values for $g_0$ are all
negative with the appropriate size, so it is justified to solve 
approximately eq.~\eqref{dis.nji} making use of UCHPT.  We think that this 
achievement is not a trivial fact and clearly indicates that our power 
counting, eq.~\eqref{ffg}, is able to map out properly the  important dynamics 
in the nuclear medium and establish a useful hierarchy to allow for systematic 
calculations. At present, the strong dependence on $g_0$ of particularly 
 the $^1 S_0$  and $^3P_0$ partial wave contributions,  see Table~\ref{tab:cont}, 
 seems to indicate that in order to go 
 forward in the application of our chiral power counting eq.~\eqref{ffg}, one 
 should consider the exact solution of $N_{JI}$, eq.~\eqref{dis.nji}, and not its 
 truncated solution  in powers of $g$, eq.~\eqref{geo.ser}, as presently done. 
 Notice that in comparison with refs.~\cite{lutz,kai1,kai3}, that also applies Baryon 
CHPT to in-medium calculations, we employ the same approach both for vacuum and nuclear 
matter calculations and are able to compare with nucleon-nucleon scattering data.
 
\section{Conclusions}
\def\theequation{\arabic{section}.\arabic{equation}}
\setcounter{equation}{0}
\label{sec:conc}

In \cite{nlou} we  derived a novel approach for an EFT in the nuclear
medium based on a chiral power counting that combines both short-range and 
pion-mediated internucleon interactions. The power counting is bounded from 
below and at a given order it requires the calculation of a finite number of
contributions, which typically implies the resummation of  an infinite string
of  two-nucleon reducible diagrams with the leading multi-nucleon CHPT
amplitudes. These resummations arise because this power counting takes into
account from the 
onset the presence of enhanced nucleon propagators and it can also be applied
to multi-nucleon forces. However, ref.~\cite{nlou} did not discuss any
specific non-perturbative method for applying this novel counting scheme to 
practical calculations. We have developed in the present paper the required
non-perturbative technique that allows us to perform these  resummations both in
scattering as well as in production processes. This non-perturbative method is 
based on Unitary CHPT, which is adapted now to the nuclear medium by
implementing the power counting of ref.~\cite{nlou}. 
We have first applied it to calculate the LO and NLO vacuum  
nucleon-nucleon interactions. There, we have also shown the tight connection 
between the UCHPT approach and dispersion relations for the 
nucleon-nucleon scattering amplitude. The former results as an approximate
solution to the dispersive treatment of nucleon-nucleon scattering in a chiral 
expansion of the imaginary part of the scattering amplitudes along the 
left-hand cut, taking advantage of the suppression of the two-nucleon 
unitarity loops along this cut. It was also shown that  the subtraction 
constant $g_0$, employed for regularizing the reducible part of the
two-nucleon reducible loops, realizes a convenient splitting between 
loops and tree-level contributions, while  the exact solution for 
nucleon-nucleon scattering in vacuum has been shown not to depend on $g_0$.   
For the in-medium case the LO nucleon-nucleon scattering is given as well. 
Then, the pion self-energy in nuclear matter up
to ${\cal O}(p^5)$  was determined together with some other contributions at
N$^2$LO. The cancellation found in ref.~\cite{nlou} between all leading
corrections to the linear density approximation for the pion self-energy is
explicitly shown here for the amplitudes calculated utilizing the non-perturbative
method developed here up to N$^2$LO,  which is a good check for the consistency of
our  approach. We have also addressed the calculation of the energy density  of
nuclear matter ${\cal E}$. The non-perturbative technique developed gives
rise to different contributions to ${\cal E}$ whose imaginary parts cancel
between each other and a real value results, as it should. We obtain
saturation in symmetric nuclear matter and repulsion for neutron matter. 
The contributions from the nucleon-nucleon interactions are of similar size 
to those from the kinetic energy term, the latter being suppressed by the 
inverse of the large nucleon mass and a dimensional factor. It is remarkable 
that we obtain for $g_0\simeq -0.62m_\pi^2$ a very good reproduction of sophisticated many-body calculations that employ 
realistic nucleon-nucleon potentials for the equation of state of neutron
matter up to rather high nuclear densities. We can also 
achieve such a good agreement for the case of nuclear matter by allowing to 
distinguish between $\widetilde{g}_0$ and $g_0$, where the former  parameter  
appears explicitly in the calculation of the energy per baryon while the
latter appears implicitly through the nucleon-nucleon scattering amplitudes 
involved. By this splitting, we can mock up the effects of higher orders.
The parameter $\widetilde{g}_0$ can be fixed from the neutron matter 
equation of state. We then obtain a very accurate reproduction to ${\cal E}$ 
as function of density  in symmetric nuclear matter for
$g_0=-0.97~m_\pi^2$, 
with saturation at $\rho=0.17$~fm$^{-3}$ and $E/A=-16$~MeV, cf. the
experimental values $\rho=0.166\pm 0.019$~fm$^{-3}$ and $E/A=-16\pm 1$~MeV. 
Furthermore, the nuclear matter incompressibility comes out with a value between 
240-250~MeV, in perfect agreement with the experimental one of $250\pm
25$~MeV.  We interpret the success of our reproduction of the nuclear matter 
equation of state for both symmetric and neutron matter in terms of one free
parameter for each of them as an indication that our power counting  is a
realistic one and that it is able to establish a useful hierarchy within 
the many contributions and complications inherent to nuclear dynamics. 
This opens the way to proceed systematically improving the calculations in a controlled way.

 Certainly, higher chiral orders should 
be worked out to obtain more precise results concerning the interesting
problems considered here. In particular, it would be desirable to solve exactly 
eq.~\eqref{disc.nji} in order to have free nucleon-nucleon partial wave amplitudes 
independent of $g_0$. In addition,  a N$^2$LO
calculation will address the interesting question about the importance of
three-nucleon forces for nuclear matter saturation. Similarly, a  N$^2$LO
calculation of the pion self-energy is a very interesting task. It will merge
important  meson-baryon mechanisms like e.g. the Ericson-Ericson-Pauli
rescattering effect \cite{ericeric}, with novel multi-nucleon contributions that
can be worked out systematically within our EFT.  More calculations and
applications of the present theory to other interesting physical problems
 should be pursued.

\section*{Acknowledgements}
We would like to thank  Andreas~Wirzba for discussions. 
This work is partially funded by the grant  MEC  FPA2007-6277, by 
BMBF grant 06BN9006,  EU-Research Infrastructure
Integrating Activity
 ``Study of Strongly Interacting Matter" (HadronPhysics2, grant n. 227431)
under the Seventh Framework Program of EU, HGF grant VH-VI-231 
(Virtual Institute ``Spin and strong QCD'') and by DFG (TR-16 ``Subnuclear
Structure of Matter'').


\appendix{}

\section{Partial wave decomposition of in-medium nucleon-nucleon amplitudes}
\label{app.pwd}
\def\theequation{\Alph{section}.\arabic{equation}}
\setcounter{equation}{0}

In this appendix, we derive the partial wave decomposition of the
nucleon-nucleon
scattering  amplitudes in the CM frame. Our states are normalized as,
\begin{align}
1-\hbox{particle state:~~~} & \langle \vp', j|\vp,
i\rangle=\delta_{ij}(2\pi)^3\delta(\vp'-\vp)\nn\\
2-\hbox{particle state:~~~} & \langle \vp',j_1 j_2|\vp,i_1 i_2\rangle=
\delta_{j_1 i_1}\delta_{j_2 i_2}(2\pi)^4\delta({\cal P}_f-{\cal P}_i)
\frac{(2\pi)^2 W}{p E_1 E_2}\delta(\Omega-\Omega')~.
\label{nor}
\end{align}
Here, ${\cal P}_f$ corresponds to the total four-momentum of the final 
state and ${\cal P}_i$ to that of the initial one, with $W={\cal P}_i^0={\cal
  P}_f^0$, the total CM energy. In addition, $E_1$ and $E_2$ are the energies 
of the particles 1 and 2, in order. The indices $i$ and $j$ refer to any 
discrete quantum number required to characterize the states. In the CM frame 
the solid angle is denoted by $\Omega$ and the modulus of the three-momentum by
$p=|\vp|$.
The two-particle states with well-defined orbital angular momentum 
are defined as,
\be
|\ell m,i_1i_2\rangle=\frac{1}{\sqrt{4\pi}}\int d\hat{\vp}\, 
Y_{\ell}^m(\hat{\vp})^* |\vp,i_1 i_2\ra~.
\ee
Taking into account eq.~(\ref{nor}) it follows then
\be
\langle \ell' m',j_1 j_2|\ell m,i_1 i_2\rangle=\frac{\pi W}{p E_1 E_2}
\delta_{\ell'\ell}\delta_{m'm}\delta_{j_1 i_1}\delta_{j_2 i_2}~. 
\ee
The decomposition into states with well-defined total angular momentum $J$, 
third component $\mu$, orbital angular momentum $\ell$ and total spin $S$ is
given by,
\be
|\vp, \sigma_1 \sigma_2\rangle=\sqrt{4\pi}\sum_{J,S,\ell,m}(\sigma_1 \sigma_2
s_3|s_1 s_2 S)
(m s_3 \mu|\ell S J)Y_\ell^m(\hat{\vp})^*|J\mu \ell S s_1 s_2\rangle~,
\label{first.sta}
\ee
where the indices  $\sigma_1$ and $\sigma_2$ refer to  the third components of
spin, $s_1$ and $s_2$ to their maximum values and $m$ to the third component 
of the orbital angular momentum $\ell$.\footnote{Strictly speaking in order to 
match with the normalization in eq.~(\ref{unitarity}) we should multiply the 
right-hand-side of eq.~(\ref{first.sta}) by $\sqrt{m/E}$. This is a
relativistic correction of ${\cal O}(p^2)$.} Next,  we introduce the isospin
indices $\alpha_1$, $\alpha_2$, for the third components,  $\tau_1$, $\tau_2$, 
for their maximum values, and decompose the free state in terms of states with 
well-defined total isospin $I$ and third component $i_3$. In addition, the
antisymmetric nature of a two-fermion (two-nucleon) state is introduced, 
\begin{align}
&\frac{1}{\sqrt{2}}\left(|\vp, \sigma_1 \alpha_1 \sigma_2 \alpha_2\rangle - 
|-\vp, \sigma_2 \alpha_2 \sigma_1
\alpha_1\rangle\right)=\frac{\sqrt{4\pi}}{\sqrt{2}}\sum\Bigl\{
(\si_1 \si_2 s_3|s_1 s_2 S)(m s_3 \mu|\ell S J)
(\alpha_1 \alpha_2 i_3|\tau_1 \tau_2 I)\nn\\
&\times Y_{\ell}^m(\hat{\vp})^*|J\mu \ell S s_3 I i_3\ra-
(\si_2 \si_1 s_3|s_2 s_1  S)(m s_3 \mu|\ell S J)(\alpha_2 \alpha_1 i_3|\tau_2
\tau_1  I)Y_{\ell}^m(-\hat{\vp})^*|J\mu \ell S s_3 I i_3\ra
\Bigr\}~,
\label{pwe.1}
\end{align}
with the repeated indices to be summed. This convention is used throughout 
this section. To simplify the notation we denote the left-hand-side of the 
previous equation as  $|\vp,\si_1 \alpha_1 \si_2 \alpha_2\ra_A$, 
with the subscript $A$ indicating that the state is antisymmetrized. 
Applying  the relations \cite{rosen} $Y_{\ell}^m(-\hat{\vp})
=(-1)^\ell Y_\ell^m(\hat{\vp})$, $(\si_2 \si_1 s_3|s_2 s_1 S)
=(-1)^{S-s_1-s_2}(\si_1 \si_2 s_3|s_1 s_2 S)$, and analogously for isospin, 
eq.~(\ref{pwe.1}) for the nucleon-nucleon case ($s_1=s_2=\tau_1=\tau_2=1/2$) 
simplifies to
\be
|\vp,\si_1\alpha_1\si_2\alpha_2\ra_A=\sqrt{4\pi}\sum_{J,S,\ell,m,I,i_3}
(\si_1\si_2 s_3|s_1 s_2 S)(m s_3 \mu|\ell S J) Y_\ell^m(\hat{\vp})^* \chi(S\ell I)|J\mu \ell S s_3 I i_3\ra~,
\label{pw.d}
\ee
with 
\be
\chi(S\ell I)=\frac{1-(-1)^{\ell+S+I}}{\sqrt{2}}=\left\{
\begin{array}{ll}
\sqrt{2} & \ell+S+I=\hbox{ odd}\\
0 & \ell+S+I=\hbox{ even}
\end{array}~.
\right.
\label{chi}
\ee
In this way, $\chi(S\ell I)$ ensures the well known rule that $S+\ell+I$ must be
 odd for any partial wave. Using the decomposition eq.~(\ref{pw.d}) we have for the
scattering amplitude,
\begin{align}
&_A\la \vp',\si_1'\alpha_1' \si_2'\alpha_2'|T(\val)|\vp,\si_1\alpha_1
\si_2\alpha_2 \ra_A=4\pi\sum 
(\si_1' \si_2' s_3'|s_1 s_2 S')(m' s_3'\mu'|\ell' S' J')(\sigma_1 \si_2 s_3|s_1
s_2 S)(m s_3 \mu|\ell S J)
\nn\\
&\times (\alpha_1' \alpha_2' i_3|\tau_1 \tau_2 I) (\alpha_1 \alpha_2 i_3|\tau_1
\tau_2 I) Y_{\ell'}^{m'}(\hat{\vp}') Y_{\ell}^m(\hat{\vp})^* \chi(S'\ell'
I)\chi(S \ell I) T_{J'JI}(\ell' S';\ell S)~.
\label{pw.d2}
\end{align}
Here, $T_{J'J I}(\ell' S';\ell S)$ is the partial wave with final total
angular momentum $J'$, initial one $J$, final total spin $S'$, initial one
$S$, isospin $I$ and final and initial orbital angular momenta $\ell'$ and 
$\ell$, respectively. Notice that in the previous equation we have
distinguished between the final and initial total angular momenta  $J'$ and 
$J$, and similarly for the total spins $S'$ and $S$. For free two nucleon
scattering we have of course $J'=J$ because of angular momentum conservation. 
This conservation law, the conservation  of parity and the rule
$S+\ell+I=\,$odd imply that $S'=S$. However, the resulting matrix elements 
in the nuclear medium depend additionally on the total three-momentum of the
two nucleons because the medium rest-frame does not coincide in general with 
their center-of-mass. This is why we have included $\val$ as an
argument in the scattering operator, with the former defined in
\eqref{change.variables}.  Employing the orthogonality properties of the
Clebsch-Gordan 
coefficients and spherical harmonics, one can invert eq.~(\ref{pw.d2}) with the
result,
\begin{align}
&4\pi \chi(S'\ell' I) \chi(S\ell I) T_{J'JI}(\ell' S';\ell S)=
\sum \int d\hat{\vp}' \int d\hat{\vp} \,_A\la
\vp',\si_1'\alpha_1'\si_2'\alpha_2'|T(\val)|\vp,
\si_1\alpha_1\si_2\alpha_2\ra_A 
(\si_1'\si_2's_3'|s_1 s_2 S')\nn\\
&\times
(m' s_3' \mu'|\ell'S' J')(\si_1\si_2 s_3|s_1 s_2 S) (m s_3 \mu|\ell S  J)
(\alpha_1'\alpha_2' i_3|\tau_1\tau_2 I)(\alpha_1\alpha_2 i_3|\tau_1\tau_2 I)
Y_{\ell'}^{m'}(\hat{\vp}')^* Y_\ell^m(\hat{\vp})~.\nn\\
\label{pw.d2.i1}
\end{align}
This expression can be further reduced by making use of properties under 
rotational invariance so that the initial relative three-momentum $\vp$ can 
be taken parallel to the $\hat{\vz}$-axis. In deriving this simplification 
we omit the isospin indices that do not play any role in the following 
considerations, and introduce the symbol
\be
T_{\si_1'\alpha_1'\si_2'\alpha_2'}^{\si_1\alpha_1\si_2\alpha_2}(\vp',\vp,\vec{
\alpha})=\, 
_A\la
\vp',\si_1'\alpha_1'\si_2'\alpha_2'|T(\val)|\vp,
\si_1\alpha_1\si_2\alpha_2\ra_A~.
\ee

Now consider the rotation $R(\hat{\vp})$, such that
$R(\hat{\vp})\hat{\vz}=\hat{\vp}$, that consists first of a rotation around the
$y$-axis with an angle $\theta$ and then a rotation around the $z$-axis with
an angle $\phi$, where $\theta$ and $\phi$ are the polar and azimuthal angles of  $\hat{\vp}$, respectively. We could also have taken first an arbitrary
rotation of angle $\gamma$ around the $\vz$-axis. Then,
\begin{align}
R(\hat{\vp})^\dagger|\vp, \si_1
\si_2;\val\ra&=\sum_{\bar{s}_1,\bar{s}_2}D_{\bar{s}_1\si_1}^{(1/2)}(R^\dagger)
D_{\bar{s}_2\si_2}^{(1/2)}(R^\dagger)|p\hat{\vz},\bar{s}_1\bar{s}_2;\val
''\ra~,\nn\\
R(\hat{\vp})^\dagger|\vp', \si_1'
\si_2';\val\ra&=\sum_{\bar{s}'_1,\bar{s}'_2}D_{\bar{s}'_1
\si'_1}^{(1/2)}(R^\dagger)
D_{\bar{s}'_2\si'_2}^{(1/2)}(R^\dagger)|\vp'',\bar{s}'_1\bar{s}'_2;\val
''\ra~,
\label{rp}
\end{align}
with $\vp''=R(\hat{\vp})^{-1}\vp'$ and
$\val''=R(\hat{\vp})^{-1}\val$. 
The dependence on the total three-momentum has been made explicit in the 
state vectors to emphasize that the total three-momentum also is rotated.  
Inserting eq.~(\ref{rp}) into  eq.~(\ref{pw.d2.i1}) we have,
\begin{align}
&4\pi \chi(S'\ell' I) \chi(S\ell I) T_{J'JI}(\ell' S';\ell S)\!=
\!\sum \int \! d\hat{\vp}'\int d\hat{\vp}
\,T_{\bar{s}'_1\bar{s}'_2}^{\bar{s}_1\bar{s}_2}(\vp'',p\hat{\vz},\val'')
D_{\bar{s}'_1 \si'_1}^{(1/2)}(R^\dagger)^*
D_{\bar{s}'_2\si'_2}^{(1/2)}(R^\dagger)^*
D_{\bar{s}_1 \si_1}^{(1/2)}(R^\dagger)
D_{\bar{s}_2\si_2}^{(1/2)}(R^\dagger)\nn\\
&\times Y_{\ell'}^{m'}(\hat{\vp}')^*  Y_\ell^m(\hat{\vp}) (\si_1'\si_2's_3'|s_1
s_2 S')
(m' s_3' \mu'|\ell'S' J')(\si_1\si_2 s_3|s_1 s_2 S)(m s_3 \mu|\ell S J)~.
\label{pwd.in3}
\end{align}
The spherical harmonics satisfy the following transformation properties under
rotations,
\begin{align}
Y_{\ell'}^{m'}(\hat{\vp}')&= \sum_{\bar{m}'} D_{\bar{m}'
m'}^{(\ell')}(R^\dagger)Y_{\ell'}^{\bar{m}'}(\hat{\vp}'')~,\nn\\ 
Y_{\ell}^m(\hat{\vp})&=\sum_{\bar{m}}
D_{\bar{m}m}^{(\ell)}(R^\dagger)Y_{\ell}^{\bar{m}}(\hat{\vz})~.
\end{align}
Inserting these equalities into eq.~(\ref{pwd.in3}) we are then left with the 
following product of rotation matrices,
\begin{align}
D_{\bar{s}'_1 \si'_1}^{(1/2)}(R^\dagger)^*
D_{\bar{s}'_2\si'_2}^{(1/2)}(R^\dagger)^*
D_{\bar{m}'m'}^{(\ell')}(R^\dagger)^*
D_{\bar{s}_1 \si_1}^{(1/2)}(R^\dagger)
D_{\bar{s}_2\si_2}^{(1/2)}(R^\dagger)
D_{\bar{m} m}^{(\ell)}(R^\dagger)~.
\label{prodm}
\end{align}
 We now take into account the Clebsch-Gordan composition of the rotation
matrices~\cite{rosen},
\be
\sum_{M'}D_{M'M}^{(L)}(R)(m'_1 m'_2 M'|\ell_1\ell_2 L)=
\sum_{m_1,m_2}D_{m'_1 m_1}^{(\ell_1)}(R)D_{m'_2 m_2}^{(\ell_2)}(R) (m_1 m_2
M|\ell_1 \ell_2 L) ~.
\ee
Since eq.~(\ref{prodm}) appears in eq.~(\ref{pwd.in3}) times Clebsch-Gordan
coefficients 
we can make use of the previous composition repeatedly.  First,
\begin{align}
&\sum D_{\bar{s}'_1 \si'_1}^{(1/2)}(R^\dagger) D_{\bar{s}'_2
\si'_2}^{(1/2)}(R^\dagger) 
(\si'_1 \si'_2 s'_3|s_1 s_2 S')=\sum D_{\bar{\si}'_3
s'_3}^{(S')}(R^\dagger)(\bar{s}'_1 \bar{s}'_2 
\bar{\si}'_3|s_1 s_2 S')~,\nn\\
&\sum D_{\bar{s}_1 \si_1}^{(1/2)}(R^\dagger)
D_{\bar{s}_2\si_2}^{(1/2)}(R^\dagger)(\si_1\si_2 s_3|s_1 s_2 S)=\sum
D_{\bar{\sigma}_3 s_3}^{(S)}(R^\dagger)(\bar{s}_1\bar{s}_2\bar{\sigma}_3|s_1 s_2
S)~.
\label{ap.rel.1}
\end{align}
The rotation matrix $D^{(S')}_{\bar{\si}'_3 s'_3}$, that appears on the
right-hand-side of the first of the previous equalities, can then be combined 
in eq.~(\ref{pwd.in3}) such that 
\be
\sum 
D_{\bar{\sigma}'_3 s_3'}^{(S')}(R^\dagger)D_{\bar{m}'m'}^{(\ell')}(R^\dagger)(m'
s'_3 \mu'|\ell' S' J')=
\sum
D_{\bar{\mu}' \mu'}^{(J')}(R^\dagger)(\bar{m}'\bar{\sigma}'_3 \bar{\mu}'|\ell'
S' J')~.
\label{ap.rel.2}
\ee
Similarly
\begin{align}
\sum 
D_{\bar{\sigma}_3 s_3}^{(S)}(R^\dagger)D_{\bar{m}m}^{(\ell)}(R^\dagger)(m s_3
\mu|\ell S J)=
\sum
D_{\bar{\mu} \mu}^{(J)}(R^\dagger)(\bar{m}\bar{\sigma}_3 \bar{\mu}|\ell S J)~.
\label{app.rel.2}
\end{align}
Incorporating eqs.~(\ref{ap.rel.2}) and (\ref{app.rel.2}) in
eq.~(\ref{pwd.in3}), the latter takes the form
\begin{align}
&4\pi \chi(S'\ell' I) \chi(S\ell I) T_{J'JI}(\ell' S';\ell S)=
\sum \int \! d\hat{\vp}' \int d\hat{\vp}
\,T_{\si_1'\si_2'}^{\si_1\si_2}(\vp'',p\vz,\val'')
Y_{\ell'}^{\bar{m}'}(\hat{\vp}'')^* Y_\ell^{\bar{m}}(\hat{\vz}) 
D_{\bar{\mu}'\mu'}^{(J')}(R^\dagger) D_{\bar{\mu}\mu}^{(J)}(R^\dagger) \nn\\
&\times (\bar{m}'\bar{\si}'_3\bar{\mu}'|\ell' S' J') 
(\bar{s}'_1 \bar{s}'_2 \bar{\si}'_3|s_1 s_2 S')(\bar{m}\bar{\si}_3\bar{\mu}|\ell
S J) 
(\bar{s}_1 \bar{s}_2 \bar{\si}_3|s_1 s_2 S)~.
\label{dev.1}
\end{align}

Let us first consider the vacuum case where the scattering amplitude does not 
depend on $\val$. In this way the integration over $\hat{\vp}$ in the 
previous equation can be done explicitly taking into account the orthogonality 
relation between two rotation matrices \cite{rosen}. For that let us recall
our previous remark about the fact that an arbitrary initial rotation over 
the $\vz$-axis and angle $\gamma$ can also be included. In this way we take 
\begin{align}
\frac{1}{2\pi}\int_0^{2\pi}d\gamma\int d\hat{\vp} 
D_{\bar{\mu}'\mu'}^{(J')}(R^\dagger)^*
D_{\bar{\mu}\mu}^{(J)}(R^\dagger)=\frac{4\pi}{2J+1}
\delta_{\bar{\mu}'\bar{\mu}}
\delta_{\mu'\mu}
\delta_{JJ'}~.
\end{align}
Inserting this back to eq.~(\ref{dev.1}) one arrives at
\begin{align}
&\chi(S\ell' I) \chi(S\ell I) T_{JI}(\ell',\ell,S)=
\frac{Y_\ell^0(\hat{\vz})\delta_{JJ'}\delta_{\mu'\mu}}{2J+1}\sum\int
d\hat{\vp}''
T_{\bar{s}'_1\bar{s}'_2}^{\bar{s}_1\bar{s}_2}(\vp'',p\vz)Y_{\ell'}^{\bar{m}'}
(\hat{\vp}'')(\bar{m}'\bar{\si}'_3\bar{\si}_3|\ell'SJ)(0\bar{\si}_3\bar{\si}
_3|\ell S J)\nn\\
&\times (\bar{s}'_1\bar{s}'_2\bar{\si}'_3|s_1s_2
S)(\bar{s}_1\bar{s}_2\bar{\si}_3|s_1s_2 S)~.
\label{lpw.exp.def}
\end{align}
In this expression we have made use that only $\bar{m}=0$ gives a contribution
to $Y_\ell^{\bar{m}}(\hat{\vz})$ and, as explained after eq.~(\ref{pw.d2}),
$S'=S$. In addition, we have also used that $d\hat{\vp}'=d\hat{\vp}''$, since 
both vectors are related by a rotation. The subscript $J'$ in $T_{J'J I}$ is 
suppressed because $J'=J$ and it is redundant. Also,  we have employed the
notation for the partial waves of section~\ref{sec:nn-int},
$T_{JI}(\ell',\ell,S)$.

We now come back to the in-medium case and keep the dependence on
$\val$. Here also $\bar{m}=0$ so that $\bar{\mu}=\bar{\sigma}_3$. 
Let us show first that a Fermi-sea with all the free three-momentum states
filled up to $\xi$ has total spin zero. This is required because for a given 
three-momentum $\vp_1$ one has two spin states that must be combined
antisymmetrically because of the  Fermi statistics so that $S=0$ for this
pair. Then, since this happens for any pair, the total spin of the Fermi 
sea must be zero. Regarding total angular momentum  we now give a
non-relativistic argument to claim that the orbital angular momentum must also
be zero.  This is due to the fact that the nuclear medium in the CM of the two 
nucleons that scatter is seen with a velocity parallel to $-\val$. 
In this way, both the CM position vector  and the total three-momentum of the 
nuclear medium are also parallel so that their cross product vanishes. As a 
result, since the intrinsic orbital angular momentum of the medium is also
zero, one expects that the total angular momentum is zero for the system 
also in the CM frame of the two nucleons. Thus, $J'=J$ also in this case and
then, because of the same reasons as in vacuum, $S'=S$. Let us recall the 
remark after eq.~(\ref{sum.mat}) to justify that $I$ is conserved also in 
the nuclear medium. In addition the third component of total angular momentum 
must be conserved, $\mu=\mu'$, and summing over $\mu$ one has
\begin{align}
\frac{1}{2J+1} \sum_\mu
D_{\bar{\mu}'\mu}^{(J)}(R^\dagger)^*D^{(J)}_{\bar{\mu}\mu}(R)=
\frac{\delta_{\bar{\mu}'\bar{\mu}}}{2J+1}~,
\end{align}
given the unitary character of the rotation matrices. Then,
\begin{align}
& \chi(S\ell' I)\chi(S \ell
I)T_{JI}(\ell',\ell,S)=\frac{Y_\ell^0(\hat{\vz})}{4\pi(2J+1)}\sum \int
d\hat{\mathbf{a}}'' \int d\hat{\vp}''
T_{\bar{s}'_1\bar{s}'_2}^{\bar{s}_1\bar{s}_2}(\vp'',p\hat{z},\val
'')(\bar{s}'_1\bar{s}'_2\bar{\sigma}'_3|s_1s_2S)(\bar{s}_1\bar{s}_2\bar{\si}_3|s_1 s_2 S)\nn\\
&\times Y_{\ell'}^{m'}(\hat{\vp}'')^*(\bar{m}'\bar{\si}'_3\bar{\si}_3|\ell'S
J)(0\bar{\si}_3\bar{\si}_3|\ell S J)~.
\label{lpw.exp2.def}
\end{align}
This expression reduces to the one in the vacuum, eq.~(\ref{pw.exp.def}), 
whenever the integral
\begin{align}
\sum\int d\hat{\vp}''
T_{\bar{s}'_1\bar{s}'_2}^{\bar{s}_1\bar{s}_2}(\vp'',p\hat{z},\val
'')(\bar{s}'_1\bar{s}'_2\bar{\sigma}'_3|s_1s_2S)(\bar{s}_1\bar{s}_2\bar{\si}_3|s_1 s_2 S)
Y_{\ell'}^{m'}(\hat{\vp}'')^*(\bar{m}'\bar{\si}'_3\bar{\si}_3|\ell'S
J)(0\bar{\si}_3\bar{\si}_3|\ell S J)
\label{operation}
\end{align} 
does not dependent on $\hat{\mathbf{a}}$. 
In that case the integration over $d\hat{\mathbf{a}}''$ is simply 
$4\pi$ and eq.~(\ref{lpw.exp.def}) is restored.

Eq.~(\ref{lpw.exp2.def}) can be further simplified because for the evaluation
of a nucleon-nucleon partial wave amplitude one only needs to consider the 
direct term in the nucleon-nucleon scattering amplitude. This follows because 
the operator $T$ is Bose-symmetric so that,
\be
T_{\si'_1\alpha'_1\si'_2\alpha'_2}^{\si_1\alpha_1\si_2\alpha_2}(\vp',\vp,\val)=
\la
\vp',\si'_1\alpha'_1\si'_2\alpha'_2|T(\val)|\vp,
\si_1\alpha_1\si_2\alpha_2\ra-
\la
-\vp',\si'_2\alpha'_2\si'_1\alpha'_1|T(\val)|\vp,
\si_1\alpha_1\si_2\alpha_2\ra~.
\label{fermi.1}
\ee
When implementing the second or exchange term  in eq.~(\ref{lpw.exp2.def}),
re-including the isospin indices as well, and using the above referred
symmetry properties of the Clebsch-Gordan coefficients and spherical
harmonics, one is left with the same expression as for the direct term in 
eq.~(\ref{fermi.1}) except for the global sign $-(-1)^{S+\ell'+I}$. Summing 
both expressions  the factor 
\be
1-(-1)^{S+\ell'+I}
\ee
arises.  Given the definition of $\chi(S\ell I)$ in eq.~(\ref{chi}) and
imposing the rule that $\ell+S+I=\,$odd and 
$\ell'+S+I=$~odd, the factor  $\chi(S\ell I)\chi(S\ell'I)$ can be simplified  
on both sides of eq.~(\ref{lpw.exp2.def}). The latter then reads 
\begin{align}
&T_{JI}^{i_3}(\ell',\ell,S)=\frac{Y_\ell^0(\hat{\vz})}{4\pi(2J+1)}\sum
(\si_1'\si_2's'_3|s_1 s_2 S)
(\si_1\si_2 s_3|s_1 s_2 S)(0 s_3 s_3|\ell S J)(m' s'_3 s_3|\ell' S J)\nn\\
&\times(\alpha'_1\alpha'_2 i_3|\tau_1\tau_2 I)(\alpha_1\alpha_2 i_3|\tau_1\tau_2
I)
\int d\hat{\mathbf{a}} \int d\hat{\vp'}\,
\la \vp',\si'_1 \alpha'_1 \si'_2 \alpha'_2 |T_d(\val)| p\hat{\vz},
\si_1\alpha_1 \si_2\alpha_2\ra
 Y_{\ell'}^{m'}(\vp')^*~,
\label{pw.mexp.def}
\end{align}
with only the direct term, as indicated by the subscript $d$ in the
scattering operator. For the particular case of the vacuum nucleon-nucleon 
scattering the previous expression simplifies to
\begin{align}
&T_{JI}(\ell',\ell,S)=\frac{Y_\ell^0(\hat{\vz})}{2J+1}\sum (\si_1'\si_2's'_3|s_1
s_2 S)
(\si_1\si_2 s_3|s_1 s_2 S)(0 s_3 s_3|\ell S J)(m' s'_3 s_3|\ell' S J)\nn\\
&\times(\alpha'_1\alpha'_2 i_3|\tau_1\tau_2 I)(\alpha_1\alpha_2 i_3|\tau_1\tau_2
I)
 \int d\hat{\vp'}\,
\la \vp',\si'_1 \alpha'_1 \si'_2 \alpha'_2 |T_d| p\hat{\vz}, \si_1\alpha_1
\si_2\alpha_2\ra
 Y_{\ell'}^{m'}(\vp')^*~.
\label{pw.exp.def}
\end{align}

\section{Explicit calculation of  $\Pi_{9}$ and $\Pi_{10}$  up to ${\cal
O}(p^6)$}
\label{app:explicit}
\def\theequation{\Alph{section}.\arabic{equation}}
\setcounter{equation}{0}

In this section we evaluate explicitly  $DL_{JI}^{(1)}$ and $L_{JI}^{(1)}$ 
since they enter for fixing $N_{JI}^{(1)}$ and $\xi_{JI}^{(1)}$,
eqs.~\eqref{eq.a1} and \eqref{fix.xi1}, respectively. We also show  with
explicit calculations some
of the steps introduced in the derivations of sections \ref{sec:sigma8} and
\ref{sec:sig.10}. 
For the calculation of these diagrams one has to recall  that the exchange of
a wiggly line corresponds to local plus one-pion exchange terms,
fig.~\ref{fig:wig}. When the nucleon loop, to which the two pion lines are
attached, includes only nucleon-nucleon local vertices, we then have $T_{10}$
and $T_{11}$ for the isovector and isoscalar cases, respectively. At the same
order, when one of the nucleon-nucleon vertices in this loop corresponds to a 
one-pion exchange then $T_{12}$ and $T_{13}$ result. Finally, when both
vertices are due to one-pion exchange  one has $T_{14}$ and $T_{15}$.

\begin{figure}[ht]
\psfrag{q}{$q$}
\psfrag{m}{$m$}
\psfrag{l}{$\ell$}
\psfrag{mp}{$m'$}
\psfrag{lp}{$\ell'$}
\psfrag{i}{$i$}
\psfrag{j}{$j$}
\psfrag{o}{$o$}
\psfrag{op}{$o'$}
\psfrag{t}{$t$}
\centerline{\epsfig{file=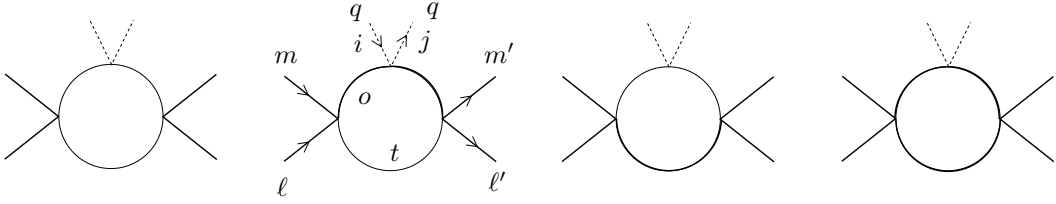,width=.8\textwidth,angle=0}}
\vspace{0.2cm}
\caption[pilf]{\protect \small
The two-nucleon reducible loop with only local vertices. The free part of the
in-medium nucleon propagator in eq.~(\ref{nuc.pro}) is indicated by a thin 
line while the in-medium part, proportional to the Dirac delta-function, is 
denoted by a thick line.
\label{fig:loop10}}
\end{figure} 

We first consider the two-nucleon reducible loop with only local vertices,
fig.~\ref{fig:loop10}. 
For the isovector case the derivative with respect to $z$ from
eq.~(\ref{pro.squ}) acts on a nucleon propagator not entering in 
eq.~(\ref{wt.mod}). 
With the four--nucleon local vertex of eq.~(\ref{feynman}) we then have for
$T_{10}$, 
\begin{align}
T_{10}&=-\frac{ \kappa  q^0\varepsilon_{ijk}}{2f^2}
\Bigl\{C_S\left(
\delta_{\alpha_{m'}\alpha}\delta_{\alpha_{\ell'}\beta}\delta_{m'o}\delta_{\ell't
}-
\delta_{\alpha_{m'}\beta}\delta_{\alpha_{\ell'}\alpha}\delta_{m't}\delta_{\ell'o
}
\right)
+C_T\left(
\vec{\sigma}_{\alpha_{m'}\alpha}\vec{\sigma}_{\alpha_{\ell'}\beta}\delta_{m'o}
\delta_{\ell't}
\right.\nn\\
&\left.-\vec{\sigma}_{\alpha_{m'}\beta}\vec{\sigma}_{\alpha_{\ell'}\alpha}
\delta_{m't}\delta_{\ell'o}
\right)\Bigr\}
\tau^k_{oo}\frac{\partial}{\partial p_1^0}\int\frac{d^4k}{(2\pi)^4} G_0(p_1-k)_o
G_0(p_2+k)_t \times 
\Bigl\{C_S\left(
\delta_{\alpha_{m}\alpha}\delta_{\alpha_{\ell}\beta}\delta_{m o}\delta_{\ell
t}\right. \nn\\
&\left. -
\delta_{\alpha_{m}\beta}\delta_{\alpha_{\ell}\alpha}\delta_{m t}\delta_{\ell o}
\right)+C_T\left(
\vec{\sigma}_{\alpha\alpha_m}\vec{\sigma}_{\beta\alpha_\ell}\delta_{m o
}\delta_{\ell t}
-\vec{\sigma}_{\beta\alpha_m}\vec{\sigma}_{\alpha \alpha_\ell}\delta_{m
t}\delta_{\ell o}
\right)\Bigr\}~,
\label{t10.1a}
\end{align}
where $
\kappa=1-g_A^2 \vq^2/q_0^2$. The spin indices are 
indicated with Greek letters.  The momentum integration is the same 
as for the function $L_{10}$, eq.~(\ref{l10.def}), so that a factor 
$\partial L_{10}^{i_3}/\partial A$ results. The previous  
equation contains both the direct and exchange terms, though the direct term 
is the one needed to evaluate the different partial waves. Notice that the 
previous equation only contributes to the S-waves. We now work out the spin
projections of the direct term in eq.~(\ref{t10.1a}) with the result
\begin{align}
\begin{array}{rrr}
&S=0 & S=1\\
\delta_{\alpha_{m'}\alpha_m}\delta_{\alpha_{\ell'}\alpha_\ell}& 1 & 1\\
\vec{\sigma}_{\alpha_{m'}\alpha_m}\vec{\sigma}_{\alpha_{\ell'}\alpha_{\ell}}& -3
& 1\\
\left(\vec{\sigma}_{\alpha_{m'}\alpha}\vec{\sigma}_{\alpha_{\ell'}\beta}\right)
\left(\vec{\sigma}_{\alpha\alpha_m}\vec{\sigma}_{\beta\alpha_\ell}\right)& 9& 1
\end{array}~.
\label{spin.pro}
\end{align}
As it should,  we have then the combinations $(C_S-3C_T)^2$ and
$(C_S+C_T)^2$ for $S=0$ and 1, respectively.  The isospin projection
corresponding to the operator  $(\delta_{\ell'\ell}\tau^3_{m'm}
+\delta_{m'm}\tau^3_{\ell'\ell})$ is $2i_3$, which excludes $I=0$ altogether. 
Keeping only the direct term in eq.~\eqref{t10.1a} , we have
\begin{align}
T_{10,d}^{i_3}&=i_3\frac{i\kappa m q^0  \epsilon_{ij3}}{f^2} 
(C_S-3C_T)^2\frac{\partial L_{10}^{i_3}}{\partial A}~.
\label{t10.3}
\end{align}
For the isoscalar contribution $T_{11,d}$ we just have to remove the
derivative $m\partial/\partial A$ and replace the vertex of
eq.~(\ref{ver:wte}) by eq.~(\ref{ver:iss}).  The spin projection is the same 
as before, eq.~(\ref{spin.pro}). However, the isospin operator now is
different and for the direct term just corresponds to twice the identity
operator.   As a result 
 \begin{align}
T_{11,d}^{i_3}&=\frac{g_A^2 \vq^2 }{f^2 q_0^2} (C_S+(4S-3)C_T)^2  L_{10}^{i_3}
 ~,
\label{t11.3}
\end{align}
and both $S=0$ and 1 contribute.  In $T_{10}$ and $T_{11}$ there is no
dependence on  $\hat{\mathbf{a}}$, they depend only on $|\val|$, so that one can
calculate the projection into the S-waves as in vacuum, making use of
eq.~\eqref{pw.exp.def}.

\begin{figure}[ht]
\psfrag{q}{{\tiny $q$}}
\psfrag{m}{{\tiny $m$}}
\psfrag{l}{{\tiny $\ell$}}
\psfrag{mp}{{\tiny $m'$}}
\psfrag{lp}{{\tiny $\ell'$}}
\psfrag{i}{{\tiny $i$}}
\psfrag{j}{{\tiny $j$}}
\psfrag{o}{{\tiny $o$}}
\psfrag{op}{{\tiny $o'$}}
\psfrag{t}{{\tiny $t$}}
\psfrag{p1}{{\tiny $p_1$}}
\psfrag{p2}{{\tiny $p_2$}}
\psfrag{pp1}{{\tiny $p'_1$}}
\psfrag{pp2}{{\tiny $p'_2$}}
\psfrag{pm}{\tiny{$p_1-k$}}
\psfrag{pp}{\tiny{$p_2+k$}}
\psfrag{r}{{\tiny $r$}}
\centerline{\epsfig{file=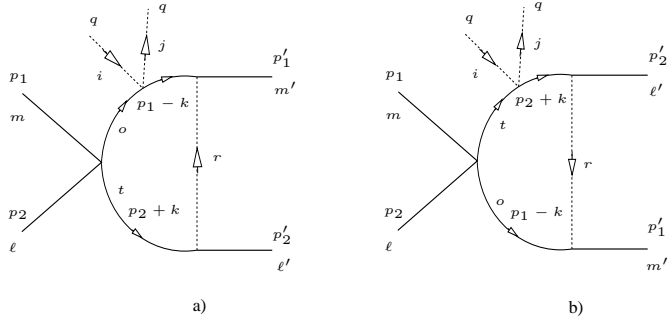,width=.5\textwidth,angle=0}}
\vspace{0.2cm}
\caption[pilf]{\protect \small
The internal four-momenta and discrete indices are indicated on the two
diagrams. The figure on the left corresponds to $T_{12}^{f,a}$ and 
the one on the right to $T_{12}^{f,b}$.  
\label{fig:self12fb}}
\end{figure} 

\begin{figure}[ht]
\psfrag{q}{{\tiny $q$}}
\psfrag{m}{{\tiny $m$}}
\psfrag{l}{{\tiny $\ell$}}
\psfrag{mp}{{\tiny $m'$}}
\psfrag{lp}{{\tiny $\ell'$}}
\psfrag{i}{{\tiny $i$}}
\psfrag{j}{{\tiny $j$}}
\psfrag{o}{{\tiny $o$}}
\psfrag{op}{{\tiny $o'$}}
\psfrag{t}{{\tiny $t$}}
\psfrag{p1}{{\tiny $p_1$}}
\psfrag{p2}{{\tiny $p_2$}}
\psfrag{pp1}{{\tiny $p'_1$}}
\psfrag{pp2}{{\tiny $p'_2$}}
\psfrag{pm}{{\tiny{$p_1-k$}}}
\psfrag{pp}{{\tiny{$p_2+k$}}}
\psfrag{k}{{\tiny $k$}}
\centerline{\epsfig{file=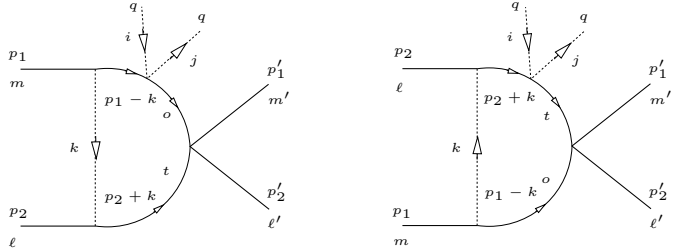,width=.5\textwidth,angle=0}}
\vspace{0.2cm}
\caption[pilf]{\protect \small
The internal four-momenta and discrete indices are shown in the figure for
$T_{12}^i$. 
\label{fig:self12ib}}
\end{figure}

We now consider  $T_{12}$ and $T_{13}$, with one local and one-pion exchange 
vertices. The isovector contribution corresponds to fig.~\ref{fig:self12fb}, 
$T_{12}^f$, and  fig.~\ref{fig:self12ib}, $T_{12}^i$, depending on whether the 
one-pion exchange is between the final or initial two nucleons, respectively.
For the sum of the diagrams in fig.~\ref{fig:self12fb} one has
\begin{align}
T_{12}^{f}&=\frac{\kappa
q^0\varepsilon_{ij3}}{2f^2}\left(\frac{g_A}{2f}\right)^2
\frac{m\partial}{\partial A}\int\frac{d^4 k}{(2\pi)^4}\Bigl\{
\left[
C_S(\vsi_{\alpha_{m'}\alpha_m}\cdot \vr)(\vsi_{\alpha_{\ell'}\alpha_\ell}\cdot
\vr)+
C_T(\vsi_{\alpha_{m'}\alpha}\cdot\vr)(\vsi_{\alpha_{\ell'}\beta}\cdot \vr)
(\vsi_{\alpha\alpha_m}\cdot\vsi_{\beta\alpha_{\ell}})
\right]\nn\\
&\times \tau^a_{m'm}\tau^a_{\ell'\ell}
-
\left[
C_S(\vsi_{\alpha_{m'}\alpha_\ell}\cdot \vr)(\vsi_{\alpha_{\ell'}\alpha_m}\cdot
\vr)+
C_T(\vsi_{\alpha_{m'}\alpha}\cdot\vr)(\vsi_{\alpha_{\ell'}\beta}\cdot 
\vr)(\vsi_{\alpha\alpha_\ell}\cdot\vsi_{\beta\alpha_{m}})
\right]\tau^a_{m'\ell}\tau^a_{\ell'm}
\Bigr\}\nn\\
&\times\left(\tau^3_{m m}
+\tau^3_{\ell \ell}\right)G_0(p_1-k)_m G_0(p_2+k)_\ell\frac{1}{\vr^2+m_\pi^2}~,
\label{t11f}
\end{align}
where the repeated indices are summed and $r=p_1'-p_1+k$.
Instead of keeping $G_0(p_1-k)_o^2$ we  take\\ $-\left.\partial
G_0(p_1-k+\lambda)_o/\partial z\right|_{z=0}$, with $\lambda=(z,\mathbf{0})$, as done 
in eq.~(\ref{pro.squ}). On the other hand, for the pion propagator we neglect 
its dependence on $r_0^2$, since it is
${\cal O}(p^4)$, while $\vr^2$ is ${\cal O}(p^2)$. Then, the energy dependence
enters in $T_{12}^f$ similarly as in $L_{10}$ and the derivative can be taken
with
respect to the variable $A$, eq.~(\ref{da.l10}), with $A\to \vp^2$ after the 
derivative is performed. Let us consider the isospin and spin projections for 
the direct term, given by  the first square bracket in the previous equation. 
The isospin operator  
$\tau^a_{m'm}\tau^a_{\ell'\ell}(\tau^3_{mm}+\tau^3_{\ell\ell})$ has a
projection between states of well defined isospin given by $2i_3$.  
The spin operator, after some algebraic manipulation in the term 
proportional to $C_T$, reads 
\begin{align}
(C_S+C_T)(\vsi_{\alpha_{m'}\alpha_m}\cdot
\vr)(\vsi_{\alpha_{\ell'}\alpha_\ell}\cdot 
\vr)+C_T\, \vr^2\left(\delta_{\alpha_{m'}\alpha_m}
\delta_{\alpha_{\ell'}\alpha_\ell}-\vsi_{\alpha_{m'}\alpha_m}
\cdot\vsi_{\alpha_{\ell'}\alpha_\ell}\right)~.
\label{newform}
\end{align}
The structures
$\delta_{\alpha_{m'}\alpha_m}\delta_{\alpha_{\ell'}\alpha_\ell}$ 
and $\vsi_{\alpha_{m'}\alpha_m}\cdot\vsi_{\alpha_{\ell'}\alpha_\ell}$ are 
already projected in eq.~(\ref{spin.pro}) for the different spin states. 
There is, however, the new structure $(\vsi_{\alpha_{m'}\alpha_m}\cdot \vr)
(\vsi_{\alpha_{\ell'}\alpha_\ell}\cdot \vr)$ whose matrix elements  are 
\begin{align}
S&=0~,~-\vr^2~,\nn\\ 
S&=1\nn\\
||B_{s'_3 s_3}||&=
\left(\begin{array}{r|rrr}
   & -1    & 0& +1\\
   \hline
-1 & r_3^2 & -\sqrt{2}r_3(r_1+ir_2) & (r_1+ir_2)^2\\ 
0  & -\sqrt{2}r_3(r_1-ir_2) & \vr^2-2r_3^2 & \sqrt{2}r_3(r_1+ir_2)\\
+1 & (r_1-ir_2)^2& \sqrt{2}r_3 (r_1-ir_2) & r_3^2
\end{array}\right)~.
\label{s1mat}
\end{align}
In this matrix the rows correspond to the final third component of 
the total spin, $s'_3$, and the columns to the initial one, $s_3$. 
Then, we can write for the direct term with total spin $S$ 
\begin{align}
T_{12,d}^{f,S=0}&=-i_3\frac{\kappa
q^0\varepsilon_{ij3}}{f^2}\left(\frac{g_A}{2f}\right)^2
(C_S-3C_T)\frac{m\partial}{\partial A}\int
\frac{d^4k}{(2\pi)^4}G_0(p_1-k)_m
G_0(p_2+k)_\ell\frac{\vr^2}{\vr^2+m_\pi^2}~,\nn\\
T_{12,d}^{f,S=1}(s'_3,s_3)&=i_3\frac{\kappa
q^0\varepsilon_{ij3}}{f^2}\left(\frac{g_A}{2f}\right)^2(C_S+C_T)
\frac{m\partial}{\partial
A}\int\frac{d^4k}{(2\pi)^4}G_0(p_1-k)_m G_0(p_2+k)_\ell\frac{1}{\vr^2+m_\pi^2}
B_{s'_3\,s_3}~,
\label{t11fs1}
\end{align}
using eqs.~(\ref{t11f}) and (\ref{s1mat}). In the integrals of
eqs.~(\ref{t11fs1}) it is convenient to perform the shift of the 
integration variable
\be
k\to \frac{p_1-p_2}{2}+k=p+k~,
\label{shq}
\ee
which implies that
\begin{align}
p_1-k&\to \frac{Q}{2}-k~,~Q=p_1+p_2\nn\\
p_2+k&\to \frac{Q}{2}+k~,\nn \\
r=p'_1-p_1+k&\to p'+k~,~p'=\frac{p'_1-p'_2}{2}~.  
\label{shi.ext}
\end{align}
For the one-pion exchange between the initial nucleons,
fig.~\ref{fig:self12ib}, 
we proceed in the same way followed for $T_{12}^f$ above. Performing the same
transformation 
in the integration variable as in eq.~(\ref{shq}), the expressions for
$T_{12}^{f,S=0,1}$  can be used for $T_{12}^{i,S=0,1}$  with the exchange of
$\vp'\to \vp$.

We now introduce the functions $L_{11}$, $L_{11}^a$ and $L_{11}^{ab}$ defined as
\begin{align}
L_{11}(\vr)&=i\int\frac{d^4
k}{(2\pi)^4}\frac{1}{(\vk+\vr)^2+m_\pi^2}G_0(Q/2-k)_m G_0(Q/2+k)_\ell
~,\nn\\
L_{11}^i(\vr)&=i\int\frac{d^4 k}{(2\pi)^4}\frac{k^i}{(\vk+\vr)^2
+m_\pi^2} G_0(Q/2-k)_m G_0(Q/2+k)_\ell=L_{11}^\alpha a^i+L_{11}^p
r^i~,\nn\\
L_{11}^{ij}(\vr)&=i\int\frac{d^4 k}{(2\pi)^4}\frac{k^i
k^j}{(\vk+\vr)^2+m_\pi^2}G_0(Q/2-k)_m G_0(Q/2+k)_\ell\nn\\
&=L_{11}^{Tg}\delta^{ij}+L_{11}^{T \alpha}a^i a^j+L_{11}^{Tp}r^i r^j+
L_{11}^{T\alpha p}(a^i r^j+a^j r^i)
~,
\label{l11s}
\end{align}
with $a=Q/2$. 
These integrals are further discussed  in Appendix~\ref{sec:l11}. 
In terms of the functions introduced in eq.~\eqref{l11s} one can write 
the different matrix elements of $T_{12}^{f,S}$ as
\begin{align}
T_{12,d}^{f,S=0}&=i_3\frac{i\kappa
q^0\varepsilon_{ij3}}{f^2}\left(\frac{g_A}{2f}\right)^2 (C_S-3C_T)
\frac{m\partial}{\partial A}(L_{10}-m_\pi^2 L_{11})~,\nn\\
T_{12,d}^{f,S=1}(1,1)&=-i_3\frac{i\kappa
q^0\ve_{ij3}}{f^2}\left(\frac{g_A}{2f}\right)^2 
\left(C_S+C_T\right)\frac{m\partial}{\partial A}\left[
{p'_3}^2\left(L_{11}+2L_{11}^p+L_{11}^{Tp}\right)+a_3^2
L_{11}^{T\alpha}\right.\nn\\
&+\left.2 a_3 p'_3\left(L_{11}^\alpha+L_{11}^{T\alpha p}\right)+L_{11}^{Tg}
\right]~,\nn\\
T_{12,d}^{f,S=1}(1,0)&=-i_3 \frac{i \kappa
q^0\ve_{ij3}}{f^2}\left(\frac{g_A}{2f}\right)^2 
\sqrt{2}(C_S+C_T)\frac{m\partial}{\partial A}\left[
(a_1-ia_2)\left\{L_{11}^{T\alpha}a_3+(L_{11}^{T\alpha
p}+L_{11}^\alpha)p'_3\right\}
\right.\nn\\
&+\left.(p'_1-ip'_2)\left\{(L_{11}^{T\alpha
p}+L_{11}^\alpha)a_3+(L_{11}^{Tp}+L_{11}+2L_{11}^p)p'_3\right\}\right]~,
\nn\\
T_{12,d}^{f,S=1}(1,-1)&=-i_3\frac{i \kappa
q^0\ve_{ij3}}{f^2}\left(\frac{g_A}{2f}\right)^2(C_S+C_T)
\frac{m\partial}{\partial A}\left[(a_1-ia_2)^2
L_{11}^{T\alpha}+(p'_1-ip'_2)^2(L_{11}^{Tp}+2L_{11}^p+L_{11})\right.\nn \\ 
&+\left. 2(p'_1-ip'_2)(a_1-ia_2)(L_{11}^{T\alpha
p}+L_{11}^\alpha)\right]~,
\end{align}
\begin{align}
T_{12,d}^{f,S=1}(0,1)&=-i_3\frac{i \kappa
q^0\ve_{ij3}}{f^2}\left(\frac{g_A}{2f}\right)^2 
\sqrt{2}(C_S+C_T)\frac{m\partial}{\partial
A}\left[(a_1+ia_2)\left\{L_{11}^{T\alpha}a_3+(L_{11}^{T\alpha
p}+L_{11}^\alpha)p'_3
\right\}\right. \nn\\
&+\left. (p'_1+ip'_2)\left\{(L_{11}^{T\alpha
p}+L_{11}^\alpha)a_3+(L_{11}^{T p}+L_{11}+2L_{11}^p)p'_3
\right\}
\right]~,\nn
\end{align}
\begin{align}
T_{12,d}^{f,S=1}(0,0)&=-i_3\frac{i \kappa
q^0\ve_{ij3}}{f^2}\left(\frac{g_A}{2f}\right)^2 (C_S+C_T)
\frac{m\partial}{\partial A}\left[L_{10}-m_\pi^2L_{11}-2 L_{11}^{Tg}
-2{a_3}^2L_{11}^{T\alpha}\right. \nn\\
&-\left.2{p'_3}^2(L_{11}+L_{11}^{Tp}+2L_{11}^p)-4a_3
p'_3(L_{11}^{T\alpha p}+L_{11}^\alpha)
\right]~,\nn\\
T_{12,d}^{f,S=1}(-1,1)&=-i_3\frac{i \kappa q^0\ve_{ij3}}{f^2}\left(\frac{g_A}{2
f}\right)^2
(C_S+C_T)\frac{m\partial}{\partial
A}\left[(a_1+ia_2)^2L_{11}^{T\alpha}+(p'_1+ip'_2)^2
(L_{11}^{Tp}+2L_{11}^p+L_{11})\right. \nn\\
&+\left.2(p'_1+ip'_2)(a_1+ia_2)(L_{11}^{T\alpha p}+L_{11}^\alpha)
\right]~,\nn\\
T_{12,d}^{f,S=1}(0,-1)&=-T_{12,d}^{f,S=1}(1,0)~,\nn\\
T_{12,d}^{f,S=1}(-1,0)&=-T_{12,d}^{f,S=1}(0,1)~,\nn\\
T_{12,d}^{f,S=1}(-1,-1)&=T_{12,d}^{f,S=1}(1,1)~.
\label{t12.s1}
\end{align}

The isoscalar contribution originates by taking the derivative of the 
intermediate nucleon propagator in fig.~\ref{fig:effective}. We have the same 
expression as for $T_{12}^{f}$, eq.~(\ref{t11f}), but removing the operator 
$m\partial/ \partial A$ and with the replacement 
\be
i\frac{\kappa q^0\ve_{ij3}}{f^2}\tau^3_{nn'}\to
\frac{g_A^2}{f^2}\frac{|\vq|^2}{q_0^2}\delta_{nn'}~.
\label{extra.2nd}
\ee
 As a result, the isospin operator changes and for the direct term it is given
now by 
\be
2\vec{\tau}_{m'm}\vec{\tau}_{\ell'\ell} G_0(Q/2-k)_m G_0(Q/2+k)_\ell~.
\label{iso.sig.13}
\ee
The projection for a state with $i_3=\pm 1$ is $ 2(4I-3)  G_0(Q/2-k)_{\pm 1/2}
G_0(Q/2+k)_{\pm
1/2}$.  
For $i_3=0$ there is now a contribution from eq.~(\ref{iso.sig.13}) given by
\begin{align}
2(4I-3)\frac{1}{2}\left[G_0(Q/2-k)_{+1/2} G_0(Q/2+k)_{-1/2}+G_0(Q/2-k)_{-1/2}
G_0(Q/2+k)_{+1/2}\right]~.
\end{align}
Hence, instead of eq.~(\ref{t11fs1}), we arrive at 
\begin{align}
T_{13,d}^{f,S=0}&=i(4I-3)\frac{g_A^2}{f^2}\frac{\vq^2}{q_0^2}\delta_{ij}
\left(\frac{g_A}{2f}\right)^2
(C_S-3C_T)\int \frac{d^4k}{(2\pi)^4}\frac{\vr^2}{\vr^2+m_\pi^2}\nn\\
& \times
\frac{1}{2}\left[G_0(Q/2-k)_{m} G_0(Q/2+k)_{\ell}+G_0(Q/2-k)_{\ell}
G_0(Q/2+k)_{m}\right]~,
\nn\\
T_{13,d}^{f,S=1}(s'_3,s_3)&=-i(4I-3)\frac{g_A^2}{f^2}\frac{\vq^2}{q_0^2}\delta_{
ij}\left(\frac{g_A}{2f}\right)^2 (C_S+C_T)
\int\frac{d^4k}{(2\pi)^4}\frac{1}{\vr^2+m_\pi^2} B_{s'_3,s_3}\nn\\
&\times
\frac{1}{2}\left[G_0(Q/2-k)_{m}
G_0(Q/2+k)_{\ell}+G_0(Q/2-k)_{\ell}G_0(Q/2+k)_{m}\right]~,
\end{align}
such that $m=\ell=\pm 1/2$ for $i_3=\pm 1$  and  $m=+1/2$, $\ell=-1/2$ for
$i_3=0$.
Similar expressions to eq.~(\ref{t12.s1}) can be written. E.g. for $S=0$ one has
now 
\begin{align}
T_{13,d}^{f,S=0}&=(4I-3)\frac{g_A^2
}{f^2}\frac{\vq^2}{q_0^2}\left(\frac{g_A}{2f}\right)^2(C_S-3C_T)
(\widetilde{L}_{10}-m_\pi^2 \widetilde{L}_{11})~.\end{align}
Here, the tilde indicates the symmetric form 
\be
\widetilde{L}_{ij}^{a b  \ldots}=\frac{1}{2}\left(L_{ij}^{a b
\ldots}(m,\ell)+L_{ij}^{a b \ldots}(\ell,m)\right)~,
\ee
with $m$ and $\ell$ given in terms of $i_3$ as explained above. The
expressions for $T_{13,d}^{i,S}$ are the same as those worked out for 
$T_{13,d}^{f,S}$ with the replacement $\vp'\to \vp$, as noted above for the
$T_{12}$ amplitudes.

Let us consider  the partial wave projection of $T_{12}$ and $T_{13}$.  As
discussed  at the end of Appendix \ref{app.pwd}, we
can still use eq.~(\ref{pw.exp.def}), valid in the vacuum, if the integral in
eq.~\eqref{operation} does not depend on $\hat{\mathbf{a}}$. For $T_{12}^f$ and 
$T_{13}^f$, with the one-pion
exchange between the final nucleons, this is clearly the case because
$T_{12}^f$ and $T_{13}^f$ only depend on $\hat{\mathbf{a}}$ through its scalar 
product with $\vp'$. Thus, there is no angular dependence left on
$\val$ 
once the integration over $d\hat{\vp}'$ is performed. 
 For the case when the 
pion is exchanged between the initial nucleons, fig.~\ref{fig:self12ib}, the 
resulting $T_{12}^i$ and $T_{13}^i$ do not depend on the final three-momentum
$\hat{\vp}'$. 
In this
way, the integration over $d\hat{\vp}'$ cannot remove the dependence on 
$\hat{\mathbf{a}}$. This also implies that this diagram only can contribute 
to partial waves with $\ell'=0$, that is, ${^3S_1}$ and ${^3D_1}\to {^3S_1}$. 
However, as remarked above after eq.~(\ref{shi.ext}), the exchange 
$\vp'\leftrightarrow \vp$ transforms $T_{12}^f$, $T_{13}^f$ into $T_{12}^i$, 
$T_{13}^i$ and vice versa. 
In addition, one has to notice the symmetry between $\vp$ and $\vp'$ in
eq.~(\ref{pw.d2.i1}) for the partial wave decomposition. It is then clear that
the same partial waves result for the diagrams of figs.~\ref{fig:self12fb} and 
\ref{fig:self12ib} with the exchange $\ell'\leftrightarrow \ell$. Thus, we can
still use eq.~(\ref{pw.exp.def}) but using the diagrams with the pion
exchanged between the final nucleons. The elastic partial wave ${^3S_1}$ is
exactly the same for both diagrams and ${^3D_1} \to {^3S_1}$ is equal to
${^3S_1}
\to {^3D_1}$ evaluated as discussed.
 We denote the corresponding partial
waves by ${\cT}_{12;JI}^f(\ell,\bar{\ell},S)$ and
${\cT}_{13;JI}^f(\ell,\bar{\ell},S)$. For
$T_{12}$, that requires $I=1$, only the partial wave ${\cal T}_{12;01}(0,0,0)
=2{\cal T}^f_{12;01}(0,0,0)$ is not zero. On the other hand, for $T_{13}$ both 
isospin combinations occur. Then one has the partial waves ${\cal
  T}_{13;01}(0,0,0)=2{\cal T}^f_{13;01}(0,0,0)$, ${\cal T}_{13;10}(0,0,1)
=2{\cal T}^f_{13;10}(0,0,1)$, ${\cal T}_{13;10}(2,0,1)={\cal
  T}^f_{13;10}(2,0,1)$ 
and ${\cal T}_{13;10}(0,2,1)={\cal T}^f_{13;10}(2,0,1)$.
 We have the following expressions for ${\cT}_{JI}^f(\ell,\bar{\ell},S)$,
omitting the subscripts 12 or 13,
\begin{align}
\cT^f_{\ell I}(\ell,\ell,0)&=\frac{Y_{\ell}^0(\hat{\mathbf{z}})}{2\ell+1}\int
d\hat{\vp}' {T_{d}^{S=0}} Y_\ell^0(\hat{\vp}')^*~,\nn\\
\cT^f_{J I}(\ell,\bar{\ell},1)&=\frac{Y_{\bar{\ell}}^0(\hat{\mathbf{z}})}{2J+1}
\left\{
\int d\hat{\vp}'
 Y_\ell^0(\hat{\vp}')\left[
T_{d}^{S=1}(0,0)(000|\bar{\ell}1J)(000|\ell 1J)\right. \right. \nn\\
&\left.\left. +
\left({T_{d}^{S=1}(+1,+1)}+{T_{d}^{S=1}(-1,-1)}\right)(011|\bar{\ell}
1J)(011|\ell 1 J) 
\right]
\right. \nn\\
&\left.- \int
d\hat{\vp}'\left(Y_{\ell}^{-1}(\hat{\vp}'){T_{d}^{S=1}(-1,0)}+Y_\ell^1(\hat{\vp}
'){T_{d}^{S=1}(+1,0)}\right)
(000|\bar{\ell}1J)(1-10|\ell 1 J)
\right.\nn\\
&\left. -\int d\hat{\vp}'\left(Y_{\ell}^{-1}(\hat{\vp}')
{T_{d}^{S=1}(0,+1)}+Y_\ell^1(\hat{\vp}'){T_{d}^{S=1}(0,-1)}\right)(011|\bar{\ell
}1 J)(101|\ell 1 J)\right.\nn\\
&\left.+\int d\hat{\vp}' \left(Y_\ell^{-2}(\hat{\vp}')
{T_{d}^{S=1}(-1,+1)}+Y_\ell^2(\hat{\vp}'){T_{d}^{S=1}(+1,-1)}\right)(011|\bar{
\ell}1J)(2-11|\ell 1 J)
\right\}~.
\label{t11.pw.s1}
\end{align}   
Let us recall that in order to apply the previous equations, the vector $\vp$
must be taken along the $\hat{\vz}$ axis. Namely, for the partial wave 
projections, we take the reference frame with the axes
\begin{align}
\hat{\mathbf{z}}&=\hat{\vp}~,\nn\\
\hat{\mathbf{x}}&=\frac{\hat{\vp}\times\hat{\mathbf{a}}}{\sin\beta}~,\nn\\
\hat{\mathbf{y}}&=\frac{\hat{\vp}\times(\hat{\vp}\times\hat{\mathbf{a}})}{\sin\beta}
=\hat{\vp}\,\hbox{ctg}\beta-\hat{\mathbf{a}}\,\hbox{csec}\beta~.
\end{align}

\begin{figure}[ht]
\psfrag{q}{{\tiny $q$}}
\psfrag{m}{{\tiny $m$}}
\psfrag{l}{{\tiny $\ell$}}
\psfrag{mp}{{\tiny $m'$}}
\psfrag{lp}{{\tiny $\ell'$}}
\psfrag{i}{{\tiny $i$}}
\psfrag{j}{{\tiny $j$}}
\psfrag{o}{{\tiny $o$}}
\psfrag{op}{{\tiny $o'$}}
\psfrag{t}{{\tiny $t$}}
\psfrag{p1}{{\tiny $p_1$}}
\psfrag{p2}{{\tiny $p_2$}}
\psfrag{pp1}{{\tiny $p'_1$}}
\psfrag{pp2}{{\tiny $p'_2$}}
\psfrag{a}{a)}
\psfrag{b}{b)}
\psfrag{pm}{{\tiny $p_1-k$}}
\psfrag{pp}{{\tiny $p_2+k$}}
\psfrag{k}{{\tiny $k$}}
\psfrag{r}{{\tiny $r$}}
\centerline{\epsfig{file=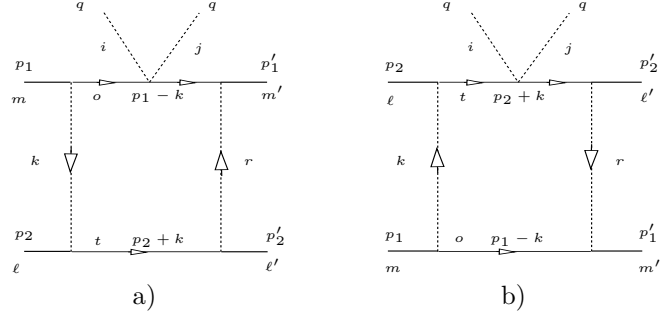,width=.5\textwidth,angle=0}}
\vspace{0.2cm}
\caption[pilf]{\protect \small
The internal four-momenta and discrete indices are indicated on the two
figures whose sum determines $T_{14,d}$. Note that the pion labels and
four-momenta are exchanged for the initial and final states separately 
between the two figures.
\label{fig:self14bis}}
\end{figure}

Let us move now to the evaluation of $T_{14}$ and $T_{15}$ where both vertices 
in the two-nucleon reducible loop at which the two pions are attached
correspond to one-pion exchange.  As in the previous cases  we start by 
calculating the isovector case. We restrict ourselves from the beginning  to 
the direct contribution, corresponding to the diagrams in
fig.~\ref{fig:self14bis}, whose sum  is 
\begin{align}
T_{14,d}&=-\left(\frac{g_A}{2f}\right)^4\frac{q^0\kappa
\ve_{ij3}}{2f^2}\left(\tau^a_{m'o}\tau^c_{om}\tau^3_{oo}
\tau^a_{\ell't}\tau^c_{t\ell}+\tau^a_{\ell't}\tau^c_{t\ell}\tau^3_{tt}\tau^a_{
m'o}\tau^c_{om}\right)\frac{m\partial}{\partial A}
\int\frac{d^4k}{(2\pi)^4}(\vec{\sigma}_{\alpha_{\ell'}\beta}\cdot \vr)
(\vec{\sigma}_{\alpha_{m'} \alpha}\cdot \vr)\nn\\
&\times (\vec{\sigma}_{\alpha\alpha_m}\cdot
\vk)(\vec{\sigma}_{\beta\alpha_\ell}\cdot\vk)
\frac{1}{\vr^2+m_\pi^2}\frac{1}{\vk^2+m_\pi^2} G_0(p_1-k)_{o} G_0(p_2+k)_{t}~.
\label{t14.sum}
\end{align}
The diagonal matrix elements of the isospin operator  
\be
\left(\tau^a_{m'o}\tau^c_{om}\tau^3_{oo}
\tau^a_{\ell't}\tau^c_{t\ell}+\tau^a_{\ell't}\tau^c_{t\ell}\tau^3_{tt}\tau^a_{
m'o}\tau^c_{om}\right)G_0(p_1-k)_{o}G_0(p_2+k)_{t}~,
\label{sig.14.iso}
\ee
present in eq.~(\ref{t14.sum}), are given between states with well defined
isospin as
\be
2i_3\, G_0(p_1-k)_{\pm\frac{1}{2}} G_0(p_2+k)_{\pm\frac{1}{2}}~.
\label{iso.14}
\ee 
which is zero for $i_3=0$. With respect  to spin we can rewrite,
\begin{align}
&(\vec{\sigma}\cdot\vr)_{\alpha_{\ell'}\beta}(\vec{\sigma}\cdot\vr)_{\alpha_{m'}
\alpha}(\vec{\sigma}\cdot\vk)_{\alpha\alpha_m}(\vec{\sigma}\cdot\vk)_{
\beta\alpha_\ell}=(\vr\cdot\vk)^2\delta_{\alpha_{\ell'}\alpha_\ell}\delta_{
\alpha_{m'}\alpha_m}-\left[(\vr\times
\vk)\cdot\vec{\sigma}_{\alpha_{\ell'}\alpha_\ell}\right]\left[
(\vr\times\vk)\cdot\vec{\sigma}_{\alpha_{m'}\alpha_m}\right]\nn\\
&+i(\vr\times\vk)\cdot\vec{\sigma}_{\alpha_{\ell'}\alpha_\ell}
(\vr\cdot\vk)\delta_{\alpha_{m'}\alpha_m}+i(\vr\times\vk)\cdot\vec{\sigma}_{
\alpha_{m'}\alpha_m}(\vr\cdot\vk)\delta_{\alpha_{\ell'}\alpha_\ell}~.
\label{sig.14.spi}
\end{align}
The matrix elements of the spin operators
$\delta_{\alpha_{m'}\alpha_m}\delta_{\alpha_{\ell'}\alpha_\ell}$ and 
$(\vec{\sigma}_{\alpha_{m'}\alpha_m}\cdot
\vv)(\vec{\sigma}_{\alpha_{\ell'}\alpha_\ell}\cdot \vv)$ between  
states with well defined total spin were already worked out in 
eqs.~(\ref{spin.pro}) and (\ref{s1mat}), respectively.  We have now in
addition the operator  
\be
\left(\delta_{\alpha_{m'}\alpha_m}\vec{\sigma}_{\alpha_{\ell'}\alpha_\ell}
+\delta_{\alpha_{\ell'}\alpha_\ell}
\vec{\sigma}_{\alpha_{m'}\alpha_m}\right)\cdot \vv~,
\ee
whose matrix elements are
\be
\left(
\begin{array}{r|rrr}
   & -1    & 0& +1\\
   \hline
-1 & -2 v_3 & \sqrt{2}(v_1+iv_2) & 0\\ 
0  & \sqrt{2}(v_1-iv_2) & 0 & \sqrt{2}(v_1+iv_2)\\
+1 & 0 & \sqrt{2} (v_1-iv_2) & 2 v_3
\end{array}
\right)~,
\label{sig.14.spi.mat}
\ee
and 0 for $S=0$. 
We follow in the previous matrix the same notation as
in eq.~(\ref{s1mat}). 
Taking into account this matrix and eqs.~(\ref{spin.pro}) and (\ref{s1mat})
one can determine the operator of eq.~(\ref{sig.14.spi}) between two-nucleon 
states with well defined third component of total spin $S$. The latter are  
inserted  into eq.~(\ref{t14.sum}) together with the isospin factor  $2 i_3
{G_0}_{\pm1/2}{G_0}_{\pm 1/2}$, eq.~(\ref{iso.14}). As a result, the amplitudes
$T_{14,d}^{S=0}$ 
and $T_{14,d}^{S=1}(s'_3,s_3)$ are determined.

The isoscalar contributions arise by taking the derivative of the intermediate
nucleon propagator in fig.~\ref{fig:effective} with respect to $z$, as discussed
in eq.~(\ref{ver:iss}). We have the same expression as for $T_{14,d}$,
eq.~(\ref{t14.sum}), but removing the derivative $m\partial/ \partial A$ and
with the replacement of eq.~(\ref{extra.2nd}). The resulting isospin operator  
is now given by
\be
2\tau^a_{m'o}\tau^c_{om}\tau^a_{\ell't}\tau^c_{t\ell}G_0(p_1-k)_o
G_0(p_2+k)_{t}~.
\label{sig.14.iso2}
\ee
One can work out straightforwardly its diagonal matrix elements between 
states with definite isospin, 
\begin{align}
2(9-8I)\frac{1}{2}\biggl\{ G_0(p_1-k)_o G_0(p_2+k)_{t}+G_0(p_1-k)_t
G_0(p_2+k)_o\biggr\}~,
\end{align}
with $o+t=i_3$. Then, instead of eq.~(\ref{t14.sum}) one has now,
\begin{align}
T_{15,d}&=
i\left(\frac{g_A}{2f}\right)^4\frac{g_A^2}{f^2}\frac{\vq^2}{q_0^2}(9-8I)
\int\frac{d^4k}{(2\pi)^4}\frac{1}{2}\left\{ G_0(p_1-k)_o
G_0(p_2+k)_t+G_0(p_1-k)_t G_0(p_2+k)_o\right\}\nn\\
&\times 
(\vec{\sigma}_{\alpha_{\ell'}\beta}\cdot \vr)
(\vec{\sigma}_{\alpha_{m'} \alpha}\cdot \vr)
 (\vec{\sigma}_{\alpha\alpha_m}\cdot
\vk)(\vec{\sigma}_{\beta\alpha_\ell}\cdot\vk)
\frac{1}{\vr^2+m_\pi^2}\frac{1}{\vk^2+m_\pi^2}~,
\label{t15.iss}
\end{align}
with $o+t=i_3$ as before. The tensor integrals required by
eqs.~\eqref{t14.sum} and \eqref{t15.iss} that involve one intermediate
two-nucleon 
state with two one-pion exchanges are calculated in Appendix~\ref{sec:int.inv}.

To determine the partial waves for $T_{14}$ and $T_{15}$ we have checked 
numerically that eq.~\eqref{operation} does not depend on $\hat{\mathbf{a}}$  at the
level of one per mil. This is similar 
to the numerical accuracy to which the in-medium integrations have been 
calculated. As a result, we can use the same equation for partial wave
projection as 
in vacuum,  eq.~(\ref{1pi.pw}), but now in terms of  $T_{14,d}$ and
$T_{15,d}$. This reduces
the calculational load because eq.~(\ref{1pi.pw}) has not the integration over 
$\hat{\mathbf{a}}$ as in eq.~(\ref{pw.mexp.def}). We denote by ${\cal
  T}_{14;JI}(\ell',\ell,S)$ and ${\cal T}_{15,JI}(\ell',\ell,S)$ the resulting 
partial waves for $T_{14}$ and $T_{15}$, in this order.

Summing over the previous partial waves it follows that
\begin{align}
DL_{JI;is}^{(1)}={\cal T}_{11,d}+{\cal T}_{13,d}+{\cal T}_{15,d}~.
\end{align}
Comparing with eq.~(\ref{fix.diss}) it is straightforward to determine
$L_{JI}^{(1)}$.

\section{The loop function $L_{10}$} 
\label{sec:l10}
\def\theequation{\Alph{section}.\arabic{equation}}
\setcounter{equation}{0}

The function $L_{10}$ is given by
\begin{align}
L_{10}&=i\int\frac{d^4k}{(2\pi)^4}\left[
\frac{1}{Q^0/2-k^0-E(\frac{\vQ}{2}-\vk)+i\epsilon}
+2\pi i \theta(\xi_1-|\frac{\vQ}{2}-\vk|) \delta(Q^0/2-k^0-E(\frac{\vQ}{2}-\vk))
\right]
\nn\\
&\times
\left[
\frac{1}{Q^0/2+k^0-E(\frac{\vQ}{2}+\vk)+i\epsilon}
+2\pi i \theta(\xi_1-|\frac{\vQ}{2}+\vk|)\delta(Q^0/2+k^0-E(\frac{\vQ}{2}+\vk))
\right]~.
\label{l10.def}
\end{align}
This integration corresponds to the loop in fig.~\ref{fig:g} with total
in-coming four-momentum $Q$. In the following we define,
\be
\val=\frac{1}{2}(\vp_1+\vp_2)=\frac{\vQ}{2}~.
\label{def.alp}
\ee

The different contributions to $L_{10}$ are calculated according to the 
number of in-medium insertions in the nucleon propagators, eq.~(\ref{nuc.pro}).
The $k^0$-integration for the free part, $L_{10,f}$, is performed  by 
applying Cauchy's theorem,
\begin{align}
L_{10,f}&=\int\frac{d^3k}{(2\pi)^3}\frac{1}{Q^0-\frac{\vk^2}{m}-\frac{\val^2}{m}
+i\epsilon}=-m\int\frac{d^3k}{(2\pi)^3}\frac{1}{\vk^2-A-i\epsilon}~,
\label{8.22}
\end{align}
 with  $A$ given in eq.~\eqref{def.a.1}. One has to
keep in mind in the following the
$+i\epsilon$ prescription in the definition of $A$. In order to emphasize this, 
we will write explicitly the combination $A+i\epsilon$ in many integrals,
though 
the $+i\epsilon$ is already contained in $A$ according to eq.~(\ref{def.a.1}). 
The result in eq.~(\ref{8.22})  corresponds to eq.~(\ref{int.g}) that is 
regularized according to the dispersion relation eq.~\eqref{dis.rel.g},
\begin{align}
L_{10,f}=g_0-i\frac{m\sqrt{A}}{4\pi}~.
\end{align}
 For the one-medium insertion, $L_{10,m}$ the $k^0$-integration is done by 
making use of the energy-conserving Dirac delta-function in the in-medium 
part of the nucleon propagator, eq.~(\ref{nuc.pro}). We are then left with
\begin{align}
L_{10,m}&=m\int \frac{d^3
k}{(2\pi)^3}\frac{\theta(\xi_1-|\vk-\val|)+\theta(\xi_2-|\vk+\val|)}{
\vk^2-A-i\epsilon}~.
\label{l10.m.def}
\end{align}
Let us concentrate on the evaluation of the integral,
\begin{align}
\label{h1}
&\ell_{10,m}(\xi_1,A,|\val|)=m
\int\frac{d^3k}{(2\pi)^3}\frac{\theta(\xi_1-|\vk-\val|)}{\vk^2-A-i\epsilon}\\
&=\frac{m}{4\pi^2}\left\{
\xi_1-\sqrt{A} \hbox{ arctanh}\frac{\xi_1-|\val|}{\sqrt{A}}
-\sqrt{A}\hbox{ arctanh}\frac{\xi_1+|\val|}{\sqrt{A}}    -
\frac{A+\val^2-\xi_1^2}{4|\val|}
\log\frac{(|\val|+\xi_1)^2-A}{(|\val|-\xi_1)^2-A}
\right\}~,\nn
\end{align}
Here, we have taken into account that the Heaviside function 
in the numerator implies the conditions,
\begin{align}
&|\val|\geq \xi_1~,\nn \\
&|\vk|\in[|\val|-\xi_1,|\val|+\xi_1]~,~\cos\theta\in 
\left[\frac{\vk^2+|\val|^2-\xi_1^2}{2|\vk||\val|},1\right]~.\nn\\
&|\val|<\xi_1~,\nn\\
&|\vk|\in [0,\xi_1-|\val|]~,~\cos\theta\in[-1,1]~,\nn\\
&|\vk|\in [\xi_1-|\val|,\xi_1+|\val|]~,~\cos\theta\in
\left[\frac{\vk^2+|\val|^2-\xi_1^2}{2|\vk||\val|},1
\right]~.
\label{1m.con}
\end{align}
Despite the separation between the cases $|\val|\geq \xi_1$ and $|\val|<\xi_1$,
both give rise to the
same expression in eq.~(\ref{h1}). In terms of the function
$\ell_{10,m}(\xi_1,A,|\val|)$, eq.~(\ref{l10.m.def}) reads 
\be
L_{10,m}(\xi_1,\xi_2,A,|\val|)=\ell_{10,m}(\xi_1,A,|\val|)+\ell_{10,m}(\xi_2,A,
|\val|)~.
\label{l10.m}
\ee

For the case with two medium insertions
\begin{align}
L_{10,d}&=\frac{-im \sqrt{A}}{ 8\pi^2}\int d\hat{\vk} \, \theta(\xi_1-|\hat{\vk}
\sqrt{A}-\val|)\theta(\xi_2- |\hat{\vk}\sqrt{A}
+\val|)~.
\label{l10d.def}
\end{align}
Here we assume that $\xi_2\geq \xi_1$. If the opposite were true one can 
use the same expressions that we derive below but with the exchange 
$\xi_1\leftrightarrow \xi_2$. This is clear after changing $\hat{\vk}\to 
-\hat{\vk}$  in the integral of eq.~(\ref{l10d.def}). 
Denoting by $\theta$ the angle between $\hat{\vk}$ and 
$\val$ the two step functions  require
\begin{align}
\cos\theta &\geq \frac{A+|\val|^2-\xi_1^2}{2|\val|\sqrt{A}}\equiv y_1~, \nn\\
\cos\theta &\leq \frac{\xi_2^2-A-|\val|^2}{2|\val|\sqrt{A}} \equiv y_2~.
\label{l10d.con}
\end{align}
One has to impose that $y_1\leq +1$ and $y_2\geq -1$, otherwise $\cos\theta$
is out of the range 
$[-1,+1]$. In addition, it is also necessary that $y_2\geq y_1$. 
\begin{align}
y_1\leq +1  & \to   |\val|-\xi_1 \leq \sqrt{A} \leq |\val|+\xi_1~,\nn\\
y_2\geq -1 & \to  |\val|-\xi_2\leq \sqrt{A} \leq |\val|+ \xi_2~,\nn\\
y_1\leq y_2& \to  A\leq \frac{\xi_1^2+\xi_2^2}{2}-|\val|^2\equiv A_{max}~.
\label{l10d.con1}
\end{align}
The simultaneous consideration of the  first and third of the previous conditions requires that  $(|\val|-\xi_1)^2\leq A_{max}$ for $|\val|\geq \xi_1$, which in turn implies
\be
|\val|\leq \frac{\xi_1+\xi_2}{2}~.
\label{l10d.alp.con}
\ee
Notice that the previous upper bound is larger than $\xi_1$ because 
$\xi_2\geq \xi_1$. From eq.~(\ref{l10d.alp.con}) it follows then that 
$|\val|- \xi_2\leq 0$, and since it is always the case that 
$(|\val|+\xi_2)^2\geq A_{max}$ the second condition in eq.~\eqref{l10d.con1} is  satisfied. On the other hand, 
\begin{align}
\hbox{if ~} |\val|\geq \frac{\xi_2-\xi_1}{2}\to A_{max}\leq
(|\val|+\xi_1)^2~,\nn\\
\hbox{if ~} |\val|\leq \frac{\xi_2-\xi_1}{2}\to A_{max} \geq (|\val|+\xi_1)^2~.
\end{align}
For the final form of $L_{10,d}$ one also has to take into account the
conditions,
\begin{align}
y_1\geq -1 &\to  \sqrt{A}\geq \xi_1-|\val|,\nn\\
y_2 \leq +1 &\to  \sqrt{A}\geq \xi_2-|\val|~.
\label{l10d.con2}
\end{align}
Gathering together the conditions in eqs.~(\ref{l10d.con})--(\ref{l10d.con2}) 
we have the following options,
\be
 y_1\leq -1~,~y_2\leq +1 ~\to ~ \xi_2-|\val|\leq \sqrt{A}\leq \xi_1-|\val|~,
\ee
which is not possible because $\xi_2\geq \xi_1$. Also
\be
y_1\leq -1~,~y_2\geq +1~\to ~\sqrt{A}\leq \xi_1-|\val|~.
\label{l10d.cos.1}
\ee
This only holds for $|\val|\leq \xi_1$. Then $\cos\theta\in [-1,+1]$ 
and $L_{10,d}=-i\, m\sqrt{A}/(2 \pi)$. Further
\be
-1\leq y_1 \leq +1~,~ y_2\geq +1~\to ~ |\xi_1-|\val||\leq \sqrt{A}\leq
\hbox{min}(\xi_1+|\val|,\xi_2-|\val|)~.  
\label{l10d.cos.2}
\ee
In this case, $\cos\theta\in [y_1,+1]$ and $L_{10,d}=-im(\xi_1^2-(\sqrt{A}
-|\val|)^2)/(8\pi|\val|)$. It follows that 
$\xi_1+|\val|\leq \xi_2-|\val|$ for $|\val|\leq (\xi_2-\xi_1)/2$ and 
$\xi_1+|\val|\geq \xi_2-|\val|$ for 
$|\val|\geq (\xi_2- \xi_1)/2$. In both cases
$[\hbox{min}(\xi_1+|\val|,\xi_2-|\val|)]^2\leq 
 A_{max}$, as can be easily seen. 
The last possibility is
\be
-1\leq y_1 \leq +1~,~y_2\leq +1~\to~\xi_2-|\val|\leq \sqrt{A}\leq \xi_1+|\val|~.
\label{l10d.cos.3}
\ee
For this case to hold, it is necessary that $|\val| \geq (\xi_2-\xi_1)/2$. 
But then $A_{max}\leq (\xi_1+|\val|)^2$ so that
the allowed upper limit for $\sqrt{A}$ is $\sqrt{A_{max}}$ not
$\xi_1+|\val|$. In this case, $\cos\theta\in [y_1,y_2]$ and  
$L_{10,d}=-i \,m(\xi_1^2+\xi_2^2-2 A-2|\val|^2)/(8\pi|\val|)$. 
In summary,
\be
L_{10,d}=\left\{
\begin{array}{ll}
 -\frac{i\, m \sqrt{A}}{2\pi}~,& \sqrt{A}\leq \xi_1-|\val|~,~|\val|\leq \xi_1 \\
-\frac{i\, m}{8\pi|\val|}(\xi_1^2-(\sqrt{A}-|\val|)^2)~, & |\xi_1-|\val||\leq
\sqrt{A} \leq \xi_1+|\val|~,~|\val|\leq \frac{\xi_2-\xi_1}{2}  \\
-\frac{i\, m}{8\pi|\val|}(\xi_1^2-(\sqrt{A}-|\val|)^2)~, & |\xi_1-|\val||\leq
\sqrt{A} \leq \xi_2-|\val|~,~\frac{\xi_2-\xi_1}{2}\leq |\val| \leq
\frac{\xi_1+\xi_2}{2}  \\
-\frac{i\, m}{8\pi|\val|}(\xi_1^2+\xi_2^2-2 A-2|\val|^2)~, & \xi_2-|\val|\leq
\sqrt{A} \leq \sqrt{A_{max}}~,~\frac{\xi_2-\xi_1}{2}\leq |\val| \leq
\frac{\xi_1+\xi_2}{2} 
\end{array}
\right.
\label{l10d.exp}
\ee
As a technical detail one can take directly the derivative of
eq.~(\ref{l10d.exp}) with respect to $A$ for a given interval. 
The different intervals do not imply the appearance of delta functions of $A$ 
when differentiating as one check by evaluating directly the 
derivative of eq.~\eqref{l10d.def}.

\section{Calculation of $L_{11}$, $L_{11}^a$ and $L_{11}^{ab}$}
\label{sec:l11}
\def\theequation{\Alph{section}.\arabic{equation}}
\setcounter{equation}{0}

We now consider  the integrals defined in eq.~(\ref{l11s}), though we
calculate here completely only the scalar integral $L_{11}$. For the calculation of
$L_{11}^a$ and $L_{11}^{ab}$ one resorts to the Passarino-Veltmann reduction 
and explicit calculations and expressions are given in ref.~\cite{techrep}. Nevertheless, we 
 evaluate at the end of this Appendix the free part 
of the tensor integral $L_{11}^{ab}$ in order to explicitly show how to handle the regularization of a two-nucleon reducible loop  different to $g(A)$ in terms of $g_0$. 

We evaluate $L_{11}$  according to the number of in-medium insertions. After
performing the $k^0$-integration by applying Cauchy's theorem we are left 
for the free part with 
\begin{align}
L_{11,f}&=-m\int\frac{d^3k}{(2\pi)^3}\frac{1}{(\vk+\vp)^2+m_\pi^2}\frac{1}{
\vk^2-A-i\epsilon}=
-\frac{m}{8\pi}\int_0^1 dx\frac{1}{\left[\vp^2 x(1-x)+m_\pi^2
x-A(1-x)-i\epsilon\right]^{1/2}}
\nn\\&=-\frac{i m }{8\pi
|\vp|}\log\frac{A-(|\vp|+im_\pi)^2}{m_\pi^2+(\sqrt{A}-|\vp|)^2}~.
\label{l11.f}
\end{align} 
In the previous equation, we have introduced a Feynman integration parameter $x$ as an intermediate step 
 and 
$\sqrt{-a\pm i\epsilon}=\pm i\sqrt{a}$ for $a>0$. For the one-medium insertion
case, the $k^0$-integration is straightforward due to the presence of the energy Dirac
delta-function
\begin{align}
L_{11,m}&=m\int\frac{d^3
k}{(2\pi)^3}\frac{\theta_m^{-}(\val-\vk)}{
(\vk^2-A-i\epsilon)((\vk+\vp)^2+m_\pi^2)}
+m\int\frac{d^3k}{(2\pi)^3}\frac{\theta_\ell^{-}(\val-\vk)}{
(\vk^2-A-i\epsilon)((\vk-\vp)^2+m_\pi^2)}~.
\label{l11.m.def}
\end{align}
Both terms in the sum can be obtained from the function
\begin{align}
\ell_{11,m}=m\int\frac{d^3
k}{(2\pi)^3}\frac{\theta(\xi_1-|\val-\vk|)}{
(\vk^2-A-i\epsilon)((\vk+\vp)^2+m_\pi^2)}~.
\label{l11.m.def.lc}
\end{align}
Let us work out the scalar product $\vk\cdot \vp$. For the integration, 
we introduce the reference frame
\begin{align}
\hat{\vz}&=\hat{\mathbf{a}}~,\nn\\
\hat{\vx}&=\frac{\val\times \vp}{|\val| |\vp| \sin\beta}~,\nn\\
\hat{\vy}&=\hat{\vz}\times \hat{\vx}=\hat{\mathbf{a}}\,\hbox{ctan}
\beta-\hat{\vp}\,\hbox{csec}\beta~.
\label{ref.l11}
\end{align}
{}From the last relation we have,
\begin{align}
\hat{\vp}&=\hat{\mathbf{a}}\,\cos\beta-\hat{\vy}\sin\beta~,\nn\\
\vk\cdot \vp&=|\vk||\vp|(\cos\theta \cos\beta- \sin\theta\sin\phi \sin \beta)~,
\end{align}
with $\theta$ and $\phi$ integration variables in eq.~(\ref{l11.m.def.lc}) 
and $\cos\beta=\hat{\vp}\cdot\hat{\mathbf{a}}$~.
Let us perform the $\phi$-integration,
\be
\frac{1}{2\pi}\int_0^{2\pi}d\phi\frac{1}{\vp^2+\vk^2+2\vk\cdot 
\vp+m_\pi^2}=\frac{1}{\sqrt{a^2-b^2}}~,
\label{intfi.2}
\ee
that also has  been checked  numerically. In eq.~(\ref{intfi.2}) we have 
\begin{align}
a&=\vk^2+\vp^2+2|\vk||\vp|\cos\beta\cos\theta+m_\pi^2~,\nn\\
b&=-2|\vk||\vp|\sin\beta\sin\theta~,
\end{align}
where $\sin\beta=\sqrt{1-\cos^2\beta}$. 
 Next, we move to the $\cos\theta$ integration of eq.~(\ref{l11.m.def.lc}). 
For this integration one has to take into account the presence of the
Heaviside function, which implies the conditions already worked out in
eq.~(\ref{1m.con}). The latter determine an interval of integration for 
$\cos\theta\in [x_1(|\vk|),x_2(|\vk|)]$ for a given value of $|\vk|$. 
Then, we can write,
\begin{align}
&\int_{x_1}^{x_2} d\cos\theta \frac{1}{\sqrt{a^2-b^2}}=\int_{x_1}^{x_2}
d\cos\theta \frac{1}{\sqrt{a'+b'\cos\theta+c'\cos^2\theta}}\nn\\
&=
\frac{1}{\sqrt{c'}}\left\{\log\left[\frac{b'+2c'\cos\theta}{\sqrt{c'}}+2\sqrt{
a'+b'\cos\theta+c'\cos^2\theta}\right]\right\}_{x_1}^{x_2}~,
\label{int.l11.m.cos}
\end{align}
with 
\begin{align}
a'&=\delta^2-4\vk^2 \vp^2\sin^2\beta~,\nn\\
b'&=4|\vk||\vp|\delta \cos\beta ~,\nn\\
c'&=4\vk^2\vp^2~,\nn\\
\delta&=\vk^2+\vp^2+m_\pi^2~.
\label{abc.d}
\end{align}
Now, we consider the final integration on $|\vk|$ in eq.~(\ref{l11.m.def}) 
and define the auxiliary function,
\begin{align}
f_{11,m}(|\vk|)=\frac{m}{8\pi^2|\vp|}\left\{\log\left[\frac{b'+2c'\cos\theta}{
\sqrt{c'}}+2\sqrt{a'+b'\cos\theta+c'\cos^2\theta}\right]\right\}_{x_1(|\vk|)}^{
x_2(|\vk|)}~,
\end{align}
in terms of which eq.~(\ref{l11.m.def.lc}) reads
\begin{align}
&|\val|\geq \xi_1~,\nn\\
&\ell_{11,m}(\xi_1,\cos\beta)=\int_{|\val|-\xi_1}^{|\val|+\xi_1}d|\vk|
\frac{|\vk|}{\vk^2-A-i\epsilon}f_{11,m}(|\vk|)~.\nn\\
&|\val| < \xi_1~,\nn\\
&\ell_{11,m}(\xi_1,\cos\beta)=\left\{\int_0^{\xi_1-|\val|}+\int_{\xi_1-|\val|}^{
|\val|+\xi_1}\right\}d|\vk| \frac{|\vk|}{\vk^2-A-i\epsilon}f_{11,m}(|\vk|)~.
\label{l11.m.final}
\end{align}
Then from eq.~(\ref{l11.m.def}) one has
\be
L_{11,m}=\ell_{11,m}(\xi_m,\cos\beta)+\ell_{11,m}(\xi_\ell,-\cos\beta)~.
\ee
Here, we have only indicated those arguments that change for each term in 
the sum of eq.~(\ref{l11.m.def}) to calculating $L_{11,m}$. Indeed,
$\ell_{11,m}$ depends also on $|\val|$ and $|\vp|$. The integration over 
$|\vk|$ when two medium insertions are represent can be done straightforwardly 
because of the additional Dirac delta-function of the energy that fixes 
$|\vk|=\sqrt{A}$. Then,
\begin{align}
L_{11,d}(\xi_1,\xi_2,\cos\beta)=-\frac{i m\sqrt{A}}{8\pi^2}\int d\hat{\vk}
\frac{\theta(\xi_1-|\hat{\vk} \sqrt{A}-\val|)\theta(\xi_2- |\hat{\vk}\sqrt{A}
+\val|)}{(\vk+\vp)^2+m_\pi^2}~.
\label{l11.d.def}
\end{align}
We have the same $\phi$-integration as in eq.~(\ref{intfi.2}), with the same 
result but now with $|\vk|=\sqrt{A}$. The integration over $\cos\theta$ is 
the same as in eq.~(\ref{int.l11.m.cos}), though with a different integration 
interval for $\cos\theta$ that is fixed by the values of $A$ and $|\val|$, 
according to the results of section~\ref{sec:l10}. They are collected
here for $\xi_1\leq \xi_2$,
 \begin{align}
\label{x1x2.l11d}
\cos\theta \in [x_1,x_2]~~\hbox{with}~~[x_1,x_2]\equiv \left\{
\begin{array}{ll}
[-1,1]~,& \sqrt{A}\leq \xi_1-|\val|~,~|\val|\leq \xi_1 \\
\left[y_1 , 1\right]~, & |\xi_1-|\val||\leq \sqrt{A} \leq
\xi_1+|\val|~,~|\val|\leq \frac{\xi_2-\xi_1}{2}  \\
\left[y_1 , 1\right]~, & |\xi_1-|\val||\leq \sqrt{A} \leq
\xi_2-|\val|~,~\frac{\xi_2-\xi_1}{2}\leq |\val| \leq \frac{\xi_1+\xi_2}{2}  \\
\left[y_1 , y_2 \right]~, & \xi_2-|\val|\leq \sqrt{A} \leq
\sqrt{A_{max}}~,~\frac{\xi_2-\xi_1}{2}\leq |\val| \leq \frac{\xi_1+\xi_2}{2} 
\end{array}
\right.
 \end{align}
with $y_1$ and $y_2$ defined in eq.~(\ref{l10d.con1}).  We also define,
similarly as was done for $\ell_{11,m}$, the auxiliary function
\begin{align}
f_{11,d}(\xi_1,\xi_2,\cos\beta)
=\frac{m}{8\pi^2|\vp|}\log\left[\frac{b'+2c'\cos\theta}{\sqrt{c'}}+2\sqrt{
a'+b'\cos\theta+c'\cos^2\theta}\right]_{x_1}^{x_2}~,
\end{align}
with $|\vk|=\sqrt{A}$ for the values of $a'$, $b'$ and $c'$ in
eq.~(\ref{abc.d}) and $x_1$ and $x_2$ according 
to eq.~(\ref{x1x2.l11d}). In this way, 
\begin{align}
L_{11,d}(\xi_1,\xi_2,\cos\beta)= -i \pi f_{11,d}(\xi_1,\xi_2,\cos\beta)~.
\end{align}
For the case  $\xi_1\geq \xi_2$ the change of variable $\vk\to-\vk$ is performed
in 
eq.~(\ref{l11.d.def}) and thus
\begin{align}
L_{11,d}=& -i \pi  f_{11,d}(\xi_2,\xi_1,-\cos\beta)~.
\end{align}

Let us now consider the calculation of $L_{11,f}^{ab}$ corresponding to
\begin{align}
L_{11,f}^{ab}&=-m\int\frac{d^3k}{(2\pi)^3}\frac{k^a k^b}{[(\vk+\vp)^2+m_\pi^2][\vk^2-A-i\epsilon]}
=L_{11,f}^{Tg}\delta^{ab}
+L_{11,f}^{Tp}p^a p^b~.
\label{l11ft.d}
\end{align}
We now multiply the previous equality by $\delta^{ab}$ and sum over repeated indices, then
\begin{align}
3L_{11,f}^{Tg}+ L_{11,f}^{Tp} \vp^2 &=-m\int\frac{d^3k}{(2\pi)^3}\frac{\vk^2}{[(\vk+\vp)^2+m_\pi^2][\vk^2-A-i\epsilon]}~.
\label{int.int}
\end{align}
Doing first the angular integration, one has
\begin{align}
3L_{11,f}^{Tg}+ L_{11,f}^{Tp} \vp^2&=-\frac{m}{8\pi^2|\vp|}\int_0^\infty d|\vk|\frac{|\vk|^3}{\vk^2-A-i\epsilon}\log\frac{(|\vk|+|\vp|)^2+m_\pi^2}{(|\vk|-|\vp|)^2+m_\pi^2}~.
\label{l11ft.2}
\end{align}
Let us consider the expansion of the last factor in the previous equation for  $|\vk|\to\infty$,
\begin{align}
\log\frac{(|\vk|+|\vp|)^2+m_\pi^2}{(|\vk|-|\vp|)^2+m_\pi^2}=\frac{4|\vp|}{|\vk|}-
\frac{4|\vp|(m_\pi^2-|\vp|^2/3)}{|\vk|^3}+{\cal O}(|\vk|^{-5})~.
\label{log.exp.1}
\end{align}
Adding and subtracting $4|\vp|/|\vk|$ to eq.~\eqref{l11ft.2} it results  
\begin{align}
3L_{11,f}^{Tg}+ L_{11,f}^{Tp} \vp^2&=-\frac{m}{8\pi^2|\vp|}\int_0^\infty d|\vk|\frac{|\vk|^3}{\vk^2-A-i\epsilon}\left(\log\frac{(|\vk|+|\vp|)^2+m_\pi^2}{(|\vk|-|\vp|)^2+m_\pi^2}
-\frac{4|\vp|}{|\vk|}\right)\nn\\
&-\frac{m}{2\pi^2}\int_0^\infty d|\vk| \frac{\vk^2}{\vk^2-A-i\epsilon}~.
\label{l11ft.3}
\end{align}
The first integral is convergent because of the expansion in eq.~\eqref{log.exp.1} while the second one corresponds to $g_0(A)$, eq.~\eqref{dis.rel.g}.\footnote{From a practical point of view it is simpler to calculate algebraically the initial integral  eq.~\eqref{int.int} in cut-off regularization than the convergent one eq.~\eqref{l11ft.3}. Then, the cut-off is sent to infinity while  keeping only the divergent linear term. The cut-off  can be expressed in terms of $g_0$ by proceeding in the same way for $g(A)$ in eq.~\eqref{dis.rel.g}.}

Next, we multiply eq.~\eqref{l11ft.d} by $p^a p^b$, sum over $a$ and $b$ and proceed in the same way as before. After performing the angular integration we have
\begin{align}
L_{11,f}^{Tg}\vp^2+L_{11,f}^{Tp}|\vp|^4&=-m\int\frac{d^3k}{(2\pi)^3}\frac{(\vk\cdot \vp)^2}{[(\vk+\vp)^2+m_\pi^2][\vk^2-A-i\epsilon]}\nn\\
&=\frac{m}{8\pi^2}\int_0^\infty d|\vk|\frac{\vk^2(\vk^2+\vp^2+m_\pi^2)}{\vk^2-A-i\epsilon}
\left(1-\frac{\vk^2+\vp^2+m_\pi^2}{4|\vk||\vp|}
\log\frac{(|\vk|+|\vp|)^2+m_\pi^2}{(|\vk|-|\vp|)^2+m_\pi^2}\right)~.
\label{int.int.1}
\end{align} 
As in eq.~\eqref{log.exp.1} we consider the limit  of the last factor in the integrand for $|\vk|\to\infty$
\begin{align}
1-\frac{\vk^2+\vp^2+m_\pi^2}{4|\vk||\vp|}
\log\frac{(|\vk|+|\vp|)^2+m_\pi^2}{(|\vk|-|\vp|)^2+m_\pi^2}&=-\frac{4\vp^2}{3\vk^2}+{\cal O}(|\vk|^{-4})~.
\end{align}
Adding and subtracting the first term on the r.h.s. of the previous equation, similarly as done 
in eq.~\eqref{l11ft.3}, we have a convergent and a divergent integral. The convergent one  is given by
\begin{align}
\frac{m}{8\pi^2}\int_0^\infty d|\vk|\frac{\vk^2(\vk^2+\vp^2+m_\pi^2)}{\vk^2-A-i\epsilon}
\left(1-\frac{\vk^2+\vp^2+m_\pi^2}{4|\vk||\vp|}
\log\frac{(|\vk|+|\vp|)^2+m_\pi^2}{(|\vk|-|\vp|)^2+m_\pi^2}+\frac{4\vp^2}{3\vk^2}\right)
-i\frac{m \vp^2(\vp^2+m_\pi^2)}{12\pi\sqrt{A}}~,
\end{align}
while the divergent integral corresponds to
\begin{align}
-\frac{\vp^2}{6\pi^2}m\int_0^\infty d|\vk|\frac{\vk^2}{\vk^2-A-i\epsilon}=\frac{\vp^2}{3}g(A)~.
\label{div.2}
\end{align}

The calculation of the integrals eqs.~\eqref{int.int} and \eqref{int.int.1}  explicitly shows how to regularize in terms of the subtraction constant $g_0$ the linear divergences present in the calculation of the two-nucleon reducible loop integrals needed for the evaluation of the loops in figs.~\ref{fig:self12fb}, \ref{fig:self12ib} and  \ref{fig:self14bis}. Since two pion exchanges are involved in the latter figure only the scalar integral with $|\vk|^4$ in the numerator, denoted by $L_{12}^{(4)}$, diverges. Its divergent part can be straightforwardly regularized in terms 
of $g(A)$, as in eqs.~\eqref{l11ft.3} and \eqref{div.2}. Indeed, it just corresponds to $g(A)$.


\section{Calculation of $L_{12}$}
\label{sec:int.inv}
\def\theequation{\Alph{section}.\arabic{equation}}
\setcounter{equation}{0}

The different integrals involved in the evaluation of $T_{14}$ and
$T_{15}$ can be expressed in terms of a set of scalar integrals. 
The tensor structure of these integrals is determined by the matrix elements 
in eq.~(\ref{sig.14.spi}). We also perform here the shift of the integration 
variable $k\to (p_1-p_2)/2+k=p+k~,$ as in eq.~(\ref{shq}), and rewrite 
eqs.~(\ref{t14.sum}) and (\ref{t15.iss}) accordingly. The integrals necessary 
for the calculation of the latter equations are evaluated in
ref.~\cite{techrep}. Here, we only discuss the calculation of the basic scalar 
integral involved, $L_{12}$. For integrals with more complicated tensor
structure one uses the Passarino-Veltmann reduction. In the expressions that 
follow it is always assumed that the $k^0$-integration has been done either by 
using Cauchy's theorem, for the free part, or employing the energy Dirac
delta-functions from the in-medium insertions. On the other hand, since now 
two pion propagators are involved we join them in one introducing 
a Feynman parameter
\be
\frac{1}{[(\vk+\vp')^2+m_\pi^2][(\vk+\vp)^2+m_\pi^2]}=\int_0^1dy\frac{1}{[
(\vk+\vec{\lambda})^2+M^2]^2}~,
\ee
with 
\begin{align}
\va&=\vp'+(\vp-\vp')y~,\nn\\
M^2&=m_\pi^2+2y(1-y)(\vp^2-\vp\cdot\vp')=m_\pi^2+2y(1-y)(1-\cos\varphi)\vp^2~,
\label{l.def}
\end{align}
where
\be
\vp\cdot\vp'=\vp^2\cos\varphi~.
\label{vfi.def}
\ee
In order to apply the results already derived in Appendix~\ref{sec:l11}, 
where only one pion propagator was involved, we take into account that
\be
\frac{1}{[(\vk+\vec{\lambda})^2+M^2]^2}
=-\frac{\partial}{\partial m_\pi^2}\frac{1}{(\vk+\vec{\lambda})^2+M^2}~,
\ee
as it follows from the definition of $M^2$ in eq.~(\ref{l.def}). 
The scalar function $L_{12}$ is defined by
\begin{align}
L_{12}&=i\int\frac{d^4k}{(2\pi)^4}\frac{1}{[(\vk+\vp')^2+m_\pi^2][
(\vk+\vp)^2+m_\pi^2]}
\left[\frac{\theta(\xi_m-|\val-\vk|)}{Q^0/2-k^0-E(\val
-\vk)-i\epsilon}
+\frac{\theta(|\val-\vk|-\xi_m)}{Q^0/2-k^0-E(\val-\vk)+i\epsilon
}\right]\nn\\
&\times
\left[\frac{\theta(\xi_\ell-|\val+\vk|)}{Q^0/2+k^0-E(\val
+\vk)-i\epsilon}
+\frac{\theta(|\val+\vk|-\xi_\ell)}{Q^0/2+k^0-E(\val
+\vk)+i\epsilon}\right]~.
\label{l12}
\end{align}

For the free part 
\begin{align}
L_{12,f}&=m\frac{\partial}{\partial m_\pi^2}\int_0^1d
y\int\frac{d^3k}{(2\pi)^3}\frac{1}{[(\vk+\va)^2+M^2](\vk^2-A-i\epsilon)}~.
\end{align}
The integration over $\vk$ was already done in eq.~(\ref{l11.f}). 
Making use of this result one has
\begin{align}
L_{12,f}=-\frac{m}{8\pi
}\int_0^1dy\frac{1}{M\left(\vp^2+m_\pi^2-A-2iM\sqrt{A}\right)}~.
\label{l12.f}
\end{align}
with $M=\sqrt{M^2}$. For the part with one-medium insertion
\begin{align}
L_{12,m}&=m\int\frac{d^3k}{(2\pi)^3}\frac{\theta_m^-(\val
-\vk)}{[(\vk+\vp')^2+m_\pi^2][(\vk+\vp)^2+m_\pi^2](\vk^2-A-i\epsilon)}\nn\\
&+
m\int\frac{d^3k}{(2\pi)^3}\frac{\theta_\ell^-(\val-\vk)}{[(\vk-\vp')^2+m_\pi^2][
(\vk-\vp)^2+m_\pi^2](\vk^2-A-i\epsilon)}
\label{L12.m.def}
\end{align}
The two terms in the sum can be obtained from the function
\begin{align}
\ell_{12,m}&=m\int_0^1dy\int\meas\frac{\theta_m^-(\val
-\vk)}{(\vk^2-A-i\epsilon)[(\vk+\va)^2+M^2]^2}~.
\label{l12.m.def}
\end{align}
This integral is analogous to  $\ell_{11,m}$ in eq.~(\ref{l11.m.def.lc}), 
but with $\vp$ and $m_\pi$ replaced by $\va$ and $M$, in that order. 
Furthermore one of the factors in the denominator is squared. 
Following the calculation of $\ell_{11,m}$, we adopt the reference system
\begin{align}
\hat{\vz}&=\hat{\mathbf{a}}~,\nn\\
\hat{\vx}&=\frac{\val\times \va}{|\val| |\va| \sin\eta}~,\nn\\
\hat{\vy}&=\hat{\vz}\times \hat{\vx}=\hat{\mathbf{a}}\,\hbox{ctan}
\eta-\hat{\lambda}\,\hbox{csec}\eta~,
\end{align}
with 
\begin{align}
\val\cdot\va&=|\vp||\val|\left[(1-y)\cos\beta'+y\cos\beta\right]
=|\val||\va|\cos\eta~,\nn\\
\cos\beta'&=\hat{\vp}'\cdot \hat{\val}~,
\label{coseta}
\end{align}
so that the scalar product 
$\vk\cdot\va=|\vk||\va|(\cos\theta\cos\eta-\sin\theta\sin\phi\sin\eta)$, 
where $\theta$ and $\phi$ are the polar and azimuthal angles of $\vk$. 
Let us perform the $\phi$-integration in eq.~(\ref{l12.m.def})
\begin{align}
&\frac{1}{2\pi}\int_0^{2\pi}d\phi\frac{1}{[(\vk+\va)^2+M^2]^2}=\frac{1}{2\pi}
\int_0^{2\pi}d\phi
\frac{1}{\left[
\vk^2+\va^2+M^2+2|\vk||\va|(\cos\theta\cos\eta-\sin\theta\sin\phi\sin\eta)\right
]^2}\nn\\
&=
\frac{-1}{2|\lambda||\vk|\cos\theta}\frac{\partial}{\partial\cos\eta}\frac{1}{
2\pi}\int_0^{2\pi}d\phi
\frac{1}{
\vk^2+\va^2+M^2+2|\vk||\va|(\cos\theta\cos\eta-\sin\theta\sin\phi\sin\eta)}~.
\label{l12.fi.1}
\end{align}
The last integral is of the type already evaluated in eq.~(\ref{intfi.2}) where
now
\begin{align}
a&=\delta+2|\va||\vk|\cos\theta\cos\eta~,\nn\\
b&=-2|\va||\vk|\sin\theta\sin\eta~,
\end{align}
with $\delta=\vk^2+\va^2+M^2=\vk^2+\vp^2+m_\pi^2$ as in eq.~(\ref{abc.d}).
 Then, eq.~(\ref{l12.fi.1}) reads
\begin{align}
&\frac{1}{2\pi}\int_0^{2\pi}d\phi\frac{1}{[(\vk+\va)^2+M^2]^2}=
\frac{\delta+2|\va||\vk|\cos\eta\cos\theta}{\left[
4\va^2\vk^2(\cos^2\theta-\sin^2\eta)+4|\va||\vk|
\delta\cos\eta\cos\theta+\delta^2\right]^{3/2}}~.
\label{l12.fi.2}
\end{align}
The $\cos\theta$ integration of the previous result includes 
the Heaviside function in eq.~(\ref{l12.m.def}) 
that fixes the limits of integration to $x_1$ and $x_2$ given by 
eq.~(\ref{1m.con}). This integration is straightforward, see e.g. 
the similar one of eq.~(\ref{int.l11.m.cos}). Then, our result for $\ell_{12,m}$
is
\begin{align}
&|\val|\geq \xi_1~,\nn\\
&\ell_{12,m}=\int_{|\val|-\xi_1}^{|\val|+\xi_1}d|\vk|
\frac{|\vk|}{\vk^2-A-i\epsilon}f_{12,m}(|\vk|)~.\nn\\
&|\val| < \xi_1~,\nn\\
&\ell_{12,m}=\left\{\int_0^{\xi_1-|\val|}
+\int_{\xi_1-|\val|}^{\xi_1+|\val|}\right\}d|\vk|\frac{|\vk|}{\vk^2-A-i\epsilon}
f_{12
,m}(|\vk|)~.
\label{l12.m.final}
\end{align}
Here we have used the function
\begin{align}
f_{12,m}(|\vk|)&=\frac{m|\vk|}{(2\pi)^2}\int_0^1 dy \left.
\frac{2|\va||\vk|\cos\eta+\delta\cos\theta}{(\delta^2-4|\va|^2|\vk|^2)\sqrt{
4\va^2\vk^2(\cos^2\theta-\sin^2\eta)+4|\va||\vk|\delta\cos\eta\cos\theta+\delta^
2}}\right|_{x_1(|\vk|)}^{x_2(|\vk|)}~.
\label{f12.m}\end{align}

For the function $L_{12,m}$ of eq.~(\ref{L12.m.def}) we have $
L_{12,m}=\ell_{12,m}(\xi_m,c\beta,c\beta')+\ell_{12,m}(\xi_\ell,-c\beta,
-c\beta')$, with $c\beta'=\cos\beta'$.
\begin{align}
L_{12,d}=\frac{-im\sqrt{A}}{8\pi^2}\int d\hat{\vk}\frac{\theta(\xi_1-|\hat{\vk}
\sqrt{A}-\val|)\theta(\xi_2- |\hat{\vk}\sqrt{A}
+\val|)}{[(\vk+\vp')^2+m_\pi^2][(\vk+\vp)^2+m_\pi^2]}~.
\label{l12.d.def}
\end{align}
The angular integrations are of the same type as already developed 
for the case of the one-medium insertion and, hence, we define
\begin{align}
f_{12,d}(\sqrt{A})&=\frac{m\sqrt{A}}{(2\pi)^2}\int_0^1 dy \left.
\frac{2|\va|\sqrt{A}\cos\eta+\delta\cos\theta}{(\delta^2-4|\va|^2 A)\sqrt{4\va^2
A
(\cos^2\theta-\sin^2\eta)+4|\va|\sqrt{A}\,\delta\cos\eta\cos\theta+\delta^2}}
\right|_{x_1(\sqrt{A})}^{x_2(\sqrt{A})}~,
\end{align}
with the integration limits given in eq.~(\ref{x1x2.l11d}), where it is 
assumed that $\xi_1\leq \xi_2$. In terms of this function $L_{12,d} =
-i\pi f_{12,d}(\xi_1,\xi_2,c\beta,c\beta')$. 
When $\xi_1\geq \xi_2$ we perform, as usual, the change of variables 
$\vk\to-\vk$ in eq.~(\ref{l12.d.def}) and then $L_{12,d}=-i\pi
f_{12,d}(\xi_2,\xi_1,-c\beta,-c\beta')$.



\begin{thebibliography}{99}
\bibitem{nlou}
  J.~A.~Oller, A.~Lacour and U.-G.~Mei{\ss}ner,
  J.\ Phys.\ G {\bf 37} (2010) 015106.
\vs
\bibitem{wein}{ S.~Weinberg,} 
Physica A {\bf 96} (1979) 327.
\vs
\bibitem{wein1}  S.~Weinberg,
  Phys.\ Lett.\  B {\bf 251} (1990) 288.
  \vs
\bibitem{wein2}  S.~Weinberg,
  Nucl.\ Phys.\  B {\bf 363} (1991) 3.
\vs
\bibitem{ordo} C.~Ordonez, L.~Ray and U.~van Kolck,
  Phys.\ Rev.\  C {\bf 53} (1996) 2086.
\vs
\bibitem{kolck}U.~van Kolck,
  Prog.\ Part.\ Nucl.\ Phys.\  {\bf 43} (1999) 337.
\vs
\bibitem{entem} D.~R.~Entem and R.~Machleidt,
  Phys.\ Rev.\  C {\bf 68} (2003) 041001.
\vs
\bibitem{epe}
E.~Epelbaum, W.~Gloeckle and U.-G.~Mei\ss ner,
 Nucl.\ Phys.\  A {\bf 671} (2000)  295;  Nucl.\ Phys.\  A {\bf 747} (2005) 
362.
\vs
\bibitem{epeprl}  E.~Epelbaum, H.~Kamada, A.~Nogga, H.~Witala, W.~Gloeckle and
U.-G.~Mei\ss ner,
  Phys.\ Rev.\ Lett.\  {\bf 86} (2001)  4787.
\vs
\bibitem{eperp}E.~Epelbaum,
  Prog.\ Part.\ Nucl.\ Phys.\  {\bf 57} (2006)  654.
 \vs
  \bibitem{Epelbaum:2008ga}
  E.~Epelbaum, H.~W.~Hammer and U.-G.~Mei{\ss}ner, Rev. Mod. Phys. {\bf 81} (2009) 1773. 
  \vs
\bibitem{kaplan} D.~B.~Kaplan, M.~J.~Savage and M.~B.~Wise,
  Nucl.\ Phys.\  B {\bf 534} (1998)  329; Phys. Lett. B {\bf 424}  (1998) 390.
\vs
\bibitem{kswnnlo} S.~Fleming,   T.~Mehen and I.~Stewart, Nucl. Phys. A{\bf
677}, 
 (2000) 313; Phys. Rev. C {\bf 61} (2000) 044005.\vs
\bibitem{bean} S.~R.~Beane, P.~F.~Bedaque, M.~J.~Savage and U.~van Kolck,
  Nucl.\ Phys.\  A {\bf 700} (2002) 377.
\vs
\bibitem{timm}  A.~Nogga, R.~G.~E.~Timmermans and U.~van Kolck,
  Phys.\ Rev.\  C {\bf 72} (2005) 054006.
\vs
\bibitem{epereply}  E.~Epelbaum and U.-G.~Mei\ss ner,
  arXiv:nucl-th/0609037.
\vs
\bibitem{pavon}M.~Pavon Valderrama and E.~Ruiz Arriola,
  Phys.\ Rev.\  C {\bf 70} (2004) 044006.
\vs
\bibitem{pera}E.~Epelbaum and J.~Gegelia, Eur. Phys. J. A {\bf 41} (2009) 341.
\vs
\bibitem{soto}J.~Soto and J.~Tarrus, Phys. Rev. C {\bf 78} (2008) 024003.
\vs
\bibitem{krew} P.~Saviankou, S.~Krewald, E.~Epelbaum and U.-G.~Mei\ss ner,
  arXiv:0802.3782 [nucl-th].
\vs
\bibitem{majo}R.~Machleidt, P.~Liu, D.~R.Entem and E.~Ruiz Arriola,
arXiv:0910.3942 [nucl-th].
\vs
\bibitem{schaefer} R.~J.~Furnstahl, G.~Rupak and T.~Sch\"afer,
 Ann. Rev. Nucl. Part. Sci. {\bf 58} (2008) 1, and references therein.
\vs
\bibitem{furni}S.~K.~Bogner, R.~J.~Furnstahl, A.~Nogga and A.~Schwenk,
arXiv:0903.33661 [nucl-th].
\vs
\bibitem{nogga} S.~K.~Bogner, R.~J.~Furnstahl, S.~Ramanan and A.~Schwenk,
  Nucl.\ Phys.\  A {\bf 773} (2006) 203;   S.~K.~Bogner, A.~Schwenk,
R.~J.~Furnstahl and A.~Nogga,
  Nucl.\ Phys.\  A {\bf 763} (2005) 59.
\vs
\bibitem{kaiser}  N.~Kaiser, M.~Muhlbauer and W.~Weise,
  Eur.\ Phys.\ J.\  A {\bf 31} (2007) 53.
\vs
\bibitem{prcoller} J.~A.~Oller,
  Phys.\ Rev.\  C {\bf 65} (2002)  025204.
\vs
\bibitem{gl1} { J.~Gasser and H.~Leutwyler,}
Ann. Phys.  {\bf 158} (1984) 142.
\vs
\bibitem{sainio}J.~Gasser, M.~E.~Sainio and A.~Svarc,
  Nucl.\ Phys.\   B {\bf307} (1988)  779.
\vs
\bibitem{annp}  U.-G.~Mei\ss ner, J.~A.~Oller and A.~Wirzba,
  Annals Phys.\  {\bf 297} (2002) 27.
\vs
\bibitem{girnalda} L.~Girlanda, A.~Rusetsky and W.~Weise,
  Annals Phys.\  {\bf 312} (2004) 92.
\vs
\bibitem{kai1}  N.~Kaiser, S.~Fritsch and W.~Weise,
  Nucl.\ Phys.\  A {\bf 697} (2002)  255.
\vs
\bibitem{kai2} N.~Kaiser, S.~Fritsch and W.~Weise,
  Nucl.\ Phys.\  A {\bf 724} (2003)  47.
\vs
\bibitem{kai3} S.~Fritsch, N.~Kaiser and W.~Weise,
  Nucl.\ Phys.\  A {\bf 750} (2005)  259.
\vs
\bibitem{wirzba} M.~Kirchbach and A.~Wirzba,
  Nucl.\ Phys.\  A {\bf 616} (1997) 648.
\vs
\bibitem{lutz}M. Lutz, B.~Friman and Ch. Appel, Phys. Lett. B {\bf 474} (2000)
7.
\vs
\bibitem{hardrock}  R.~Rockmore,
  Phys.\ Rev.\  C {\bf 40} (1989) 13.
\vs
\bibitem{osetdo} M.~D\"oring and E.~Oset,
  Phys.\ Rev.\  C {\bf 77} (2008) 024602.
\vs
\bibitem{osetnie} E.~Oset, C.~Garcia-Recio and J.~Nieves,
  Nucl.\ Phys.\  A {\bf 584} (1995) 653;
 J.~Nieves, E.~Oset and C.~Garcia-Recio,
  Nucl.\ Phys.\  A {\bf 554} (1993) 509.
\vs
\bibitem{kalas} M.~M.~Kaskulov, E.~Oset and M.~J.~Vicente-Vacas,
arXiv:nucl-th/0506031 [nucl-th].
\vs
\bibitem{korean} 
T.~S.~Park, H.~Jung and D.~P.~Min,
  J.\ Korean Phys.\ Soc.\  {\bf 41} (2002) 195.
\vs
\bibitem{npa}J.~A.~Oller and E.~Oset,  Nucl.\ Phys.\  A {\bf620} (1997) 438;
(E)$-ibid$  A {\bf 652}(1999) 407.
\vs
\bibitem{nd} J.~A.~Oller and E.~Oset,
  Phys.\ Rev.\  D {\bf 60} (1999) 074023.
\vs
\bibitem{higgs}J.~A.~Oller,
  Phys.\ Lett.\  B {\bf 477} (2000) 187.
\vs
\bibitem{meis}J.~A.~Oller and U.-G.~Mei\ss ner,
  Phys.\ Lett.\  B {\bf 500} (2001) 263.
\vs
\bibitem{ericeric}M. Ericson and T.~E.~O. Ericson, Annals Phys.\  {\bf 36}
(1966)  323.
\vs
\bibitem{galrep}  E.~Friedman and A.~Gal,
  Phys.\ Rept.\  {\bf 452} (2007) 89, and references therein.
\vs
\bibitem{chanfray}  G.~Chanfray, M.~Ericson and M.~Oertel,
  Phys.\ Lett.\  B {\bf 563} (2003) 61.
\vs
\bibitem{weiprl} E.~E.~Kolomeitsev, N.~Kaiser and W.~Weise,
  Phys.\ Rev.\ Lett.\  {\bf 90} (2003) 092501.
\vs
\bibitem{panda}V.~R.~Pandharipande and R.~B. Wiringa, Rev. Mod. Phys. {\bf 51} (1979) 821.
\vs
\bibitem{day}B.~D.~Day, Phys. Rev. C {\bf 24} (1981) 1203.
\vs
\bibitem{wal}B.~D.~Serot and J.~D.~Walecka, Adv. Nucl. Phys. {\bf 16} (1986) 1.
\vs
\bibitem{polls}A.~Ramos, W.~H.~Dickhoff and A.~Polls, Phys. Lett. B {\bf 219} (1989) 15; Nucl. Phys. A {\bf 503} (1989) 1.
\vs
\bibitem{majo2}R.~Brockman and R.~Machleidt, Phys. Rev. C {\bf 42} (1990) 1965 and references therein.
\vs
\bibitem{urbana}A. Akmal, V.R.~Pandharipande, D.G.~Ravenhall, Phys. Rev. C {\bf
58} (1998) 1804.
\vs
\bibitem{pollmach}L.~Engvik, M.~Hjorth-Jensen, R.~Machleidt, H.~M\"uther and   A.~Polls, Nucl. Phys. A {\bf 627} (1995) 4396.
\vs
\bibitem{Borasoy:2006qn}
  B.~Borasoy, E.~Epelbaum, H.~Krebs, D.~Lee and U.-G.~Mei{\ss}ner,
  Eur.\ Phys.\ J.\  A {\bf 31} (2007) 105.
\vs
\bibitem{fetter}A.~L.~Fetter and J.~D.~Walecka, ``Quantum Theory of
  Many-Particle Systems'',  Dover Publications, Inc., Mineola, New York. 1971
edition. 
\vs
\bibitem{ulfrev} V.~Bernard, N.~Kaiser and U.-G.~Mei\ss ner,
  Int.\ J.\ Mod.\ Phys.\  E {\bf  4} (1995) 193. 
\vs
\bibitem{peripheral}N. Kaiser, R. Brockmann and W.~Weise, Nucl. Phys. A {\bf
625} (1997) 758.
\vs
\bibitem{selfkaiser} N.~Kaiser and W.~Weise, Phys. Lett. B {\bf 512} (2001) 283.
\vs
\bibitem{aspects}  V.~Bernard, N.~Kaiser and U.-G.~Mei{\ss}ner,
  Nucl.\ Phys.\  A {\bf 615} (1997) 483.
\vs
\bibitem{kreb1} H.~Krebs, E.~Epelbaum  and U.-G.~Mei{\ss}ner, Eur. Phys. J. A
{\bf 32} (2007) 127; Nucl. Phys. A {\bf 806} (2008) 65. 
\vs
\bibitem{techrep} J.~A.~Oller and A.~Lacour, ``{\it Chiral Effective Field
Theory for Nuclear Matter: Technical Report }''\\
http://www.um.es/oller/talks/techrep.pdf
\vs
\bibitem{report1} J.~A.~Oller, E.~Oset and A.~Ramos,
  Prog.\ Part.\ Nucl.\ Phys.\  {\bf 45} (2000) 157.
\vs
\bibitem{alba} M.~Albaladejo and J.~A.~Oller,
  Phys.\ Rev.\ Lett.\  {\bf 101} (2008) 252002.
\vs
\bibitem{ramos} E.~Oset and A.~Ramos,
  Nucl.\ Phys.\  A {\bf 635} (1998) 99~.
\vs
\bibitem{prl1}B.~Borasoy, R.~Ni{\ss}ler and W.~Weise,
  Phys.\ Rev.\ Lett.\  {\bf 94} (2005) 213401; {\it ibid}  {\bf 96} (2006)
199201.
\vs
\bibitem{prl2}J.~A.~Oller, J.~Prades and M.~Verbeni,
  Phys.\ Rev.\ Lett.\  {\bf 95} (2005) 172502;  {\it ibid} {\bf 96} (2006)
199202.
\vs
\bibitem{Borasoy:2006sr}
  B.~Borasoy, U.-G.~Mei{\ss}ner and R.~Ni{\ss}ler,
  Phys.\ Rev.\  C {\bf 74} (2006) 055201.
\vs
\bibitem{ojpa} J.~A.~Oller,
  Eur.\ Phys.\ J.\  A {\bf 28} (2006) 63.
\vs
\bibitem{nn}J.~A.~Oller,   Nucl.\ Phys.\  A {\bf 725} (2003) 85.
\vs
\bibitem{spearman} A.~D.~Martin and T.~D.~Spearman, ``Elementary Particle Theory", North-Holland Publishing Company, Amsterdam. 1970.
\vs
\bibitem{multikaiser} N.~Kaiser, Phys. Rev. C ; {\bf 61} (2000) 014003;  {\bf 62} (2001) 024001; {\bf 63} (2001) 044010; {\bf 64} (2001) 057001
\vs
\bibitem{iteratedk}N.~Kaiser, Phys. Rev. C {\bf 74} (2006) 014002.
\vs
\bibitem{landau} L.~D.~Landau, Nucl. Phys. {\bf 13} (1959) 181.
\vs
\bibitem{cutkosky} R.~E.~Cutkosky, J. Math. Phys. {\bf 1} (1960) 429.
\vs
\bibitem{barton} G. Bargon, ``Introduction to Dispersion Techniques in Field Theory", W.~A.~Benjamin, Inc, New York, Amsterdam, 1965.
\vs
\bibitem{castillejo} L.~Castillejo, R.~H.~Dalitz and F.~J.~Dyson, Phys.  Rev. {\bf 101} (1956) 453.
\vs
\bibitem{mandelstam}S.~Mandelstam, Phys. Rev. {\bf 112} (1958) 1344.
\vs
\bibitem{goldberger} R.~Blankenbecler, M.~L.~Goldberger, N.~N.~Khuri and S.~B.~Treiman, Ann. Phys. {\bf 10} (1960) 62.
\vs
\bibitem{nijmegen} V.~G.~J.~Stoks, R.~A.~M.~Klomp, M.~C.~M.~Rentmeester and
J.~J.~de Swart,
  Phys.\ Rev.\  C {\bf 48} (1993) 792; 
 V.~G.~J.~Stoks, R.~A.~M.~Klomp, C.~P.~F.~Terheggen and J.~J.~de Swart,
  Phys.\ Rev.\  C {\bf 49} (1994) 2950.
\vs
\bibitem{nijweb}NN-Online program, M.~C.~M.~Rentmeester {\it et al.},
http://nn-online.org/
\vs
\bibitem{brown} G.~E.~Brown and R.~Machleidt, Phys.  Rev. C {\bf 50} (1994)
1731.
 \vs
\bibitem{rosen}M.~E.~Rose, ``Elementary Theory of Angular Momentum'', Dover, New York,
 1995.
\vs
\bibitem{globo1} J.~P.~Blaizot, Phys. Rep. {\bf 64} (1980) 171.
\vs
\bibitem{globo2} D.~Vretenar, G.~A.~Lalazissis, R.~Behnsch, W.~P\"oschl, P.~Ring, Nucl. Phys. A {\bf 621} (1997) 853.
   \end{thebibliography}
\end{document}